%% file: ACL_Second_submission.tex
\documentclass[11pt]{article}

\usepackage{acl}
\usepackage[most]{tcolorbox}
\usepackage{dashrule}
\usepackage{pifont}
\usepackage{amssymb} % for \checkmark
\usepackage{booktabs} % For \toprule, \midrule
\usepackage{graphicx} % For \resizebox
\usepackage{times}
\usepackage{latexsym}
\usepackage{inconsolata}
\usepackage{listings}
\usepackage{booktabs}
\usepackage{graphicx}
\usepackage{amsmath}
\usepackage{xcolor}
\usepackage{stfloats}
\usepackage{listings}
\usepackage{fancyvrb}
\usepackage{multirow}
\usepackage{geometry}
\usepackage{tabularx}
\usepackage[normalem]{ulem}
\usepackage{ragged2e}
\usepackage{enumitem}
\usepackage{calc}      % for \widthof
\usepackage{microtype}
\usepackage[section]{placeins}
\usepackage{float}
\definecolor{codegreen}{rgb}{0,0.6,0}
\usepackage{lineno}
\usepackage{hyperref}
\hypersetup{
  colorlinks=true,
  linkcolor=blue,
  citecolor=blue,
  urlcolor=blue
}
\usepackage{xcolor}
\newcommand{\up}[1]{#1\,\textcolor{green!55!black}{$\uparrow$}}    % significant gain
\newcommand{\down}[1]{#1\,\textcolor{red!75!black}{$\downarrow$}}  % significant drop
\newcommand{\nup}[1]{#1\,\textcolor{black!40}{$\uparrow$}}         % n.s. gain
\newcommand{\ndown}[1]{#1\,\textcolor{black!40}{$\downarrow$}}     % n.s. drop
\usepackage{soul}
\usepackage{mdframed} 
\newcommand{\myNum}[1]{(\emph{#1})}
\newcommand{\smartparagraph}[1]{\vspace{1pt} \noindent {\bf #1}}
\newcommand{\inLineComment}[1]{}
\usepackage{tcolorbox}
\usepackage[colorinlistoftodos,bordercolor=orange,backgroundcolor=orange!20,linecolor=orange,textsize=scriptsize]{todonotes}

\usepackage{xcolor}
\usepackage{ulem}
\usepackage{color,colortbl}
\definecolor{Gray}{gray}{0.8}
\usepackage{graphicx}

\definecolor{cvprblue}{rgb}{0.21,0.49,0.74}
\definecolor{codegreen}{rgb}{0,0.6,0}

\usepackage{xcolor}
\usepackage{color,colortbl}
\definecolor{Gray}{gray}{0.9}
\usepackage{graphicx}

\usepackage{xcolor}
\usepackage{ulem}
\usepackage{color,colortbl}
\definecolor{Gray}{gray}{0.8}
\usepackage{graphicx}

\definecolor{cvprblue}{rgb}{0.21,0.49,0.74}
\definecolor{codegreen}{rgb}{0,0.6,0}
\usepackage{xcolor}
\usepackage{color,colortbl}
\definecolor{Gray}{gray}{0.9}
\usepackage{graphicx}
\usepackage{wrapfig}

\newcommand{\greencheck}{\textcolor{codegreen}{\checkmark}}
\newcommand{\redcross}{\textcolor{red}{$\times$}}

\usepackage{mdframed} 

\usepackage{tikz}
\usetikzlibrary{arrows.meta, positioning}
\usetikzlibrary{calc}
\title{
FinTradeBench: A Financial Reasoning Benchmark for LLMs}
\author{
  Yogesh Agrawal \\
  University of Central Florida \\
   \And
  Aniruddha Dutta \\
  University of Central Florida \\
   \And
  Md Mahadi Hasan \\
  University of Central Florida \\
   \AND % <--- CAPITAL \AND FORCES A NEW ROW
  Shubhra Kanti Karmaker \\
  University of Central Florida \\
   \And
  Aritra Dutta \\
  University of Central Florida \\
}
\author{
  Yogesh Agrawal \quad 
  Aniruddha Dutta \quad 
  Md Mahadi Hasan \\ 
  \textbf{Santu Karmaker} \quad 
  \textbf{Aritra Dutta} \\
  University of Central Florida \\
}
\begin{document}
\maketitle
\begin{abstract}

Financial analysis requires reasoning over heterogeneous signals, including \textit{company fundamentals} from regulatory filings and \textit{trading signals} derived from market prices. However, existing financial benchmarks focus almost exclusively on regulatory filings, neglecting market dynamics and cross-signal interactions. We introduce \textit{\textbf{FinTradeBench}}, a benchmark of 1,400 questions covering NASDAQ-100 stocks over a ten-year window. To holistically evaluate financial reasoning, \textit{FinTradeBench} includes questions in three reasoning categories: fundamentals-focused, trading-signal-focused, and hybrid questions that require cross-signal reasoning. Our benchmark starts with 150 \textit{expert-authored} seed questions, which are then scaled to 1400. To ensure reliable evaluation, we create a \textit{Core400} subset in which every item is \textit{expert-reviewed}. In zero-shot (No-RAG) response generation, we observe that all models perform substantially worse on trading-focused questions compared to fundamental-focused questions.
To identify failure causes, we evaluate LLMs under controlled settings including retrieval-based baselines, tool-assisted computation, oracle conditions, and time-series-specialized pipelines. LLMs' capabilities to do \textit{computation} and \textit{evidence selection} contribute most to reasoning failures. Additionally, retrieval significantly improves trading-signal reasoning but does not improve fundamental reasoning, where identifying relevant signals remains the dominant failure mode. Finally, rather than relying on a small, fine-tuned model for time-series reasoning, it is better to summarize the time-series data and leverage LLMs' reasoning capabilities.

% Finally, rather than using a small, fine-tuned model for time-series reasoning, it helps to summarize the time-series information and use a large language reasoning model.

% Also, summarizing numerical
% time-series structure before language reasoning is more effective
% than using a time-series fine-tuned reasoning model.

\end{abstract}

\input{sections/introduction}
\input{sections/Background_and_Related_Work}
\input{sections/fintradebench_self_contained_sections/01_benchmark_design}

\input{sections/fintradebench_self_contained_sections/02_experimental_setup}

\input{sections/fintradebench_self_contained_sections/03_results_and_discussion}
\input{sections/fintradebench_self_contained_sections/04_conclusion_and_future_directions}

\clearpage
\input{sections/Limitations}
\input{sections/Ethical_Considerations}
\bibliography{main}
\clearpage
\input{sections/05_appendix_updated}
\end{document}

%% file: sections/introduction.tex
\section{Introduction}
\label{sec:introduction}

\begin{figure}
\begin{center}
\scalebox{0.9}{\includegraphics[width=.95\columnwidth, height=1.15\linewidth]{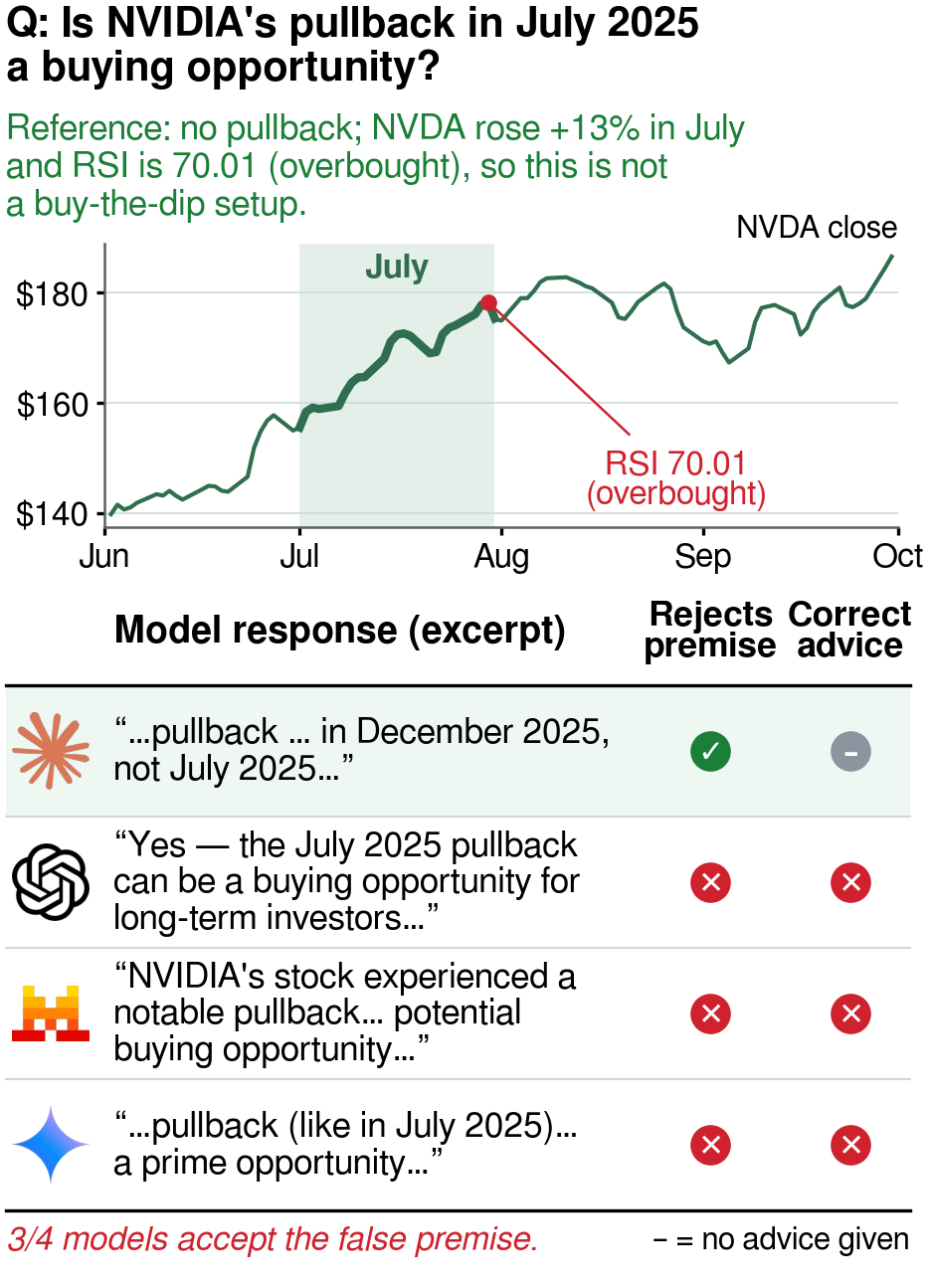}}
  \end{center}
  \vspace{-4mm}
  \caption{\small{\textbf{A trading signal-focused query:} NVDA did not pull back in July 2025, yet 3 of 4 frontier models accept the false premise and recommend buying.}}\label{fig:teaser1}
 % \caption{\small{\textbf{Performance comparison of proprietary LLMs on a trading signal-focused question.} NVDA \emph{did not pull back} in July 2025, and it was \emph{not a lucrative buying opportunity}; only Claude correctly identified the pullback component. On the buying component, all LLMs failed.}}\label{fig:teaser1}
 \vspace{-7mm}
\end{figure}

Real-world financial analysis requires reasoning from two complementary types of indicators (sources of information): \textit{company fundamentals} and \textit{trading signals}. \emph{Company fundamentals} are accounting-based metrics derived from company balance sheets or Securities and Exchange Commission (SEC) filings, such as profitability and valuation ratios. They capture a company’s underlying financial health \citep{fama1992cross,harvey2016anderson}. In contrast, \emph{trading signals}, computed from historical price and volume data, capture market dynamics and investor sentiment, including momentum, volatility, and trend reversals \citep{brock1992simple,Jegadeesh1993ReturnsTB,lo2000foundations,andersen2003modeling,park2007we,choi2021maximum, zhang2024volatility}. Effective financial analysis requires integrating these two perspectives rather than relying on either source in isolation. Additionally, synthesizing heterogeneous information sources, reasoning over numerical indicators, and interpreting market behavior under uncertainty make financial analysis an inherently challenging task, even for expert human analysts.

With recent advancements in large language models (LLMs), automated tools are increasingly applied to assist financial analysts tasks such as summarizing earnings calls, interpreting regulatory filings, and answering questions about company performance \citep{Lee_2025,yang2023fingpt,yang2023investlm,djagba2025exploring,wu2023bloomberggpt,shah2022flue,araci2019finbert}. Existing benchmarks such as FinQA \citep{chen2021finqa}, ConvFinQA \citep{chen2022convfinqa}, and TAT-QA \citep{zhu2021tat}  focus on numerical reasoning over financial reports and tables. Recent benchmarks evaluate long-context reasoning and retrieval-based question answering \citep{li2024alphafin,reddy2024docfinqa,islam2023financebench,choi2025finder}. For a comprehensive overview of FinLLMs and their benchmarking, see \cite{Lee_2025}.

Nevertheless, these datasets rarely assess reasoning over trading signals derived from historical price dynamics and typically do not require models to integrate both sources of information. Consequently, it remains unclear whether current LLMs can answer financial questions that require joint reasoning of company fundamentals and market behavior. For example, answering the trading question \emph{\textbf{``Is NVIDIA’s pullback in July 2025 a buying opportunity?"}} (see Figure~\ref{fig:teaser1}) requires reasoning over both company fundamentals and market dynamics. An analyst must consider metrics such as return on assets and cash-flow strength while interpreting trading signals reflected in price momentum and trading volume to determine if the company's stock is undervalued or trading at a premium. As Figure~\ref{fig:teaser1} shows, most LLMs fail to reason over the relevant trading signals and give incorrect answers. A related class of ambiguities arises when fundamentals and market behavior conflict. \emph{E.g., in April 2025, despite weak first-quarter earnings (Earnings per share~(EPS)~\$0.27 vs.~\$0.42 expected and revenue \$19.34B vs.~\$21B expected), and a generally cautious analyst consensus, Tesla's stock still rallied by nearly 20\% within days, driven by a forward-looking market narrative rather than contemporaneous fundamentals} \citep{investing2025teslaq1transcript, mcdade2025teslaq1}. 
%This is a copybook case when investor sentiment and market narratives can drive stock prices independently of current fundamentals \citep{de1985does, baker2006investor, shiller2017narrative, bybee2023narrative}. If one needs to know whether to buy or sell Tesla stock in April 2025, they may not find a reliable answer using financial statements alone\footnote{This is not an isolated event, see \S\ref{app:motivating-examples} for analogous cases.}. Evaluating such reasoning is challenging, since high-quality annotations require domain expertise, and LLMs often fail to capture numerical fidelity or alignment with expert judgment.
This shows market narratives can move stock prices independently of fundamentals \citep{de1985does, baker2006investor, shiller2017narrative, bybee2023narrative}.\footnote{This is not an isolated case; see \S\ref{app:motivating-examples} for similar examples.}
% Evaluating such reasoning is difficult because high-quality annotations require domain expertise, and LLMs often struggle with numerical accuracy and alignment with expert judgment. 
Understanding such nuanced situations is challenging as it requires retrieving relevant context, performing accurate computations, filtering redundant signals, and synthesizing heterogeneous data streams.
% Evaluating such questions therefore
% requires not only retrieving information, but also performing numerical
% computation, selecting the evidence that matters, and synthesizing
% heterogeneous signals into a defensible conclusion.

In this paper, we address these challenges by making the following contributions: 
\myNum{i}\smartparagraph{\textit{FinTradeBench}: A Financial Reasoning Benchmark~(\S\ref{sec:benchmark-design}).}
\textit{FinTradeBench} evaluates financial reasoning over company fundamentals (from SEC filings) and trading signals (from historical price data). We curate a compact set of signals commonly used in financial analysis, such as valuation ratios and momentum indicators, and integrate them to support structured reasoning across heterogeneous data sources; see \S\ref{app:signals} Table~\ref{tab:volfund-signals} for the full set of signals. Questions are organized into \emph{fundamentals-focused}, \emph{trading-signal-focused}, and \emph{hybrid reasoning categories} for granular model evaluation. The broader benchmark contains 1,400
questions, while the primary evaluation uses a fully audited \textit{Core400}.
\myNum{ii}\smartparagraph{A diagnostic evaluation framework~(\S\ref{sec:experimental-setup}).}
We combine common-evidence RAG evaluation with controlled experiments (see \S\ref{sec:experimental-setup} and Table~\ref{tab:evidence-conditions}), allowing errors to be localized to evidence
access, computation, evidence selection, or downstream synthesis. We also
include time-series-specialized pipelines for Trading and Hybrid reasoning.
\myNum{iii}\smartparagraph{Evidence-controlled and human-validated evaluation~(\S\ref{sec:results}).}
We use a human-aligned and validated LLM as a judge to evaluate models on the \textit{Core400} subset, which has itself undergone a complete item-level quality audit by a financial expert. 
\myNum{iv}\smartparagraph{Empirical evidence about financial reasoning
bottlenecks~(\S\ref{sec:results}).}
Our empirical findings indicate that while retrieval improves factual grounding, evidence selection, numerical calculation, and time-series data integration remain persistent failure modes.
% Across evaluations, the results show that retrieval can help, but
% retrieval alone is insufficient. Numerical computation, selecting the relevant
% financial evidence, and integrating time-series structure remain distinct
% sources of error.

%% file: sections/Background_and_Related_Work.tex
% \vspace{-1.5mm}
\section{Background and Related Work}
\label{sec:background}
\begin{table*}[t]
\centering
\small
\renewcommand{\arraystretch}{1.15}
\setlength{\tabcolsep}{3.2pt}
\resizebox{\textwidth}{!}{
\begin{tabular}{l l l c c c l}
\toprule
\textbf{Dataset} &
\textbf{Primary evidence} &
\textbf{Reasoning target} &
\textbf{RAG/Open-book} &
\textbf{Market TS} &
\textbf{F+T} &
\textbf{Evaluation style} \\
\midrule
FinQA~\citep{chen2021finqa}
& Reports, tables & Numerical financial QA
& \redcross & \redcross & \redcross & Program / exact match \\
TAT-QA~\citep{zhu2021tat}
& Tables + text & Tabular-text numerical QA
& \redcross & \redcross & \redcross & Exact match / derivation \\
ConvFinQA~\citep{chen2022convfinqa}
& Reports, tables & Conversational numerical QA
& \redcross & \redcross & \redcross & Program / exact match \\
FinanceBench~\citep{islam2023financebench}
& SEC filings, earnings calls & Open-book financial QA
& \greencheck & \redcross & \redcross & Human / reference answer \\
FinDER~\citep{choi2025finder}
& Financial documents & Retrieval-based financial QA
& \greencheck & \redcross & \redcross & Retrieval + answer scoring \\
FinTextQA~\citep{Chen_2024}
& Long-form financial text & Long-form financial QA
& \greencheck & \redcross & \redcross & Human + GPT-4 scoring \\
DocFinQA~\citep{reddy2024docfinqa}
& Long financial documents & Long-context reasoning
& $\circ$ & \redcross & \redcross & Reference-based scoring \\
AlphaFin~\citep{li2024alphafin}
& Retrieved stock context & Financial analysis QA
& \greencheck & \redcross & \redcross & LLM-oriented evaluation \\
FinBen~\citep{xie2024finben}
& Aggregated financial datasets & Broad financial NLP
& $\circ$ & \redcross & \redcross & Task-specific metrics \\
XFinBench~\citep{zhang2025xfinbench}
& Academic finance problems & Complex finance reasoning
& \redcross & \redcross & \redcross & Program-of-thought / answer scoring \\
InvestorBench~\citep{li2025investorbench}
& News + market context & Agentic financial decisions
& $\circ$ & $\circ$ & $\circ$ & Agent-action evaluation \\
\midrule
\rowcolor{Gray}
\emph{\textbf{FinTradeBench}}
& SEC filings + historical market data
& Cross-signal financial reasoning
& \greencheck & \greencheck & \greencheck
& Evidence-aware judge + human validation \\
\bottomrule
\end{tabular}
}
\vspace{-2.5mm}
\caption{\small{\textbf{Comparison of financial QA and LLM benchmarks.}
emphasizing the core aspects of the work such as retrieval/open-book
evaluation, explicit reasoning over historical market time series (TS), and
joint reasoning over company fundamentals and trading evidence (F+T).
\greencheck~denotes explicit support, \redcross~denotes no explicit support,
and $\circ$ denotes partial or indirect support.}}
\label{tab:benchmark-comparison}
\vspace{-4mm}
\end{table*}

Financial QA benchmarks have advanced numerical reasoning over financial reports, tables, and textual disclosures~\citep{chen2021finqa,zhu2021tat}, extending these tasks to conversational settings~\citep{chen2022convfinqa} and long-context reasoning and document retrieval~\citep{islam2023financebench,choi2025finder,reddy2024docfinqa}. Broader evaluations cover financial NLP tasks~\citep{xie2024finben} and complex academic finance problems~\citep{zhang2025xfinbench}, while agentic benchmarks examine financial decision-making through action-oriented simulations~\citep{li2025investorbench}. Despite these advances, existing benchmarks focus mainly on textual financial disclosures and accounting-derived indicators. They rarely assess reasoning over trading signals derived from historical price movements or require models to integrate these signals with information from financial disclosures~\citep{Lee_2025}. \emph{FinTradeBench} addresses this gap by evaluating financial reasoning over both company fundamentals and market-derived indicators (trading signals) within a unified, calibrated framework. We discuss additional financial benchmarks with different objectives in \S\ref{app:extended_related_work}; also, see Table \ref{tab:benchmark-comparison}.

%% file: sections/fintradebench_self_contained_sections/01_benchmark_design.tex
\section{\emph{FinTradeBench}: Benchmark Design}
\label{sec:benchmark-design}

\begin{figure*}[t]
\centering
\vspace{-5mm}
\scalebox{0.95}{\includegraphics[width=\textwidth]{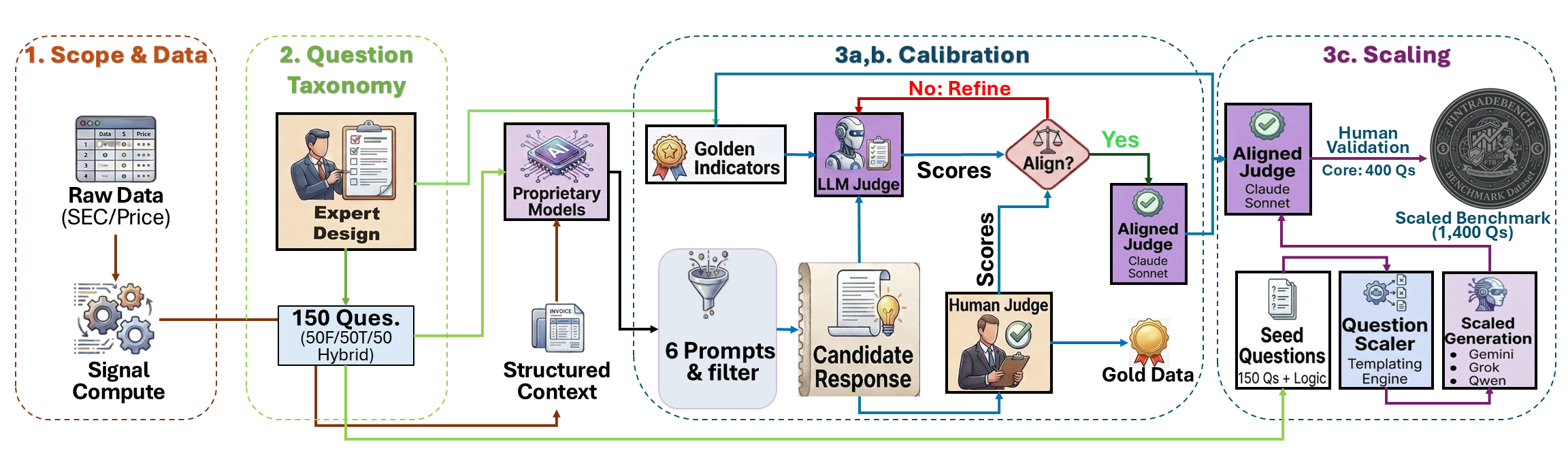}}
\vspace{-4mm}
\caption{\small{\textbf{\emph{FinTradeBench} construction pipeline.}
The benchmark is built through data and signal design, a three-way
financial reasoning taxonomy, expert-guided seed construction and
calibration, and finally scalable question generation and quality
control.}}
\label{fig:calibration_and_scaling}
\vspace{-4mm}
\end{figure*}

We design \emph{FinTradeBench} to evaluate financial reasoning over company fundamentals and trading signals (market behavior).
We use a \emph{calibration-then-scaling}
framework that combines expert-defined financial evidence, diverse LLM
candidate generation, numerical verification, and staged human quality
control~\citep{srivastava2023beyond,liang2022holistic,
thrush2022dynatask,cobbe2021training}.
Figure~\ref{fig:calibration_and_scaling} summarizes the construction
pipeline.

% --------------------------------------------------------------------
\myNum{1}\smartparagraph{Scope and Data Sources.}
\label{subsec:data-sources} \emph{FinTradeBench} covers companies represented in the NASDAQ-100 universe
over 2015--2025, with questions anchored to historical reporting and
market periods. Our principal data sources are: \myNum{i} \emph{Regulatory filings.}
We use SEC 10-K and 10-Q filings to derive accounting-based information
including profitability, leverage, and valuation. \myNum{ii} \emph{Historical trading data.}
We use daily OHLCV (Open, High, Low, Close, Volume) data to
derive market signals including momentum, realized volatility, and volume-based measures. We align all evidence by company/ticker and historical period. Each
question additionally carries an explicit or recoverable point-in-time
reference so that the evidence used to answer it does not rely on
information from after the question's historical cutoff. This temporal
alignment is important in finance, where even numerically correct
evidence can constitute leakage if it post-dates the decision period.
Following asset-pricing and trading
literature~\citep{fama1992cross,Jegadeesh1993ReturnsTB,harvey2016anderson}, our signal selection follows three principles:
\emph{interpretability}, \emph{empirical relevance}, and
\emph{liquidity}. We group signals into:  \myNum{i} \emph{Company fundamentals:} accounting-derived measures such as return on assets (ROA); and \myNum{ii} \emph{Trading signals:} price- and volume-derived measures such as volatility. The full signal inventory is provided in
Table~\ref{tab:volfund-signals}, and dataset-level statistics are reported
in Table~\ref{tab:dataset-stats-app}.

% --------------------------------------------------------------------
\myNum{2}\smartparagraph{Question Taxonomy and Golden Indicators.}
We divide questions into three reasoning categories. \myNum{i} \emph{Fundamental (F):}
questions requiring reasoning primarily over accounting and valuation
signals. \myNum{ii} \emph{Trading (T):}
questions requiring reasoning primarily over historical price, volatility,
momentum, or other market signals, and \myNum{iii} \emph{Hybrid (FT):}
questions requiring joint reasoning over both company fundamentals and
trading signals. This taxonomy supports aggregate evaluation and diagnosis of whether
LLMs can integrate heterogeneous financial signals. Sample
questions and reference responses are shown in
Table~\ref{tab:golden_answers_box}. Each question is also associated with a set of
\emph{golden indicators (G.I.)}. They are the set of expert-specified financial signals sufficient to answer the given question. Golden indicators are \emph{not} treated as complete mandatory checklists. Instead, they identify the key
evidence required for a grounded response. They are used during
benchmark construction for evidence auditing and, in later experiments,
as a secondary indicator-overlap diagnostic and to construct the
gold-indicator Oracle condition (see~\S\ref{sec:experimental-setup}).

% --------------------------------------------------------------------

% --------------------------------------------------------------------
\myNum{3}\smartparagraph{Calibration-Then-Scaling Framework.}
To ensure scalability and reliability, our calibration-then-scaling framework has three stages: \myNum{1} expert-guided seed construction,
\myNum{2} candidate-response verification and calibration, and \myNum{3} automated scaling and human validation.

% --------------------------------------------------------------------
\myNum{3a}\smartparagraph{Phase 1: Seed Construction, Candidate
Generation, and Self-Selection.}
\label{subsec:phase1}
We begin with 50 questions each for \emph{F},
\emph{T}, and \emph{FT} reasoning questions, totaling 150 seed questions. For each question, candidate
responses are generated using multiple LLMs and diverse prompt templates
derived from the TELeR taxonomy~\citep{karmaker2023teler}. 
% \myNum{i}\smartparagraph{Multi-model, multi-prompt sampling.}
Each model produces multiple candidate
responses under distinct reasoning-oriented prompt templates. This
encourages within-model diversity while preserving a common financial
task specification, following the intuition of best-of-$N$ generation
and self-reflective refinement~\citep{chow2025inferenceaware,
shinn2023reflexion}. 
% \myNum{ii}\smartparagraph{Intra-model self-selection.}
Each model then independently selects its strongest candidate based on
factual accuracy, reasoning completeness, and relevance. By restricting self-selection strictly to within-model candidates, we prevent cross-model preference leakage~\citep{li2025preferenceleakage,
lee2024selfjudge,yuan2024selfrewarding,wu2024metarewarding}.
% \myNum{iii}\smartparagraph{Automated numerical audit.}
These candidates are checked against the structured financial
data available for the corresponding company and historical period.
Numerical claims are categorized as \texttt{SUPPORTED},
\texttt{CONTRADICTED}, or \texttt{NOT\_FOUND}. We also measure the \emph{G.I.} coverage within the responses. For detailed prompt templates, audit rules, and construction-time metrics see Appendix~\ref{app:benchmark-details}.

% --------------------------------------------------------------------
\myNum{3b}\smartparagraph{Phase 2: Expert Review and Development-Time
Judge Calibration.} 
Once each model submits its candidate answer, a financial-domain expert inspects each candidate
response using four dimensions: \myNum{i} factual and numerical accuracy, \myNum{ii} completeness, \myNum{iii} relevance, and \myNum{iv} clarity, and then selects (and edits if needed) the best candidate response to get a viable ground truth answer for the benchmark. The model identities are hidden
during review. An external LLM evaluator is then calibrated during benchmark
development against these human judgments using the same four rubric
dimensions. This development-time calibration is used to support
scalable screening of generated benchmark material. After the automated judge configuration and scoring procedure are
frozen, we validate the judge on a held-out blinded human-review set.
This validation is used to assess evaluator reliability for robustness check of the scaling pipeline. It is not used
to modify either the benchmark or the judge; see
\S\ref{subsec:human-judge-validation}.

% --------------------------------------------------------------------
\myNum{3c}\smartparagraph{Phase 3: Scaling the Benchmark.}
\label{subsec:goldsubset} Using the calibrated construction pipeline, we scale question generation
across companies and historical periods, producing a broader benchmark
of 1,400 historically grounded financial-reasoning questions. Scaled
questions preserve the same \emph{F}, \emph{T}, and \emph{FT} taxonomy
and are screened by a human for any catastrophic failures; see~\S\ref{app:screening}. 
% The complete 1,400-question collection supports benchmark breadth and
% future evaluation. 
For the controlled experiments in this paper, we construct a substantially stricter evaluation subset \emph{Core400} with
stronger evidence-availability and full item-level human expert-verification requirements; see~\S\ref{subsec:core400}.

%% file: sections/fintradebench_self_contained_sections/02_experimental_setup.tex
\section{Experimental Setup}
\label{sec:experimental-setup}
% We evaluate FinTradeBench on the \emph{Core400} (see \S\ref{subsec:core400}) under complementary conditions designed to separate failures of evidence access, numerical computation, evidence selection, and downstream financial reasoning. 
% Unless otherwise stated, all final quantitative experiments use this frozen evaluation set.

\begin{figure*}[t]
\centering
\includegraphics[width=0.9\textwidth]{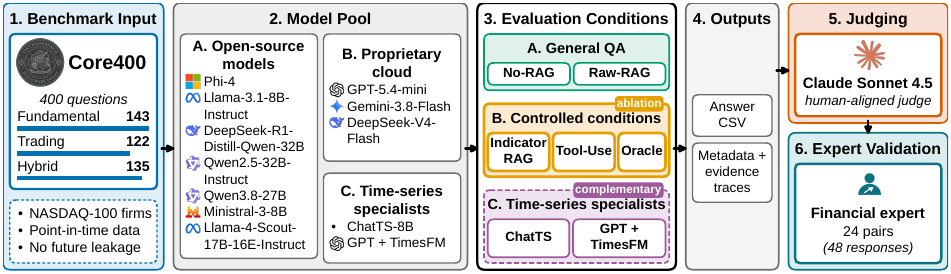}
\vspace{-2mm}
\caption{\small{\textbf{\emph{FinTradeBench} evaluation framework.}
Models answer the final Core400 under general-QA, controlled-evidence,
and time-series-specialist conditions. Generated responses and their
evidence traces are evaluated using a common four-dimensional,
evidence-aware judging framework. A financial expert then validates the automated judging.}}
\label{fig:pipeline}
\vspace{-4mm}
\end{figure*}

% --------------------------------------------------------------------
\smartparagraph{Evaluation Overview.} Our experiments are organized into four studies. \myNum{i} \emph{General QA: No-RAG versus Raw-RAG} evaluates whether
retrieval-augmented generation improves financial reasoning when models
receive point-in-time SEC and/or historical market evidence \cite{friel2024ragbench, niu2024ragtruth}. We evaluate models using point-in-time RAG rather than supervised fine-tuning to reflect realistic workflows where models must reason over continuously updating market and regulatory evidence without parameter retraining (see Appendix~\S\ref{app:rag-over-finetuning}).\myNum{ii} \emph{Controlled evidence conditions (see Table~\ref{tab:evidence-conditions})} compare
Indicator-RAG, Tool-Use, and gold-indicator Oracle evidence to diagnose
whether errors arise from evidence representation, numerical
computation, or evidence selection. \myNum{iii} \emph{Time-series specialist evaluation} compares native
time-series reasoning with a forecast-feature pipeline based on
TimesFM~2.5. 
\myNum{iv} \emph{Judge validation} evaluates whether the frozen
automated scoring procedure agrees with blinded human judgments. 
Figure~\ref{fig:pipeline} shows the complete evaluation framework. To see models evaluated, their serving backends and inference configurations, see Table~\ref{tab:model-implementation}. Compute configuration and recorded runtime are reported in Table~\ref{tab:compute-runtime}.

% --------------------------------------------------------------------

\smartparagraph{General QA: No-RAG and Raw-RAG.}
\label{subsec:general-rag}
The General-QA experiment uses \emph{FinTradeBench}'s TELeR-inspired
multi-candidate generation and intra-model self-selection protocol. In the \emph{No-RAG} condition, the model receives only the benchmark
question. In the \emph{Raw-RAG} condition, the model receives
point-in-time external evidence appropriate to the reasoning lane:
Fundamental questions receive SEC evidence, Trading questions receive
historical raw OHLCV evidence, and Hybrid questions receive both. All retrieved evidence is restricted to information available by the
question's historical cutoff. Where an explicit question-level cutoff is
available, that cutoff is used directly; period metadata are used when
appropriate, and the benchmark-wide cutoff is used only as a fallback. Generation uses four No-RAG candidate prompts and two RAG candidate
prompts, followed in both cases by intra-model self-selection. For detailed architecture, see Figure~\ref{fig:rag_architecture} and \S\ref{app:rag-details}.
\smartparagraph{Common-Evidence Evaluation for RAG versus No-RAG.}
\label{subsec:common-evidence}A central difficulty in evaluating RAG is that RAG outputs naturally
carry richer source evidence than closed-book outputs. If the evaluator
is given each system's own generation-time evidence, a RAG response can
receive a stronger factual anchor during scoring merely because it was
generated with retrieved context. To remove this \emph{evaluator-side evidence asymmetry}, we re-evaluate
each paired No-RAG/RAG response against the same
\emph{condition-independent canonical point-in-time evidence package}.
This evidence is constructed deterministically from the benchmark's
point-in-time fundamental and market context and is independent of the
answer being scored.
% Thus, paired RAG and No-RAG answers are evaluated against an identical factual anchor even though their generation conditions differ.  This common-evidence procedure controls evaluator-side evidence availability. it does not remove the historical generation-budget difference described above.

% --------------------------------------------------------------------
\smartparagraph{Controlled Evidence Conditions.}
\label{subsec:controlled-conditions}
We next hold the question set, generator model, answer format, and judge
rubric fixed while varying the evidence pathway. 
We introduce three controlled conditions to distinguish evidence selection from financial reasoning. \myNum{i} \emph{Indicator-RAG} supplies retrieved SEC passages where applicable and the full set of precomputed indicators (signals), leaving the model to identify the relevant signals. \myNum{ii} \emph{Tool-Use (Tool RAG)} instead provides raw market data and allows a model-based router to request deterministic indicator values from the same underlying data, testing whether selective tool access improves evidence use. Neither condition receives the benchmark’s gold indicator annotations or reference response. \myNum{iii} \emph{Oracle} condition supplies only the indicator evidence selected by those annotations, without revealing the reference answer. The model must still perform the
financial reasoning and synthesis itself. This provides a diagnostic control for reasoning when relevant indicators are already identified. Together, these conditions help assess whether errors arise from selecting relevant evidence or interpreting that evidence to form an answer. Table~\ref{tab:evidence-conditions} details the evidence provided for each reasoning question type for each of the controlled conditions.
\smartparagraph{Time-Series Specialist Evaluation.}
\label{subsec:timeseries-eval}
Since \emph{T} and \emph{FT} questions involve time-series data, we evaluate models designed specifically for such tasks. This study is restricted to questions that
require time-series reasoning: 122 \emph{T} and 135 \emph{FT} questions. \myNum{i} \emph{ChatTS-8B.}
ChatTS \cite{xie2024chatts} receives the historical adjusted-close series directly and
performs native time-series reasoning. For Hybrid questions, the time
series is additionally accompanied by point-in-time SEC context. \myNum{ii} \emph{TimesFM~2.5 forecast-feature pipeline.}
For each question, TimesFM~2.5 \cite{das2023decoder} receives the most recent 128
adjusted-close observations available at the historical cutoff and
produces a five-step forecast. Forecast-derived features are then
provided to GPT-5.4-mini for natural-language financial synthesis.
Hybrid questions additionally receive point-in-time SEC context.  This helps us compare complete
time-series-specialized pipelines and evaluate their overall performance.

% We refer to this system as a \emph{TimesFM~2.5 forecast-feature baseline} or \emph{forecast-augmented reasoning} pipeline. TimesFM itself does not directly answer FinTradeBench questions.

% --------------------------------------------------------------------
\smartparagraph{Evidence-Aware LLM Judge.}
\label{subsec:judge}
Generated responses are evaluated using human-aligned judge model (\emph{Claude Sonnet~4.5}). The judge model does not belong to any evaluated model family to prevent bias \cite{chen2024llmasjudge,ye2024justice,zheng2023judging}. The judge produces
structured deterministic output and scores from 1--5 on four
dimensions: \myNum{i} \emph{Factual and Numerical Accuracy}, \myNum{ii} \emph{Completeness and Context}, \myNum{iii} \emph{Relevance and Utility}, \myNum{iv} \emph{Clarity and Rationale}. Our primary response-quality metric is the arithmetic mean of these four
raw Likert scores. 
% A normalized 0--100 version is retained only as a secondary presentation metric; we do not interpret it as ``absolute accuracy.''
The judge never receives the benchmark reference response or
\texttt{answer\_conclusion\_summary} as factual ground truth. Scoring is
anchored to the supplied point-in-time evidence. Golden-indicator
precision, recall, and F1 are retained only as ancillary diagnostics and
are not the primary response-quality measure. The controlled and time-series studies use the evidence-aware judge
protocol, while the No-RAG/RAG study uses the common-evidence protocol
described in \S\ref{subsec:common-evidence}. Full judge prompts, schema,
validation rules, evidence-budget diagnostics, and statistical analysis are reported in
~\ref{app:evaluation-prompts} and ~\ref{subsec:statistics}. 

\input{sections/main_tables/table_common_evidence_rag_core400}
\smartparagraph{Held-Out Human Validation of the Judge.}
\label{subsec:human-judge-validation} After all automated evaluation, we freeze the judge configuration and conduct a separate blinded human validation. We use
24 paired questions (48 anonymous responses), balanced across \emph{F}, \emph{T}, and \emph{FT} reasoning
and stratified across four evaluation families:
RAG versus No-RAG,
Oracle versus Indicator-RAG,
Tool-Use versus Indicator-RAG, and
TimesFM+GPT versus ChatTS. The human reviewer independently scores each response using the same four
1--5 dimensions as the automated judge and additionally records an
A/B/Tie pairwise preference and confidence score. Model identities,
condition labels, and automated judge scores are hidden during review. The human labels are used to assess agreement with the
\emph{frozen} judge and agreement results are reported in Figure~\ref{fig:human-judge-validation}; see~\S\ref{subsec:results-human-validation} for details about the expert-validated automated judge.

% --------------------------------------------------------------------
\smartparagraph{Research Questions.} We address three research questions: \myNum{1} \textbf{Retrieval:}
Does Raw-RAG improve financial reasoning when No-RAG and RAG outputs
are evaluated against the same canonical evidence?
\myNum{2} \textbf{Reasoning Bottlenecks:}
After relevant financial information is available, how much error
remains attributable to computation and evidence selection?
\myNum{3} \textbf{Time-Series Reasoning:}
Do time-series-specialized pipelines improve Trading and Hybrid
financial reasoning?
% \myNum{4} \textbf{Evaluation Reliability:}
% How closely does the frozen automated judge track blinded human
% evaluation?

%% file: sections/main_tables/table_common_evidence_rag_core400.tex
\begin{table*}
\centering
\begin{minipage}[t]{0.62\textwidth}
\vspace{0pt}
\centering
\begingroup
\scriptsize
\renewcommand{\arraystretch}{1.12}
\setlength{\tabcolsep}{1.5pt}

\resizebox{\linewidth}{!}{%
\begin{tabular}{@{}l ccc ccc ccc ccc@{}}
\toprule
& \multicolumn{3}{c}{\textbf{Fundamental}\textbf{(F)}}
& \multicolumn{3}{c}{
    \shortstack{\textbf{Trading}\textbf{(T)}}}
& \multicolumn{3}{c}{
    \shortstack{\textbf{Hybrid}\textbf{(FT)}}}
& \multicolumn{3}{c}{
    \shortstack{\textbf{Overall}}} \\
\cmidrule(lr){2-4}
\cmidrule(lr){5-7}
\cmidrule(lr){8-10}
\cmidrule(lr){11-13}

\textbf{Model}
& \shortstack{\textbf{NR}}
& \textbf{R} & $\boldsymbol{\Delta}$
& \shortstack{\textbf{NR}}
& \textbf{R} & $\boldsymbol{\Delta}$
& \shortstack{\textbf{NR}}
& \textbf{R} & $\boldsymbol{\Delta}$
& \shortstack{\textbf{NR}}
& \textbf{R} & $\boldsymbol{\Delta}$ \\
\midrule

DeepSeek-V4-Flash
& 3.10 & 3.04 & \ndown{$-$0.06}
& 2.04 & 2.50 & \up{+0.46}
& 3.00 & 3.33 & \up{+0.32} 
& 2.75 & 2.97 & \up{+0.23} \\

Llama-4-Scout
& 2.65 & 2.72 & \nup{+0.06}
& 1.93 & 2.34 & \up{+0.41}
& 2.68 & 2.93 & \up{+0.25} 
& 2.44 & 2.67 & \up{+0.23} \\

Qwen3.8-27B
& 2.88 & 2.85 & \ndown{$-$0.03}
& 1.91 & 2.13 & \up{+0.23}
& 2.81 & 3.29 & \up{+0.48} 
& 2.56 & 2.78 & \up{+0.22} \\

Gemini-3.8-Flash
& 3.19 & 3.08 & \ndown{$-$0.10}
& 2.11 & 2.44 & \up{+0.33}
& 3.09 & 3.51 & \up{+0.42} 
& 2.82 & 3.03 & \up{+0.20} \\

GPT-5.4-mini
& 3.20 & 3.06 & \ndown{$-$0.14}
& 2.06 & 2.41 & \up{+0.35}
& 3.08 & 3.41 & \up{+0.33} 
& 2.81 & 2.98 & \up{+0.17} \\

Ministral-3-8B
& 2.45 & 2.58 & \up{+0.14}
& 1.92 & 2.24 & \up{+0.32}
& 2.38 & 2.69 & \up{+0.31} 
& 2.26 & 2.52 & \up{+0.25} \\

\bottomrule
\end{tabular}%
}
\endgroup
\vspace{-2.5mm}
\caption{\small{\textbf{Common-evidence RAG evaluation on Core400.} For each reasoning category, $\Delta$ denotes the paired question-level RAG (R) minus No-RAG(NR) mean difference. Coloured arrows mark changes that are significant under a two-sided Wilcoxon signed-rank test, Holm-corrected ($p_{\mathrm{Holm}}<0.05$); gray arrows are not significant. The test depends on how consistently questions move in one direction, not on the size of the mean change alone.}}
% For example, on Fundamental, Ministral-3-8B's $+0.14$ is significant (RAG improves 49\% of questions and worsens 29\%; $p_{\mathrm{Holm}}=.013$), whereas GPT-5.4-mini's $-0.14$ is not (37\% improve, 45\% worsen; $p_{\mathrm{Holm}}=.48$).}}
\label{tab:common-evidence-rag-core400}
\end{minipage}\hfill\hfill
\begin{minipage}[t]{0.35\textwidth}
\vspace{0pt}
\centering
\includegraphics[width=\linewidth,  height=1.2in]{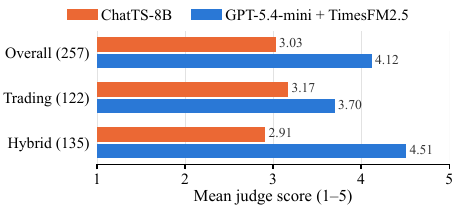}

\captionof{figure}{\small{Time-series-specialized evaluation on the final
257-question eligible subset(122 Trading (T), 135 (FT)). Scores are raw 1--5 judge Likert means.}}
\label{fig:time_series}
\end{minipage}
\vspace{-4mm}
\end{table*}

%% file: sections/fintradebench_self_contained_sections/03_results_and_discussion.tex
\section{Results and Discussion}
\label{sec:results}

We organize the results based on the research questions mentioned in \S\ref{sec:experimental-setup}. Unless otherwise stated, reported scores
are raw 1--5 mean judge Likert scores, and condition effects are paired
question-level differences. Statistical conclusions use 5,000-sample
paired bootstrap confidence intervals together with Holm-corrected
Wilcoxon signed-rank tests. 
% See~\S\ref{subsec:results-human-validation} for details regarding the expert-validated automated judge, and \S\ref{subsec:results-core400-sensitivity} to study the robustness of evaluation to final \emph{Core400} questions.

% ====================================================================

\begin{figure*}[t]
  \centering
\includegraphics[width=\textwidth, height=1.7in]{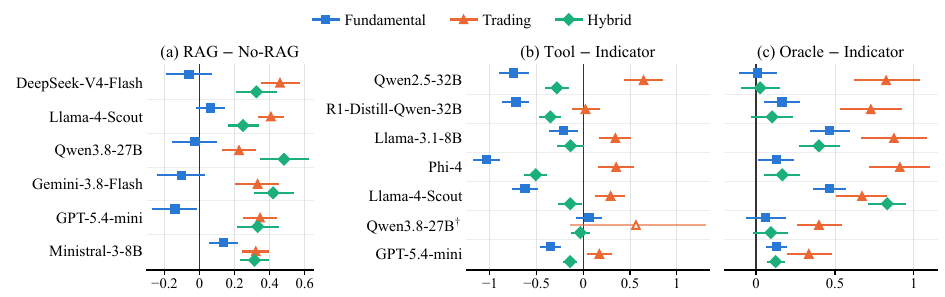}
\vspace{-7mm}
  \caption{\small{\textbf{Lane-level paired effects: (a) RAG$-$No-RAG on the common-evidence Core400; (b) Tool$-$Indicator and (c) Oracle$-$Indicator for the seven controlled models.} Points are mean paired changes in 1--5 judge Likert score with 95\% bootstrap CIs; the vertical line marks no change. $^{\dagger}$Hollow marker: only 12 paired Trading items under Tool.}}
  \label{fig:lane-effects}
  \vspace{-2mm}
\end{figure*}

\smartparagraph{RQ1: Does Retrieval Improve Financial Reasoning?}
\label{subsec:results-rag}
Under common-evidence judging, RAG improves overall response quality for
all six General-QA models (Table~\ref{tab:common-evidence-rag-core400}
and Figure~\ref{fig:lane-effects}). The overall paired gains
% from $+0.17$ to $+0.25$ Likert points and are 
 are statistically reliable
after Holm correction.
% The category-level pattern is substantially more informative than the aggregate gain.
RAG improves \emph{T} performance for every model,
with paired effects between approximately $+0.23$ and $+0.46$, and
improves \emph{FT} performance by approximately $+0.25$ to $+0.48$.
All \emph{T} and \emph{FT} effects remain statistically significant after
Holm correction. In contrast, \emph{F} effects are comparatively small and
model-dependent; see Table~\ref{tab:common-evidence-rag-core400}.
After multiple-comparison correction, only the Ministral Fundamental
effect is statistically reliable. 

\emph{Retrieval benefit depends on the reasoning modality.} With an identical factual anchor for paired outputs, retrieval is most
consistently beneficial on \emph{T} and \emph{FT} questions, where answers must
connect historical market evidence to a financial conclusion. The common
evidence package ensures that these gains are not explained by giving the
RAG response a richer factual basis at scoring time. \emph{F} questions show a different behavior. Providing additional SEC
evidence does not reliably improve the final answer for already-capable
models.
This suggests that for fundamental queries, performance is constrained by evidence selection and cross-statement synthesis rather than raw evidence availability.
% One possible interpretation is that, for these questions, simple
% evidence availability is less limiting than selecting and synthesizing
% the relevant quantities. 
We test this explanation directly with the
controlled evidence conditions next.
% Because the General-QA RAG and No-RAG conditions use different candidate
% budgets, these results should be interpreted as a benchmark comparison
% rather than a pure causal retrieval ablation. The
% common-evidence protocol controls the \emph{evaluation-side} factual
% anchor, not the generation budget.
% ====================================================================

\input{sections/main_tables/table_controlled_all7_core400}
\smartparagraph{RQ2: What Limits Performance After Evidence Is Available?}
\label{subsec:results-controlled}
\emph{Gold-selected evidence consistently improves reasoning.} Figure~\ref{fig:contolled_experiment_result} and Table~\ref{tab:controlled-all7-core400} compares Indicator-RAG,
Tool-Use, and gold-indicator Oracle evidence across seven models spanning
multiple model families and capability levels. The most consistent result is the gap between Oracle and Indicator-RAG.
Oracle evidence improves the overall mean score for \emph{all
models}. Overall Oracle--Indicator improvements
remain statistically significant after Holm correction; see Figure~\ref{fig:lane-effects}. The pattern is particularly strong on \emph{T} questions. Every model
receives a positive Oracle--Indicator effect in this lane, ranging from
approximately $+0.33$ for GPT-5.4-mini to more than $+0.9$ for Phi-4,
with statistically reliable improvements across all models.
This bottleneck shows that performance does not
plateau once structured financial evidence is given as context.
Models improve further when the relevant evidence is selected for them.
This performance delta indicates that \emph{models frequently access sufficient context but fail to isolate the decisive signals}.
% This gap therefore reflects \emph{current systems often possess
% sufficient information but fail to identify which subset should drive
% the financial conclusion}. 
Here, Oracle is not a perfect-reasoning condition. It supplies
gold-selected indicators but not the benchmark reference answer. The residual gap
between Oracle performance and the maximum score, therefore, reflects
remaining errors in interpretation, synthesis, and answer generation.
% \begin{figure*}[hbtp!]
%   \centering
% \includegraphics[width=\textwidth, height=1.8in]{images/fig_lane_effects_row_ragfirst.pdf}
% \vspace{-6mm}
%   \caption{Lane-level paired effects: (a) RAG$-$No-RAG on the common-evidence Core400; (b) Tool$-$Indicator and (c) Oracle$-$Indicator for the seven controlled models. Points are mean paired changes in 1--5 judge Likert score with 95\% bootstrap CIs; the vertical line marks no change. $^{\dagger}$Hollow marker: only 12 paired Trading items under Tool.}
%   \label{fig:lane-effects}
%   \vspace{-4mm}
% \end{figure*}
\begin{figure}[t]
\vspace{-1mm}
\small
  \centering
  \scalebox{.9}{\includegraphics[width=\columnwidth]{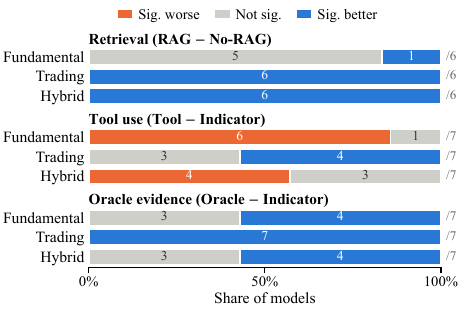}}
  \vspace{-2mm}
  \small
  \caption{\small{Number of models whose score changes significantly under each intervention, by lane (two-sided Wilcoxon signed-rank test on per-question paired differences, Holm-corrected; $p_{\mathrm{Holm}}<0.05$). Numbers inside bars are model counts. For details, see Tables~\ref{tab:rag-common-core400} and \ref{tab:controlled-effects-full}}}
  \label{fig:significance-tally}
  \vspace{-5.5mm}
\end{figure}

% \begin{figure}[htbp!]
%     \centering
% \includegraphics[width=0.75\linewidth]{images/figures/A_dotplot.pdf}
%     \vspace{-3mm}
%     \caption{\small{\textbf{Oracle--Indicator effects across seven
%     controlled models.} Positive values indicate improved performance
%     when question-specific gold-indicator evidence replaces the broader
%     Indicator-RAG evidence-selection process.}}
%     \label{fig:contolled_experiment_result}
%     \vspace{-5mm}
% \end{figure}

% 

% \begin{figure}[t]
%     \centering
%     \includegraphics[width=\columnwidth]{images/figures/fig_oracle_minus_indicator_all7_core400.pdf}
%     \vspace{-3mm}
%     \caption{\small{\textbf{Oracle--Indicator effects across seven
%     controlled models.} Positive values indicate improved performance
%     when question-specific gold-indicator evidence replaces the broader
%     Indicator-RAG evidence-selection process.}}
%     \label{fig:oracle-indicator}
%     \vspace{-4mm}
% \end{figure}

% --------------------------------------------------------------------
\emph{Tool access helps computation selectively, not
universally.} 
% Tool-Use produces a different pattern. 
At the aggregate level, Tool-Use is below Indicator-RAG for six of the seven models, while Llama-3.1 is approximately unchanged.
Thus, simply giving a system selective access to deterministic tools
does not improve overall financial reasoning. The \emph{T} lane, however, shows the opposite tendency. All six
complete-coverage models have non-negative Trading Tool--Indicator point
estimates, and Qwen2.5, Llama-3.1, Phi-4, and Llama-4 show statistically
reliable gains after multiple-comparison correction. In contrast,
Tool-Use reduces Fundamental performance for every complete-coverage
model; see Figure~\ref{fig:lane-effects}. This lane asymmetry clarifies the role of tools. Explicit computation is
useful when the task requires deriving quantities from historical market
data, but selective tool routing can remove broader context required for
fundamental or cross-signal synthesis. Tool access should therefore be
viewed as a task-dependent computational interface rather than a
universally stronger form of retrieval.

% Qwen3.8 Tool-Use is excluded from complete-coverage conclusions because only 290 outputs are valid and only 12 Trading questions are available for the paired Tool comparison.

% Optional main figure; otherwise move to appendix.
%
% \begin{figure}[t]
%     \centering
%     \includegraphics[width=\columnwidth]
%     {images/fig_tool_minus_indicator_all7_core400.pdf}
%     \caption{\small{Tool--Indicator effects by reasoning lane.}}
%     \label{fig:tool-indicator}
% \end{figure}

% ====================================================================
\smartparagraph{RQ3: Do Time-Series Specialists Improve Financial Reasoning?}
\label{subsec:results-timeseries}
% \input{sections/main_tables/table_timeseries_core400}
% \begin{figure}
%     \centering
%     \includegraphics[width=0.8\linewidth]{images/F_timeseries_bars.pdf}
%     \vspace{-4mm}
%     \caption{Time-series-specialized evaluation on the final 257-question eligible subset.}
%     \label{fig:time_series}
%     \vspace{-5mm}
% \end{figure}
\emph{Forecast-derived features substantially improve the
time-series pipeline.} This experiment evaluates the 257 \emph{Core400} questions requiring time-series reasoning.
Figure~\ref{fig:time_series} compares direct ChatTS reasoning with
the GPT-5.4-mini + TimesFM~2.5 forecast-feature pipeline. 
% ChatTS obtains an overall mean score of $3.03$, compared with $4.12$ for the forecast-feature pipeline. The paired overall improvement is $+1.09$ Likert points (95\% bootstrap CI $[0.94,1.25]$). 
The gain appears in both reasoning categories but is substantially
larger for \emph{FT} questions. 
% On \emph{T} questions, the paired
% TimesFM+GPT--ChatTS effect is $+0.52$ ($95\%$ CI $[0.28,0.77]$); on
% \emph{FT} questions it rises to $+1.60$ ($95\%$ CI $[1.47,1.74]$).
All comparisons are statistically significant. The result suggests that forecast-derived numerical features can provide
a useful intermediate representation between historical prices and
language-based financial synthesis. The particularly large Hybrid gain
is consistent with the value of combining summarized market dynamics
with point-in-time fundamental evidence. Table~\ref{tab:gpt54mini-conditions} shows the performance of \emph{GPT-5.4-mini} across all evaluation conditions (for qualitative responses, see Tables~\ref{tab:qual-gpt54mini-fund} --
% \ref{tab:qual-gpt54mini-trad}
\ref{tab:qual-gpt54mini-hyb}).
% Table \ref{tab:gpt54mini-conditions} compares the performance of \emph{GPT-5.4-mini} across all conditions evaluated in this paper; see \S\ref{subsec:results-discussion} for a summarized discussion of the results.
\begin{table}[t]
\scalebox{0.85}
\centering
\small
\setlength{\tabcolsep}{3pt}
\begin{tabular}{@{}lcccc@{}}
\toprule
Condition & Overall & F & T & FT \\
\midrule
% \multicolumn{5}{@{}l}{\textit{Judge: common evidence}} \\
\quad No-RAG & 2.81 & \textbf{3.20} & 2.06 & 3.08 \\
\quad RAG & \textbf{2.98} & 3.06 & \textbf{2.41} & \textbf{3.41} \\
\midrule
% \multicolumn{5}{@{}l}{\textit{Judge: per-condition evidence}} \\
\quad Indicator & 4.69 & 4.77 & 4.44 & 4.83 \\
\quad Tool & 4.57 & 4.42 & 4.61 & 4.69 \\
\quad Oracle & \textbf{4.88} & \textbf{4.90} & \textbf{4.77} & \textbf{4.95} \\
\midrule

\quad GPT-5.4-mini + TimesFM2.5 & 4.12$^{\dagger}$ & -- & 3.70 & 4.51 \\
\bottomrule
\end{tabular}
\vspace{-2mm}
\caption{\small{\textbf{GPT-5.4-mini across all experimental conditions.}
$^{\dagger}$TimesFM2.5 is evaluated only on the time-series-eligible questions, so its Overall excludes Fundamental.}}
\label{tab:gpt54mini-conditions}
\vspace{-5mm}
\end{table} 

%% file: sections/main_tables/table_controlled_all7_core400.tex
% \begin{table}[htbp!]
% \centering
% \scriptsize
% \renewcommand{\arraystretch}{1.12}
% \setlength{\tabcolsep}{2.6pt}
% \small
% \begin{tabular}{lrrrrr}
% \toprule
% Model & Ind. & Tool & Orcl. & $\Delta_{\mathrm{TI}}$ & $\Delta_{\mathrm{OI}}$ \\
% \midrule
% Qwen2.5-32B&3.72 & 3.55 & 3.98 & -0.17 & +0.26 \\
% R1-Distill-Qwen-32B & 3.38 & 3.00 & 3.69 & -0.37 & +0.31 \\
% Llama-3.1-8B&2.91 & 2.89 & 3.47 & -0.02 & +0.57 \\Phi-4 & 3.68 & 3.25 & 4.06 & -0.44 & +0.38 \\
% % \midrule
% Llama-4-Scout&3.53 & 3.35 & 4.19 & -0.18 & +0.65 \\
% Qwen3.8-27B&4.34 & 4.53 & 4.52 & +0.04 & +0.17 \\
% GPT-5.4-mini&4.69 & 4.57 & 4.88 & -0.12 & +0.19 \\
% \bottomrule
% \end{tabular}
% \caption{Controlled-condition results on \emph{Core400}. Scores are 1--5 judge Likert means on valid responses for Indicator (Ind.), Tool and Oracle (Orcl.) conditions; deltas are paired mean differences between Tool- Indicator ($\Delta_{TI}$) and Oracle-Indicator ($\Delta_{OI}$) conditions.}
% \label{tab:controlled-all7-core400}
% \vspace{-4mm}
% \end{table}

\begin{table*}[t]
\centering
\begin{minipage}[t]{0.475\textwidth}
\vspace{0pt}
\centering
\begingroup
\tiny
\renewcommand{\arraystretch}{1.12}
\setlength{\tabcolsep}{2.6pt}

\resizebox{\linewidth}{!}{%
\begin{tabular}{@{}lrrrrr@{}}
\toprule
Model & Ind. & Tool & Orcl.
& $\Delta_{\mathrm{TI}}$ & $\Delta_{\mathrm{OI}}$ \\
\midrule
Qwen2.5-32B
& 3.72 & 3.55 & 3.98 & \down{$-$0.17} & \up{+0.26} \\
R1-Distill-Qwen-32B
& 3.38 & 3.00 & 3.69 & \down{$-$0.37} & \up{+0.31} \\
Llama-3.1-8B
& 2.91 & 2.89 & 3.47 & \ndown{$-$0.02} & \up{+0.57} \\
Phi-4
& 3.68 & 3.25 & 4.06 & \down{$-$0.44} & \up{+0.38} \\
Llama-4-Scout
& 3.53 & 3.35 & 4.19 & \down{$-$0.18} & \up{+0.65} \\
Qwen3.8-27B$^{\dagger}$
& 4.34 & 4.53 & 4.52 & \nup{+0.04} & \up{+0.17} \\
GPT-5.4-mini
& 4.69 & 4.57 & 4.88 & \down{$-$0.12} & \up{+0.19} \\
\bottomrule
\end{tabular}%
}
\endgroup
\vspace{-2mm}
\caption{\small{\textbf{Controlled-condition results on \emph{Core400}.}
Scores are 1--5 judge Likert means for
Indicator (Ind.), Tool and Oracle (Orcl.) conditions; deltas
are paired mean differences between Tool--Indicator
($\Delta_{\mathrm{TI}}$) and Oracle--Indicator
($\Delta_{\mathrm{OI}}$) conditions.  Colored arrows are significant (Wilcoxon signed-rank, Holm-corrected, $p_{\mathrm{Holm}}<0.05$), gray arrows are not. $^{\dagger}$Hollow marker: only 12 paired Trading items under Tool.}}
\label{tab:controlled-all7-core400}
\end{minipage}\hfill%
\begin{minipage}[t]{0.475\textwidth}
\vspace{-2mm}
\centering
\includegraphics[width=\linewidth]
    {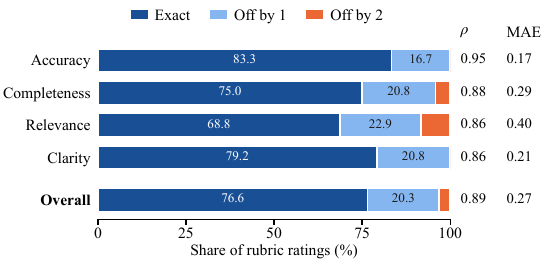}

\captionof{figure}{\small{\textbf{Held-out human--judge agreement on
48 responses (192 rubric ratings), stratified across 3 reasoning lanes and 4 evaluation families.}
Bars give the share of ratings where the judge matches
the human score exactly, differs by 1, or differs by
2 points (no larger differences occur); $\rho$ is
Spearman's correlation and MAE the mean absolute error.}}
\vspace{-1mm}
\label{fig:human-judge-validation}
\end{minipage}
\vspace{-4mm}
\end{table*}

%% file: sections/fintradebench_self_contained_sections/04_conclusion_and_future_directions.tex
\vspace{-1mm}
\section{Conclusion}\label{conclusion}
\vspace{-1mm}
We introduce \emph{FinTradeBench} for evaluating financial
reasoning across company fundamentals and trading signals. We use
a decade of \textit{NASDAQ-100} financial data to curate our benchmark using a calibration-then-scaling
construction pipeline and human validation. Evaluation uses human-validated \emph{Core400} questions. We find that financial reasoning failures cannot be
explained by evidence access alone. Under common-evidence evaluation,
RAG improves overall performance with the consistent gains on \emph{T} and \emph{FT} questions, while \emph{F}
effects are smaller and model-dependent. Controlled experiments further
separate downstream bottlenecks. Tool-Use is beneficial when
explicit numerical computation is required, particularly for \emph{T}
questions, but does not universally improve \emph{F} or \emph{FT}
reasoning. In contrast, Oracle evidence improves all
controlled models, showing that
\emph{evidence selection} remains a persistent bottleneck even when
relevant financial information is available. The remaining gap
under Oracle evidence further indicates that evidence selection is not
the only limitation; interpretation and financial synthesis also remain
imperfect. Our time-series experiments provide complementary evidence that the
representation of numerical market history matters. Transforming historical price sequences into
forecast-derived features can provide a useful interface between
time-series models and language-based financial reasoning. Overall, financial reasoning errors
cannot be reduced to a single retrieval problem. Retrieval remains useful, but providing more evidence alone is insufficient. \emph{Computation} and \emph{evidence selection} also cause financial reasoning errors. In time series pipelines, powerful reasoning model with summarized time series data outperforms smaller fine-tuned models for time-series reasoning.
% Because the
% two systems differ in their downstream reasoning components, this result
% should be interpreted as a comparison of complete pipelines rather than
% a pure architectural ablation.
We validate benchmark quality and evaluator reliability through
separate human-review procedures. 
% A full item-level audit of the 403-question support-screened Core excludes three items and yields Core400.
% Independently, held-out validation of the frozen Claude Sonnet~4.5 judge
% shows strong agreement with human scoring`, supporting scalable automated evaluation while not replacing expert assessment. 
Overall, we find that to make progress in financial LLMs, we must build systems that retrieve the right
evidence, perform the required computation, identify which signals
matter, and synthesize them into a grounded financial conclusion.
\emph{FinTradeBench} provides a framework for measuring each of these stages.

%% file: sections/Limitations.tex
\section{Limitations}
\label{sec:limitations}

\myNum{i}\smartparagraph{Market and asset-class coverage.}
FinTradeBench focuses on the NASDAQ-100 universe over 2015--2025. These firms
are predominantly large, liquid, U.S.-listed companies, so the benchmark may
not represent small-cap equities, international or emerging markets, or other
asset classes such as fixed income, commodities, derivatives, and currencies.

\myNum{ii}\smartparagraph{Broader benchmark versus audited Core.}
The broader collection contains 1,400 questions, but the full item-level human
quality audit is performed on the filtered Core403, producing Core400.
The 99.3\% acceptance rate therefore characterizes an already-screened
evaluation subset and should not be interpreted as a quality estimate for all
1,400 questions. We perform a catastrophic error screening on the overall benchmark.

% \myNum{iii}\smartparagraph{Single-reviewer benchmark audit.}
% The full Core403 audit is performed by one reviewer. It provides complete
% item-level quality control but not inter-annotator reliability. A future release
% would benefit from independent secondary review of a larger stratified subset.

\myNum{iii}\smartparagraph{Golden indicators are not exhaustive.}
Golden indicators specify an expert-defined minimal evidence set. Financial
questions can admit defensible reasoning paths using additional or alternative
evidence, especially for subjective valuation or investment-style questions.
For this reason, golden-indicator F1 is treated as an ancillary diagnostic rather
than the primary quality score.

\myNum{iv}\smartparagraph{LLM-as-a-Judge limitations.}
The final evaluator is automated. Held-out human validation shows strong
agreement, but the study uses financial expert only for validation of the evaluation and a PhD student to screen catastrophic errors.
Relevance exhibits a modest positive judge bias, and automated scores remain an
approximation of expert assessment rather than ground truth.

\myNum{v}\smartparagraph{No-RAG/RAG generation-budget asymmetry.}
The historical General-QA protocol uses four No-RAG candidates and two RAG
candidates before self-selection. Common-evidence re-judging removes
evaluator-side evidence asymmetry but does not make generation budgets
identical. RAG--No-RAG is therefore a benchmark comparison, not a strict causal
ablation.

% \myNum{vii}\smartparagraph{Pipeline-specific missingness.}
% Although Core400 contains 400 human-validated questions, some automatic
% pipelines have lower coverage. Typical controlled/common-evidence conditions
% have 399 valid judgments, Gemini No-RAG has 398, and Qwen3.8 Tool-Use has 290.
% These failures are reported rather than imputed.

% \myNum{viii}\smartparagraph{Tool-Use coverage.}
% Qwen3.8 Tool-Use has especially incomplete coverage, including only 12 valid
% Trading questions. It is therefore excluded from complete-coverage conclusions
% about the Tool-Use condition.

\myNum{vi}\smartparagraph{Time-series comparison is not a pure ablation.}
\emph{ChatTS} and the \emph{TimesFM~2.5} forecast-feature pipeline differ in their downstream
language components. Their comparison evaluates complete systems and does not
isolate the causal contribution of the forecasting architecture alone.

\myNum{vii}\smartparagraph{Point-in-time reconstruction.}
FinTradeBench uses question-specific historical cutoffs and point-in-time
evidence whenever available, with a benchmark-wide cutoff as a fallback.
Historical financial data can nevertheless contain complications such as filing
availability dates, fiscal-period conventions, ticker changes, and market
reaction timing. The benchmark should therefore be interpreted as a
point-in-time historical reasoning benchmark rather than a live trading
simulator.

\myNum{viii}\smartparagraph{Temporal contamination and model knowledge.}
Because the benchmark covers historical periods, some facts may already exist
in a model's pretraining or post-training data. Closed-book performance cannot
be interpreted as proof that the model reconstructed an answer solely from
reasoning. The controlled evidence conditions are intended to diagnose
evidence-use behavior rather than to establish absence of memorization.

\myNum{ix}\smartparagraph{Hosted-model reproducibility.}
Several evaluated systems are accessed through hosted APIs. Even when model
identifiers are fixed, providers may update infrastructure or serving behavior.
We therefore preserve frozen generated outputs, judge outputs, model identifiers,
and analysis scripts with the release.

\myNum{x}\smartparagraph{Task scope.}
FinTradeBench evaluates historical evidence-grounded question answering. It
does not measure portfolio optimization, execution quality, risk management,
market impact, profitability, or personalized investment suitability.

%% file: sections/Ethical_Considerations.tex
\section{Ethical Considerations}
\label{sec:ethics}
\myNum{i} \smartparagraph{Financial decision-making risk.} FinTradeBench evaluates language model reasoning over real financial data for named public companies. Scores on this benchmark should not be interpreted as endorsements of any model for live trading, investment advisory, or automated financial decision-making. Even the highest-performing models in our evaluation exhibit substantial error rates, and financial decisions based on LLM outputs carry real economic risk to end users.
% \paragraph{Data sourcing and licensing.} All company fundamentals are derived from SEC regulatory filings, which are public records released under the EDGAR system. Daily trading data is sourced from publicly available market feeds. No proprietary, paywalled, or non-public data was used in benchmark construction. We verify that our use of these data sources is consistent with their respective terms of access.
\myNum{ii} \smartparagraph{Benchmark misuse.} While FinTradeBench is designed for research evaluation, we acknowledge that fine-tuning models specifically to maximise FinTradeBench scores without genuine improvement in financial reasoning could inflate reported performance. We encourage the community to treat benchmark results as one signal among many and to complement automated evaluation with human expert review before drawing strong conclusions about financial reasoning capability.
\myNum{iii} \smartparagraph{Annotator and expert involvement.}
Human financial experts involved in seed question authoring were part of the research team. Evaluation and calibration were conducted with anonymized model outputs to reduce rater bias. Scaled benchmark responses were additionally screened for fatal flaws as a quality-control step rather than used as independent graded labels. 
\myNum{iv} \smartparagraph{Societal impact.} Improvements in LLM financial reasoning could benefit retail investors and analysts by democratising access to structured financial analysis. However, the same capabilities could be exploited to automate misleading financial narratives or market manipulation at scale. We call for responsible disclosure norms and human-in-the-loop oversight in any deployment of LLM-based financial analysis tools.
\myNum{v} \smartparagraph{Misinformation and market manipulation.}
Because FinTradeBench evaluates models on real public companies, there is a risk that financial reasoning systems trained or evaluated on similar data could be used to generate persuasive but misleading market narratives. This risk is especially important when LLM outputs are presented as authoritative financial analysis. We therefore recommend that any deployment of LLM-based financial analysis tools include source attribution, uncertainty disclosure, numerical verification, and human review before dissemination.

%% file: sections/05_appendix_updated.tex
\twocolumn[
  \begin{center}
    \LARGE \textbf{FinTradeBench: A Financial Reasoning Benchmark for LLMs}\\
    \vspace{4mm}
    \Large Supplementary Material\\
    \vspace{8mm}
  \end{center}
]

\appendix

% placeins[section] flushes all pending floats at every \section, which strands large
% full-width tables on their own pages and leaves the previous page half empty.
% For sections that directly follow such tables we let the text keep flowing.
\newcommand{\sectionNoBarrier}[1]{%
  \let\savedFloatBarrier\FloatBarrier \let\FloatBarrier\relax
  \section{#1}%
  \let\FloatBarrier\savedFloatBarrier}

% Let several wide tables share the top of one page (default: two) and fill float pages better.
\setcounter{dbltopnumber}{3}
\renewcommand{\dbltopfraction}{0.85}
\renewcommand{\dblfloatpagefraction}{0.6}
\renewcommand{\topfraction}{0.85}
\renewcommand{\floatpagefraction}{0.6}

% ============================================================
\section{Organization of the Supplementary Material}
\label{app:organization}

This supplementary material expands the main paper in four parts:
\emph{benchmark construction} (\S\ref{app:motivating-examples}--\S\ref{app:benchmark-details}),
\emph{experimental implementation} (\S\ref{app:experimental-details}),
\emph{evaluation evidence} (\S\ref{app:results}--\S\ref{app:full-results}), and
\emph{illustrations and responsible-use information}
(\S\ref{app:case-studies}--\S\ref{app:Dt_and_all}).
Limitations and ethical considerations appear in the main paper.

\begin{itemize}[leftmargin=*, nosep, itemsep=2pt]
\item[\S\ref{app:motivating-examples}] \textbf{Motivating examples.}
Extended historical cases of fundamentals--market divergence that
motivate joint reasoning over both signal types (cf.\ \S\ref{sec:introduction}).
\item[\S\ref{app:extended_related_work}] \textbf{Extended related work.}
Trading signals, LLM-as-a-judge evaluation, retrieval and tool use, and a
comparison with existing financial benchmarks.
\item[\S\ref{app:benchmark-details}] \textbf{Benchmark design addendum.}
Validation gates, dataset statistics, the financial-signal reference, the
TELeR prompt taxonomy, sample questions and golden answers, the
450$\rightarrow$403$\rightarrow$400 curation process with its provenance,
item-level audit, and fatal-flaw screening, and benchmark quality control.
\item[\S\ref{app:experimental-details}] \textbf{Experimental setup addendum.}
Models and serving details, the RAG architecture, evidence conditions,
tool routing, time-series specialist pipelines, the evidence-aware judge,
metrics and statistical testing, and compute resources.
\item[\S\ref{app:results}] \textbf{Results addendum.}
Held-out judge validation, robustness to the final Core400 curation, and
discussion, as referenced from the main-text results.
\item[\S\ref{app:validation}] \textbf{Human review and evaluator validation.}
The four human-review stages, held-out validation of the frozen judge, and
pointers to the full annotation instruments (\S\ref{app:annotation-instruments}).
\item[\S\ref{app:evaluation-integrity}] \textbf{Evaluation integrity and
sensitivity checks.}
Evidence coverage and QC, and sensitivity of all results to the
Core403$\rightarrow$Core400 transition.
\item[\S\ref{app:full-results}] \textbf{Complete quantitative results.}
Full paired effects, bootstrap intervals, and Holm-adjusted tests for the
common-evidence, controlled, and time-series studies.
\item[\S\ref{app:case-studies}] \textbf{Qualitative case studies.}
Per-condition response examples, presented as illustrations rather than as
primary evidence.
\item[\S\ref{app:Dt_and_all}] \textbf{Transparency and responsible use.}
Data transparency and privacy, safety and broader impact, data
maintenance, and disclosure of AI assistance.
\item[\S\ref{app:annotation-instruments}] \textbf{Annotation instruments.}
Reviewer rubrics, label sets, and record fields for the four human-review
stages (reference material).
\end{itemize}

\section{Extended Motivating Examples: Fundamentals--Market Divergence}
\label{app:motivating-examples}

\begin{table*}
\centering
\small
\renewcommand{\arraystretch}{1.2}

\begin{tabular}{@{}l l l l@{}}
\toprule
\textbf{Company} & \textbf{Period} & \textbf{Fundamental Signal} & \textbf{Dominant Signal} \\
\midrule
Amazon   & 1999--2001 & Persistent losses, negative EPS      & Market (narrative) \\
Nvidia   & 2016--2017 & Moderate earnings, GPU revenue lag   & Market (forward expectation) \\
Tesla    & 2020       & Modest profitability, high P/E       & Market (sentiment) \\
Tesla    & Apr.\ 2025 & EPS miss (\$0.27 vs.\ \$0.42), revenue miss & Market (narrative rally) \\
\bottomrule
\end{tabular}
\caption{Summary of fundamentals--market divergence episodes. ``Dominant Signal'' refers to which signal better characterised the market outcome at the time.}
\label{tab:divergence-summary}
\end{table*}

The Tesla April 2025 case discussed in \S\ref{sec:introduction} is illustrative of a broader and well-documented phenomenon in financial markets: sustained divergences between company fundamentals and market price dynamics driven by investor sentiment, narrative momentum, and forward-looking expectations. We present three additional historical examples that further motivate the need for benchmarks capable of reasoning over both types of signals jointly.

\myNum{i} \smartparagraph{Amazon (1999--2001).}
During the dot-com era, Amazon sustained extremely high market valuations despite persistent operating losses and negative earnings. Rather than anchoring on contemporaneous accounting metrics, investors priced in long-run platform dominance and e-commerce adoption narratives. This example illustrates how growth narratives can sustain valuations that are entirely disconnected from current fundamental signals \citep{shiller2000irrational}. A system reasoning solely from financial statements would have consistently flagged Amazon as financially distressed during this period, while the market priced in the opposite trajectory.

\myNum{ii} \smartparagraph{Nvidia (2016--2017).}
NVIDIA's stock appreciated substantially in 2016-2017, well before its earnings reports fully reflected the revenue impact of GPU adoption in deep learning workloads. The rally was driven primarily by forward-looking narratives around artificial intelligence and autonomous vehicles, with price movements leading rather than following the fundamental confirmation \citep{greenwood2018asset,bybee2023narrative}. This case illustrates the temporal asymmetry that motivates hybrid reasoning: trading signals captured the market's forward expectation long before quarterly filings reflected it, meaning neither signal alone would have been sufficient to correctly make the investment decision.

\myNum{iii} \smartparagraph{Tesla (2020): Sentiment-Driven Rally Under Modest Profitability.}
Tesla's stock rose dramatically in 2020 despite modest profitability at the time, as investors priced in narratives around electric vehicle adoption and autonomous driving at scale \citep{shiller2017narrative, baker2006investor}. This example predates the main example we give in \S\ref{sec:introduction} (the April 2025 earnings miss) and hence establishes that the fundamentals--market divergence is a recurring feature of Tesla's price history rather than an isolated anomaly. It also shows how investor sentiment \citep{baker2006investor} and narrative economics \citep{shiller2017narrative} can sustain multi-year divergences, not merely short-term reactions.

\smartparagraph{Takeaway.}
From these examples, we can see a consistent pattern emerging: price dynamics reflect investor expectations and narrative momentum that often precede or contradict the signal available in the company's financial statements. Table~\ref{tab:divergence-summary} summarises the key characteristics of each case.

These cases collectively motivate FinTradeBench's dual-signal design. A benchmark that evaluates reasoning over fundamentals alone would reward models that correctly identify Amazon, Nvidia, and Tesla as weak or fairly valued, missing the market trading signal entirely. A benchmark that evaluates trading signals alone would capture momentum but provide no mechanism for assessing whether that momentum is anchored in improving fundamentals or purely sentiment-driven. Only by evaluating both types of signals jointly, and by including hybrid questions that require reconciling conflicting signals, can we create a benchmark that reflects the reasoning demands of real-world financial analysis.
% ============================================================
\section{Extended Related Work}
\label{app:extended_related_work}

% retrieval bottleneck where RAG succeeds on fundamentals but struggles on
% market dynamics" with the paragraph below.

To understand the performance of LLMs across niche domain requires a comprehensive benchmark that covers different aspects of the domain \cite{hendrycks2020measuring, liu2024mathbench, fei2024lawbench, campos2025gaea,liu2024large,kader2025survey,liang2022holistic, 10658434,ghazal2013bigbench,mialon2024gaia,singhal2023large,mazeika2024harmbench}. Recent financial-AI work includes text-centric financial QA benchmarks,
chart- and table-reasoning benchmarks, as well as agentic and tool-use
frameworks \cite{finance_agent_v2,chen2021finqa,xie2024finben,choi2025finder,chen2022convfinqa,yang2023fingpt,xie2023pixiu,klimaszewski2025avenibench,peng2026multifinben,peng2025plutus,larsen2026v}. FinTradeBench is complementary to these efforts. Its
evaluation is designed to diagnose distinct stages of a financial-reasoning
pipeline which includes evidence access, computation, evidence selection, and synthesis, 
under a shared Fundamental/Trading/Hybrid taxonomy. The controlled conditions
therefore focus on mechanism-level differences rather than treating a single
RAG-versus-No-RAG score as a complete characterization of financial reasoning.

\smartparagraph{Trading Signals and Quantitative Finance.}
Trading signals derived from price and volume data play a key role in understanding stock market behavior and risk dynamics~\citep{lo2000foundations}.~Indicators such as momentum, volatility, moving averages, and drawdowns have been widely studied in asset pricing and quantitative finance~\citep{fama1992cross, Jegadeesh1993ReturnsTB, lo2000foundations}.~Volatility measures are also used to capture perceived market risk and regime changes~\citep{engle2004risk,ang2012regime,bollerslev2015stock,bollerslev2018risk}.~While machine learning approaches have recently been applied to forecasting trading signals in finance~\citep{han2025can,mishra2024volatility, moreno2024deepvol,li2024volatility}, there remains a gap in accurate financial analysis from multi-modal data and existing NLP benchmarks rarely evaluate whether LLMs can reason about these signals for financial analysis.

% \smartparagraph{Retrieval-Augmented Generation.}
% Retrieval-augmented generation (RAG) has emerged as a widely used approach for grounding LLM outputs in external knowledge sources~\citep{lewis2021retrievalaugmentedgenerationknowledgeintensivenlp}. 
% Recent work has proposed systematic evaluations of RAG systems, including RAGBench~\citep{friel2025ragbenchexplainablebenchmarkretrievalaugmented} and RAGTruth~\citep{niu2024ragtruthhallucinationcorpusdeveloping}, which study retrieval faithfulness, attribution, and hallucination reduction. 
% However, these benchmarks primarily focus on textual retrieval tasks and rarely evaluate heterogeneous knowledge sources such as structured financial indicators or time-series signals.

\smartparagraph{LLM Evaluation and Benchmark Design.} Carefully designed benchmarks are essential for evaluating the reasoning capabilities of LLMs. Recent LLM benchmarks increasingly rely on calibrated automatic evaluation for open-ended outputs that are difficult to score with exact-match metrics. Judging LLM-as-a-judge with MT-Bench and Chatbot Arena~\citep{zheng2023judging} established LLM-as-a-judge as a scalable evaluation paradigm for multi-turn, preference-oriented model outputs, while also identifying important evaluator biases such as position, verbosity, and self-enhancement bias. Subsequent work further studies reliability, bias, and calibration in LLM judging~\citep{chen2024llmasjudge,ye2025justice,hossain2025llmmeta,gu2024survey}. Financial QA presents additional challenges due to numerical fidelity, domain expertise, and the integration of heterogeneous data sources \citep{yang2023investlm, ran2019numnet,zhang2024careful}.
Open-ended answers often require factual grounding, numerical consistency, and domain-specific reasoning, making reference-only or lexical metrics insufficient. FinTextQA~\citep{Chen_2024}, for example, combines human ranking, automatic metrics, and GPT-4 scoring to evaluate long-form financial QA systems. These works provide a design of scalable evaluation pipeline while preserving domain-grounded controls over factuality, numerical fidelity, and reasoning quality thus motivating us to use a calibrated LLM-as-a-judge framework rather than treating LLM judgments as ground truth.

\smartparagraph{Comparison with Existing Benchmarks.}
% Table~\ref{tab:benchmark-comparison} summarizes key differences between existing financial benchmarks and ours. Prior financial QA benchmarks, \cite{islam2023financebenchnewbenchmarkfinancial,chen2022convfinqaexploringchainnumerical} emphasize reasoning over textual financial documents, and quantitative finance research \citep{oberlechner2001importance, Jegadeesh1993ReturnsTB}, focuses on predictive modeling of trading signals. None explicitly evaluates reasoning over trading signals or the joint interaction between fundamentals and market dynamics. FinTradeBench bridges these two areas by introducing a benchmark that evaluates financial reasoning across both company fundamentals and trading signals within a unified evaluation framework. By explicitly modeling these two signal types, we perform financial reasoning tasks that closely reflect real-world financial analysis. Further discussions of recent financial benchmarks that do not directly impact our work but are highly relevant in the domain are provided in \S\ref{app:extended_related_work}.
Table~\ref{tab:benchmark-comparison} summarizes key differences between existing financial benchmarks and FinTradeBench. Prior financial QA benchmarks~\citep{islam2023financebench,chen2022convfinqa,chen2021finqa,Chen_2024} primarily emphasize reasoning over textual financial documents, while broad financial LLM benchmarks such as FinBen~\citep{xie2024finben} aggregate diverse NLP-style financial tasks. In parallel, quantitative finance research studies trading signals for forecasting, risk modeling, and asset pricing~\citep{oberlechner2001importance, Jegadeesh1993ReturnsTB}, and recent agentic benchmarks evaluate simulated financial actions~\citep{li2025investorbench}.
\smartparagraph{Retrieval, Tools, and Heterogeneous Financial Evidence.}
Retrieval-augmented financial QA systems typically ground answers in external
documents or structured context. FinanceBench, FinDER, FinTextQA, and AlphaFin
~\citep{islam2023financebench,choi2025finder,Chen_2024,li2024alphafin}
illustrate the importance of grounding financial answers in retrieved
information. Financial reasoning, however, also requires operations that are
not purely retrieval problems, such as computing quantities from historical series,
selecting among competing indicators, and reconciling evidence across
modalities. FinTradeBench therefore evaluates several evidence pathways rather
than a single RAG condition. Raw-RAG exposes point-in-time evidence,
Indicator-RAG supplies precomputed structured signals, Tool-Use provides
deterministic computation over raw market data, and Oracle supplies
question-specific gold-selected indicators without providing the reference
answer. This design makes retrieval, computation, and evidence selection
separately observable.

\smartparagraph{LLM-as-a-Judge and Open-Ended Evaluation.}
Open-ended financial answers are difficult to score with exact match because
quality depends on factual grounding, numerical consistency, completeness,
relevance, and the quality of the reasoning used to connect evidence to a
conclusion. LLM-as-a-judge work such as MT-Bench and Chatbot Arena
~\citep{zheng2023judging} established scalable model-based evaluation while
also documenting biases such as position, verbosity, and self-enhancement.
Subsequent work further examines judge reliability, bias, and calibration
~\citep{chen2024llmasjudge,ye2025justice,hossain2025llmmeta,gu2024survey}.
Financial QA adds domain-specific concerns because plausible language can mask
incorrect quantities or unsupported interpretations
~\citep{yang2023investlm,ran2019numnet,zhang2024careful}. FinTextQA
~\citep{Chen_2024}, for example, combines human and model-based evaluation for
long-form financial answers. Motivated by these concerns, FinTradeBench uses an
evidence-aware LLM judge whose factual anchor is point-in-time benchmark
evidence rather than the benchmark reference response. FinTradeBench separately validates
the frozen judge on blinded human ratings to ensure the robustness of judge.

% ============================================================
\section{Addendum to Benchmark Design}
\label{app:benchmark-details}

\subsection{Benchmark Summary and Validation Gates}

Table~\ref{tab:robustness-checks} summarizes the distinct validation
stages used in FinTradeBench. We distinguish construction-time checks from the
final human audit of the Core evaluation set and from held-out validation of the
automated judge.

\begin{table*}[tbph]
\centering
\small
\begin{tabular}{p{3.2cm} p{7.2cm} p{4.0cm}}
\toprule
\textbf{Check} & \textbf{Purpose} & \textbf{Stage} \\
\midrule
Golden indicators &
Provide an expert-defined minimal evidence set for each question; used for
construction auditing and controlled Oracle evidence, not as a complete answer
template. &
Question design / controlled evaluation \\
Numerical audit &
Flag unsupported, contradicted, or unavailable numerical claims in
construction-time candidate responses. &
Benchmark construction \\Human--LLM alignment & Verify that automated judge scores approximate expert scores & Judge calibration \\
Aligned judge validation & Validate the human-aligned LLM judge for benchmark scaling & Benchmark Scaling\\ 
Scaled-set fatal-flaw screen &
Identify catastrophic construction failures in the broader 1,400-question
collection; 42 responses required regeneration during benchmark scaling. &
Broader benchmark construction \\
Canonical-data support screen &
Remove Core candidates whose required company/time-period evidence is absent
from the frozen canonical data universe. &
450 $\rightarrow$ 403 \\
Full item-level human audit &
Review question clarity, reference-answer correctness, canonical PIT evidence,
golden-indicator adequacy, and entity/time alignment for every retained Core
item. &
403 $\rightarrow$ 400 \\
Common-evidence judging &
Score paired RAG/No-RAG responses against the same canonical point-in-time
evidence to remove evaluator-side evidence asymmetry. &
Final General-QA evaluation \\
Held-out human judge validation &
Evaluate the frozen Claude judge on blinded responses that were not used to
modify its scoring prompt. &
Final evaluator validation \\
\bottomrule
\end{tabular}
\caption{\small{\textbf{Construction and validation checks in the
FinTradeBench pipeline.} The checks operate at different stages and should not
be interpreted as interchangeable forms of validation.}}
\label{tab:robustness-checks}
\end{table*}

\subsection{Dataset Statistics}
\label{app:dataset-stats}

Table~\ref{tab:dataset-stats-app} summarizes the key FinTradeBench dataset and final evaluation statistics.
\begin{table}[tbhp]
\centering
\small
\begin{tabular}{lr}
\toprule
\textbf{Statistic} & \textbf{Value} \\
\midrule
Broader FinTradeBench collection & 1,400 \\
Expert-authored seed questions & 150 \\
Programmatically scaled questions & 1,250 \\
Core candidate pool & 450 \\
After canonical-data support screen & 403 \\
Final human-validated Core & 400 \\
Core400 Fundamental & 143 \\
Core400 Trading & 122 \\
Core400 Hybrid & 135 \\
Core400 seed-origin & 147 \\
Core400 company-scaled origin & 85 \\
Core400 year-scaled origin & 168 \\
Historical window & 2015--2025 \\
Fundamental indicators & 10 \\
Trading indicators & 9 \\
\bottomrule
\end{tabular}
\caption{\small{\textbf{FinTradeBench dataset and final evaluation-set
statistics.} The 1,400-question collection provides benchmark breadth, whereas
the human-validated Core400 is the frozen set used for the final controlled
evaluation.}}
\label{tab:dataset-stats-app}
\vspace{-5mm}
\end{table}

\subsection{Financial Signal Reference}
\label{app:signals}

Table~\ref{tab:volfund-signals} summarizes the company-fundamental and trading
signals used in question design. Trading signals are derived from historical
OHLCV data, while fundamental signals originate from SEC 10-K/10-Q filings.

% Verify formulas/definitions against the released code before camera-ready.
\begin{table*}
\centering
\renewcommand{\arraystretch}{1.3} 
\small
\begin{tabular}{p{4cm} p{3.8cm} p{5.2cm}}
\toprule
\textbf{Signal} & \textbf{Formula} & \textbf{Definition / Description} \\
\midrule
\multicolumn{3}{l}{\textbf{Trading Signals}}{\scriptsize \textit{(Notation: $P_t$ = Price at time $t$, $V_t$ = Volume, $R_t$ = Return, $N, k$ = Lookback periods, $\alpha$ = Smoothing factor)}} \\
\midrule
\textbf{MA (Moving Average)} & $\frac{1}{N} \sum_{i=1}^N P_{t-i}$ & Average price of stock over fixed lookback window. Smooths out short-term fluctuations to reveal longer-term trends. \\
\textbf{EMA (Exp. Moving Average)} & $\alpha P_t + (1-\alpha)\text{EMA}_{t-1}$ & Weighted moving average emphasizing recent prices; reacts faster to new information ($\alpha$ is the smoothing factor). \\
\textbf{MACD} & $\text{EMA}_{\text{short}} - \text{EMA}_{\text{long}}$ & Momentum indicator based on difference between short- \& long-term EMAs; captures trend strength \& reversal signals. \\
\textbf{RSI (Relative Strength Index)} & $100 - \frac{100}{1 + \text{RS}}$ \newline \scriptsize ($\text{RS} = \frac{\text{Avg Gain}}{\text{Avg Loss}}$) & Scaled measure of recent gains vs.\ losses; high RSI suggests overbought conditions, low RSI indicates oversold conditions. \\
\textbf{OBV (On-Balance Volume)} & $\text{OBV}_{t-1} + V_t \cdot \text{sgn}(P_t - P_{t-1})$ & Cumulative volume measure linking price movement and trading volume; rising OBV indicates accumulation, falling OBV suggests distribution. \\
\textbf{One-Day Reversal} & $\frac{P_t - P_{t-1}}{P_{t-1}}$ & Daily return from previous close to current close; captures immediate short-term reversals or shocks. \\
\textbf{Max Return (20-day)} & $\max_{1 \le i \le 20} \left( \frac{P_{t-i} - P_{t-i-1}}{P_{t-i-1}} \right)$ & Maximum single-day return observed over the past 20 trading days; indicates short-term volatility extremes. \\
\textbf{Medium-Term Momentum} & $\prod_{i=1}^k (1 + R_{t-i}) - 1$ & Price persistence over weeks or months; positive values indicate sustained trends. \\
\textbf{Long-Term Mean Reversal} & $-(P_t - \bar{P}_{long})$ & Tendency of price to revert toward historical average, representing equilibrium-seeking behavior in markets. \\
\midrule
\multicolumn{3}{l}{\textbf{Company Fundamentals}} \\
\midrule
\textbf{Cash Flow / Assets} & $\frac{\text{Operating Cash Flow}}{\text{Total Assets}}$ & Operating cash flow divided by total assets; measures how efficiently assets generate cash. \\
\textbf{Book / Price (Quarterly)} & $\frac{\text{Book Value of Equity}}{\text{Market Capitalization}}$ & Ratio of book value to market price; high values may indicate undervaluation. \\
\textbf{Earnings / Price (Quarterly)} & $\frac{\text{Earnings Per Share (EPS)}}{\text{Price Per Share}}$ & Inverse of price-to-earnings ratio; higher values imply cheaper valuations relative to earnings. \\
\textbf{Forecast Earnings / Price} & $\frac{\text{Expected Future EPS}}{\text{Price Per Share}}$ & Forward-looking E/P ratio using analyst forecasts; reflects expected profitability. \\
\textbf{Sales / Assets (Quarterly)} & $\frac{\text{Total Sales}}{\text{Total Assets}}$ & Asset turnover ratio; measures how effectively a company uses assets to generate revenue. \\
\textbf{Debt / Assets (Quarterly)} & $\frac{\text{Total Debt}}{\text{Total Assets}}$ & Leverage ratio showing the proportion of assets financed by debt. \\
\textbf{Debt / Equity (Quarterly)} & $\frac{\text{Total Debt}}{\text{Shareholders' Equity}}$ & Ratio of total debt to shareholders’ equity; higher values indicate greater financial leverage. \\
\textbf{Dividend Yield (Quarterly)} & $\frac{\text{Dividends Per Share}}{\text{Price Per Share}}$ & Dividend per share divided by stock price; represent cash return to shareholders. \\
\textbf{Return on Assets (Quarterly)} & $\frac{\text{Net Income}}{\text{Total Assets}}$ & Net income divided by total assets; gauges profitability relative to firm size. \\
\textbf{Return on Equity (Quarterly)} & $\frac{\text{Net Income}}{\text{Shareholders' Equity}}$ & Net income divided by shareholders’ equity; measures profitability relative to owners’ capital. \\
\bottomrule
\end{tabular}
\caption{
Summary of trading signals and company fundamentals used in the question design.
}
\label{tab:volfund-signals}
\end{table*}

\begin{table*}
\centering
\small
\renewcommand{\arraystretch}{1.15}
\scalebox{0.9}{\begin{tabular}{p{1.3cm} p{4cm} p{9cm}}
\toprule
\textbf{ID} & \textbf{Attributes} & \textbf{Description (Example Behaviour)} \\
\midrule
L1 & Single-turn, low detail, instruction-style, role-specified & Instructs the model in simple one-sentence instructions focusing on the high-level goal. \\
L2 & Single-turn, moderate detail, instruction-style, role-specified & Paragraph-style instructions expressing the high-level goal and sub-tasks that need to be performed to achieve the goal\\
L3 & Step-by-step reasoning, moderate detail, decomposed & adds a structured reasoning template (bulleted-list-style): clarify goal, decompose into sub-questions, answer sub-parts using both contexts, and synthesize a final answer. \\
L4 & Step-by-step reasoning, moderate detail, decomposed & ``Level 3 Prompt" + ``It provides a guideline on how LLMs will be evaluated." \\
L5 & Step-by-step reasoning, high detail, decomposed & ``Level 4 Prompt" + ``Provide additional relevant gathered via RAG"\\
L6 & Step-by-step reasoning, Maximalist, high detail, evidence-citing, justified & ``Level 5 Prompt" + ``Provide explicit statement asking LLM to explain its own output."\\
\bottomrule
\end{tabular}}
\caption{TELeR-inspired prompt variants used for multi-prompt generation during benchmark creation \S\ref{sec:benchmark-design}.
}
\label{tab:teler-prompts}
\end{table*}

\subsection{TELeR Prompt Taxonomy}
\label{app:teler}

FinTradeBench uses a TELeR-inspired prompt taxonomy for multi-prompt
candidate generation. Table~\ref{tab:teler-prompts} describes the six TELeR-inspired prompt variants used for multi-prompt candidate generation during benchmark construction \S\ref{sec:benchmark-design}. Prompts vary systematically in instruction richness and reasoning explicitness, from Level~0 (minimal context) to Level~6 (maximal justification).

Table~\ref{tab:teler_taxonomy} describes the  TELeR taxonomy used during the General-QA study (\S\ref{sec:experimental-setup}). No-RAG uses four candidate
prompt variants, and RAG uses two, followed by intra-model self-selection. Prompts vary systematically in instruction richness and reasoning explicitness, from Level 1 (single sentence high-level directive) to Level 6 (maximalist, evidence-citing, self-justified). Levels 1--4 are used without RAG; Levels 5--6 have RAG context.

\begin{table*}
    \centering
    % \scriptsize
    \small
    \begin{tabular}{@{}l l p{8.2cm} c@{}}
        \toprule
        \textbf{Lvl} & \textbf{Type} & \textbf{Strategy} & \textbf{RAG} \\ 
        \midrule
        L1 & Baseline & Single-sentence high-level directive. & No \\
        L2 & Focus & Role specification + Paragraph style breakdown. & No \\
        L3 & CoT & Bulleted Step-by-Step Chain-of-Thought. & No \\
        L4 & Auditor & Adds explicit evaluation criteria (e.g., coherence). & No \\
        \midrule
        L5 & Context & Injects retrieved evidence and citations. & \textbf{Yes} \\
        L6 & Explain & Adds self-justification for data usage. & \textbf{Yes} \\
        \bottomrule
    \end{tabular}
    \vspace{1mm}
    \caption{The TELeR Prompt Taxonomy used for Benchmarking and evaluation for the RAG and No-RAG setting in \S\ref{sec:experimental-setup}. The framework systematically increases prompt complexity, with Levels 5 and 6 designed to leverage RAG.}
    \label{tab:teler_taxonomy}
    \vspace{-2mm}
\end{table*}

% every final experiment. Controlled conditions use a single deterministic final
% answer, with Tool-Use additionally using a deterministic router.

\subsection{Sample Questions, Reference Answers, and Golden Indicators}
\label{app:golden}

Table~\ref{tab:golden_answers_box} presents representative seed questions from each of the three FinTradeBench categories alongside their expert-authored golden answers. Each answer cites the specific golden indicators required for a complete response; these indicators serve as the reference set for Golden Indicator F1 scoring.

\begin{table}
    \centering
    \begin{tcolorbox}[
        enhanced,
        colback=blue!2!white,      % Very light blue background
        colframe=black!70,         % Dark gray frame
        arc=2pt,                   % Slightly rounded corners
        boxrule=0.5pt,             % Thin border
        left=4pt, right=4pt, top=4pt, bottom=4pt, % Padding
        title=\textbf{Sample Golden Answers by Category},
        coltitle=black,
        colbacktitle=gray!15,      % Light gray title background
        attach boxed title to top center={yshift=-2mm},
        boxed title style={size=small,colback=gray!15,colframe=black!70}
    ]
    \small
    
    {\color{blue!60!black}\textbf{1. Company Fundamentals}} \\[2pt]
    \textbf{Q:} \textit{Is Nvidia's profitability sustainable with its valuation in September 2025?} \\
    \textbf{A:} Nvidia's profitability is highly sustainable and supported by strong operational efficiency. Its exceptional financial health is characterized by a ROE of 33.38\% and a ROA of 21.31\%, achieved with low leverage (Debt/Equity of 0.102). However, the valuation is extremely premium, evidenced by a Book/Price ratio of 0.023.
    
    \vspace{-1mm}
    \hdashrule{\linewidth}{0.5pt}{1pt 2pt} 
    % \vspace{1mm}

    {\color{blue!60!black}\textbf{2. Trading Signals}} \\[2pt]
    \textbf{Q:} \textit{Based on trading in H1 2025, which stocks show the strongest EMA support?} \\
    \textbf{A:} INTC, MU, and LRCX show the strongest support, trading approx. 18.5\%, 16.8\%, and 14.4\% above their 20-day EMAs. However, LRCX provides the most sustainable profile, as its technical trend is backed by strong fundamentals (18.7\% ROE), whereas INTC's support coincides with negative profitability metrics.
    
     \vspace{-1mm}
    \hdashrule{\linewidth}{0.5pt}{1pt 2pt}
    % \vspace{1mm}

    {\color{blue!60!black}\textbf{3. Hybrid Reasoning}} \\[2pt]
    \textbf{Q:} \textit{Is Microsoft overvalued based on Q3 2025 prices despite strong performance?} \\
    \textbf{A:} Microsoft is trading at a premium valuation, but it is not necessarily overvalued. While the Book/Price ratio of 0.0929 indicates the stock is expensive relative to its book value, this premium is supported by high operational efficiency (Cash Flow/Assets of 0.0689) and strong profitability (ROE of 0.0890). Technical indicators show strong momentum (RSI 65.42, MACD $>$ Signal) without yet reaching extreme "overbought" conditions ($>$70 RSI).
    
    \end{tcolorbox}
    % \vspace{-5mm}
    \caption{{Sample seed questions \& corresponding golden answers; see details about key financial terms in Table \ref{tab:volfund-signals}.}}
    \label{tab:golden_answers_box}
\end{table}

\subsection{From 450 Candidates to the Human-Validated
Core400.}\label{subsec:core400}
For the main controlled evaluation, we first form a 450-item candidate
Core pool. The pool contains the 150 seed questions together with
company- and year-scaled questions selected for stronger historical
coverage. We then apply two independent quality gates: \myNum{i}\smartparagraph{Canonical-data support screening.}
We verify that the frozen canonical financial-data universe can support each
question and that the candidate contributes distinct evaluation coverage.
47 questions did not pass this screening because either
the required company or historical evidence was insufficient or the question was
redundant/repetitive with the retained pool. 
% This gave \emph{Core403}, containing
% 143 Fundamental, 124 Trading, and 136 Hybrid questions.
\myNum{ii}\smartparagraph{Full item-level human quality audit.}
An expert then audits all 403 retained questions to evaluate \myNum{i} question clarity and answerability, \myNum{ii} correctness of the reference answer, \myNum{iii} support from canonical point-in-time evidence, \myNum{iv} adequacy of the golden-indicator set, and \myNum{v} consistency of company identity and historical cutoff.
The reviewer accepts 400 of 403 items (99.3\%) thus giving us the final \emph{Core400}: 143 Fundamental, 122 Trading, and
135 Hybrid questions. Table~\ref{tab:core-curation} summarizes this curation process.
% The retained set contains 147 seed questions,
% 85 company-scaled questions, and 168 year-scaled questions.
% \input{sections/main_tables/table_core_curation_450_403_400}
The human audit provides an additional content-quality check beyond the
initial automated support screen. Across Core403, 400/403 questions
receive a clean clarity pass, 401/403 reference answers are marked
correct, 402/403 golden-indicator sets are judged adequate, and all 403 pass
entity/time alignment. The full audit breakdown is reported in~\S\ref{app:core-human-audit}.
Unless otherwise stated, all final quantitative experiments in this
paper use the human-validated \emph{Core400}. See~\S\ref{app:quality-control} for detailed steps taken for benchmark quality control.
% Automated evaluation coverage can occasionally be slightly below 400 when a model-specific evidence pipeline cannot construct a valid input; such cases are marked \texttt{SOURCE\_INVALID} rather than assigned an artificial score.

% --------------------------------------------------------------------
% \smartparagraph{Benchmark Quality Control.}
% FinTradeBench applies quality control at multiple distinct
% levels. First, expert-defined golden indicators specify a minimal evidence set
% for grounded reasoning and reduce reliance on generic financial explanations. Second, candidate responses undergo structured numerical auditing against
% historical financial data. Third, the construction-time LLM evaluator is calibrated against human
% judgments on the expert seed material. Fourth, the 450-item Core candidate pool undergoes canonical-data support
% screening before experimental use. Fifth, all 403 support-screened Core items undergo a full human
% item-level audit, giving the final Core400 used in this paper. Finally, the human-aligned LLM judge used for model evaluation is frozen and
% validated separately on a held-out blinded human review.

% Together, these checks separate evidence availability, annotation
% quality, benchmark-content validation, and evaluator reliability rather
% than treating them as a single quality-control stage. Detailed audit
% criteria, excluded-item documentation, numerical-audit definitions, and
% judge-validation statistics are provided in the appendix.
\subsection{Core Candidate Provenance}
\label{app:core-provenance}

\begin{table}
\centering
\small
\begin{tabular}{lrrrr}
\toprule
Stage & Total & F & T & FT \\
\midrule
Candidate pool & 450 & 163 & 124 & 163 \\
 Canonical-data screening & 403 & 143 & 124 & 136 \\
 Human audit & \textbf{400} & \textbf{143} & \textbf{122} & \textbf{135} \\
\bottomrule
\end{tabular}
\caption{\small{\textbf{FinTradeBench Core curation}. We begin with 450 candidate items, then apply canonical-data support screening, followed by item-level human audit, producing the final human-validated Core400.}}
\label{tab:core-curation}
% \vspace{-5mm}
\end{table}

The auditable 450-item candidate pool consists of 150 seed-origin,
114 company-scaled, and 186 year-scaled items. Canonical-data support screening
removes 47 items (29 company-scaled and 18 year-scaled), yielding \emph{Core403}.

\subsection{Full Human Quality Audit of \emph{Core403}}
\label{app:core-human-audit}

After the canonical-data support screen, a financial expert audited all items in \emph{Core403} one at a time. For each item, the reviewer inspected the question,
reference answer, expert-defined golden indicators, canonical point-in-time
evidence, and entity/time alignment. The audit was a benchmark-quality review,
not a model-response evaluation.

\begin{table}
\centering
\small
\begin{tabular}{lrr}
\toprule
Human-audit criterion & Count & Rate \\
\midrule
Question clarity: Pass & 400/403 & 99.3\% \\
Question clarity: Pass or Minor & 403/403 & 100.0\% \\
Reference answer: Correct & 401/403 & 99.5\% \\
Canonical evidence: Full support & 399/403 & 99.0\% \\
Golden indicators: Adequate & 402/403 & 99.8\% \\
Entity/time alignment: Pass & 403/403 & 100.0\% \\
No fatal flaw & 401/403 & 99.5\% \\
Accepted into final Core & \textbf{400/403} & \textbf{99.3\%} \\
\bottomrule
\end{tabular}
\caption{\small{\textbf{Full human quality audit of Core403.} High pass rates
reflect the preceding canonical-data support screen. The audit verifies the
filtered Core and should not be interpreted as a quality estimate for the
broader 1,400-question collection.}}
\label{tab:core403-human-audit}
\end{table}

Three items are excluded after audit: FTB-0061 (Hybrid), FTB-0135 (Trading),
and FTB-0149 (Trading). The final evaluation set therefore contains 143
Fundamental, 122 Trading, and 135 Hybrid (see Table~\ref{tab:core-curation}). We preserve the pre-audit
counts and exclusion log in the released benchmark metadata rather than
silently overwriting the audit history. Automated evaluation coverage can occasionally be slightly below 400 when a model-specific evidence pipeline cannot construct a valid input; such cases are marked \texttt{SOURCE\_INVALID} rather than assigned an artificial score.

% One Core400 item cannot be assembled by the automated evidence pipeline,
% although the human audit verified that the underlying evidence exists in the
% dataset. We keep the benchmark item and treat the missing automated context as
% a pipeline-coverage failure (\texttt{SOURCE\_INVALID}), not as a
% benchmark-content error.

\subsection{Construction-Time Scaled-Set Screening}
\label{app:screening}

Benchmark scaling includes a separate fatal-flaw screen of the broader
1,400-question collection as a final screening before full benchmark release. For this, a Computer Science PhD student went through all the question-answer pairs. The idea was to screen for any catastrophic generation failures (e.g., wrong company, incomplete output, or
abrupt truncation), not to serve as a full expert validation of financial
reasoning. Any ambiguous responses were passed on to the financial expert for screening. Forty-two responses (3\%) were regenerated after this screen. Some qualitative samples are provided in Table~\ref{tab:screening}.

\begin{table}
    \centering
    \begin{tcolorbox}[
        enhanced,
        colback=blue!2!white,      % Very light blue background
        colframe=black!70,         % Dark gray frame
        arc=2pt,                   % Slightly rounded corners
        boxrule=0.5pt,             % Thin border
        left=4pt, right=4pt, top=4pt, bottom=4pt, % Padding
        title=\textbf{Sample bad responses by severity},
        coltitle=black,
        colbacktitle=gray!15,      % Light gray title background
        attach boxed title to top center={yshift=-2mm},
        boxed title style={size=small,colback=gray!15,colframe=black!70}
    ]
    \small
    
    {\color{blue!60!black}\textbf{1. Catastrophic failure}} \\[2pt]
    \textbf{Q:} \textit{Is Walmart a good defensive stock for September 2025?} \\
    \textbf{Response:} ..no direct assessment of its performance or defensive characteristics can be made. Proxy consumer staples and retail stocks (e.g., COST, PEP, KHC) exhibited mixed or negative returns during the period, with no clear defensive outperformance...\\
    \textbf{Reason:} Does not talk about Walmart. Starts comparing different proxy stocks instead of focusing on Walmart.
    
    \vspace{-1mm}
    \hdashrule{\linewidth}{0.5pt}{1pt 2pt} 
    % \vspace{1mm}

    {\color{blue!60!black}\textbf{2. Incomplete answer}} \\[2pt]
    \textbf{Q:} \textit{What companies have the most reliable moving average signals, based on 2025 back-testing?} \\
    \textbf{A:} ...The company displays a bearish Moving Average signal configuration. The last price (45.7400) is trading significantly below both the 20-day Simple Moving Average (49.6850) and the 20-day Exponential Moving Average (49.3665). This suggests that the short-term trend is negative. This technical weakness is corroborated by the fundamental data,\\
    \textbf{Reason:} Focuses on one company only.
    
     \vspace{-1mm}
    \hdashrule{\linewidth}{0.5pt}{1pt 2pt}
    % \vspace{1mm}

    {\color{blue!60!black}\textbf{3. Sudden break}} \\[2pt]
    \textbf{Q:} \textit{During July 2025, what stocks are breaking out of long-term downtrends?} \\
    \textbf{A:} ........Additionally, **AMD**, **Synopsys (SNPS)**, and **The Trade Desk (TTD)** are breaking out of *medium-term* corrections. These stocks showed negative momentum over the 2-to-12-month period but reversed sharply in July with gains ranging from 18\% to 29\%, supported by strong MACD crossover signals......\\
    \textbf{Reason:} Breaks off suddenly without answering the main question.
    \end{tcolorbox}
    % \vspace{-5mm}
    \caption{{Sample bad responses caught during screening.}}
    \label{tab:screening}
    \vspace{-4mm}
\end{table}

\subsection{Benchmark Quality Control.}\label{app:quality-control}
FinTradeBench applies quality control at multiple distinct
levels. First, expert-defined golden indicators specify a minimal evidence set
for grounded reasoning and reduce reliance on generic financial explanations. Second, candidate responses undergo structured numerical auditing against
historical financial data. Third, the construction-time LLM evaluator is calibrated against human
judgments on the expert seed material and then validated by a human using a scaled subset. Fourth, the 450-item Core candidate pool undergoes canonical-data support
screening before experimental use. Fifth, all 403 support-screened Core items undergo a full human
item-level audit, giving the final Core400 used in this paper. Finally, the human-aligned LLM judge used for model evaluation is frozen and
validated separately on a held-out blinded human review.

Together, these checks separate evidence availability, annotation
quality, benchmark-content validation, and evaluator reliability rather
than treating them as a single quality-control stage. Detailed audit
criteria, excluded-item documentation, numerical-audit definitions, and
judge-validation statistics are provided in the appendix.

% ============================================================
\section{Addendum to Experimental Setup and Evaluation}
\label{app:experimental-details}

% \subsection{Evaluation Prompts and Human Annotation Rubric}
% \label{app:evaluation_prompts}

\subsection{Models Evaluated.}
\label{subsec:models}
For the General-QA study, we evaluate six contemporary systems:
\emph{Gemini-3.8-Flash},
\emph{DeepSeek-V4-Flash},
\emph{Llama-4-Scout-17B-16E-Instruct},
\emph{Qwen3.8-27B},
\emph{GPT-5.4-mini}, and
\emph{Ministral-3-8B-Instruct-2512}. At the time of research, \emph{Gemini 3.8 Flash} leads Vals Finance Agent v2 with 61.44 $\%$ accuracy \cite{finance_agent_v2}.
These systems span API-hosted and open-weight models and provide a
cross-model comparison without relying on parameter-count groupings.

For the controlled evidence study, we evaluate seven models spanning
multiple model families and generations:
\emph{Qwen2.5-32B-Instruct},
\emph{DeepSeek-R1-Distill-Qwen-32B},
\emph{Llama-3.1-8B-Instruct},
\emph{Phi-4},
\emph{Llama-4-Scout},
\emph{Qwen3.8-27B}, and
\emph{GPT-5.4-mini}.
This broader controlled set tests whether evidence-pathway effects are
consistent across heterogeneous model architectures and capability levels.

The time-series study evaluates
\emph{ChatTS-8B} and a
\emph{GPT-5.4-mini + TimesFM~2.5} forecast-feature pipeline.
Exact model identifiers, provider endpoints, inference parameters,
hardware, and runtime details are reported in
Appendix~\ref{app:compute}.

\subsection{Why Retrieval and Feature Augmentation over Parametric Fine-Tuning?}
\label{app:rag-over-finetuning}

We prioritize retrieval-augmented generation and modular feature injection over parameter fine-tuning because financial analysis demands reasoning over exact, point-in-time evidence rather than static stylistic adaptation. Financial conclusions hinge on non-stationary numerical values across SEC filings and historical OHLCV series; relying strictly on parametric memory risks hallucination and prevents verifiable evidence auditing. Grounding parametric models with non-parametric external memory preserves provenance and supports factuality in knowledge-intensive domains~\citep{lewis2020retrieval}, while allowing continuous ingestion of newly released filings and market ticks without prohibitive retraining costs~\citep{gao2023retrieval}. Recent empirical studies confirm that retrieval provides more dependable factual injection than supervised fine-tuning, which often adapts conversational tone without ensuring precise numerical recall~\citep{ovadia2024fine}. 

Our empirical results in the time-series specialist evaluation (\S\ref{subsec:results-timeseries}, RQ3) reinforce this distinction: a fine-tuned open-weight reasoning model on time-series representations, \emph{ChatTS} underperformed a modular pipeline coupling a dedicated forecasting model \emph{TimesFM} with a frozen frontier reasoning model \emph{GPTmini} by 1.09 Likert points overall and by 1.60 points on Hybrid tasks (Figure~\ref{fig:time_series}). End-to-end parametric fine-tuning of smaller models often degrades complex cross-signal synthesis and numerical reasoning capabilities compared to modular pipelines that supply structured, up-to-date context to state-of-the-art general-purpose reasoners. Hence, our benchmark focuses on evaluating how effectively models reason over point-in-time financial evidence supplied dynamically at inference time.
% --------------------------------------------------------------------

\subsection{RAG architecture explained}
\label{app:rag-details}
% \smartparagraph{Point-in-Time Dual-Track Retrieval.}
% \label{app:rag-details}

Here, we provide details for the architecture used for raw RAG vs No-RAG evaluation study described in \S\ref{subsec:general-rag}.
\begin{figure*}[htbp!]
\centering
\includegraphics[width=\textwidth]{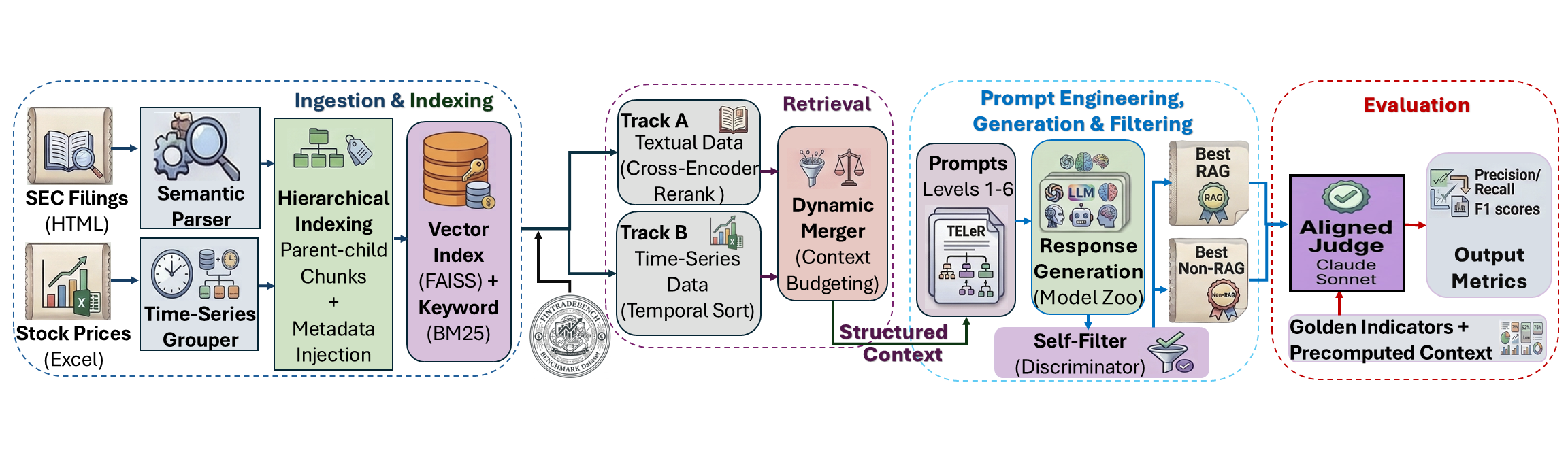} 
\vspace{-5.5mm}
\caption{\small{\textbf{Overview of the RAG.} The pipeline features a dual-track retrieval engine that processes unstructured text and structured time-series separately. The generation phase utilizes TELeR taxonomy to produce multiple candidate responses across the model zoo, which are filtered via self-selection before final evaluation.}}
\label{fig:rag_architecture}
\vspace{-4.5mm}
\end{figure*}

\myNum{1} \smartparagraph{Domain-Aware Hybrid Retrieval via RAG.}\label{sec:rag-architecture}
We design a RAG-based architecture for the dual nature of financial analysis by integrating text and tabular data with numerical time-series data. Our four-part RAG implements a Dual-Track Retrieval Engine followed by TELeR-guided generation and self-filtering before evaluation; see Figure~\ref{fig:rag_architecture}. This design follows the general RAG principle of grounding generation in external evidence~\citep{lewis2020retrieval}, while adapting retrieval to the financial setting. We use a domain-aware dual-track retrieval architecture because
financial evidence has two structurally different forms. \myNum{i} \emph{SEC track.}
Regulatory filings are parsed from HTML; non-content elements are
removed, financial tables are converted into text-preserving
representations, and filing sections are indexed hierarchically.
Retrieval combines semantic and lexical matching followed by reranking,
while larger parent sections are returned for generation to preserve
financial context. \myNum{ii} \emph{Market-data track.}
Historical OHLCV observations are retrieved through a separate temporal
path prioritizing ticker and date alignment rather than semantic
similarity. This prevents natural-language retrieval models from
underweighting structured numerical evidence.

\myNum{2} \smartparagraph{Data Ingestion and Indexing.} Financial documents often contain high token density and complex tables that standard chunking corrupts. For regulatory filings, we parse SEC HTML documents, remove non-content elements, preserve section boundaries, and convert embedded financial tables into markdown-formatted text. To handle the complexities of the domain, we adopt two strategies: \myNum{a}
\emph{Hierarchical Indexing:}A parent-child strategy segments documents by logical boundaries (e.g., SEC ``Item 7'') \cite{shaukat2026systematic, zhou2026beyond, lewis2020retrieval}. Each filing is segmented into parent sections (max 2,000 tokens) and child retrieval chunks of 300 tokens with overlap. Retrieval is performed over child chunks, while generation receives the corresponding parent context to preserve document-level coherence. \myNum{b}
\emph{Metadata Injection:} Structured metadata is prepended to every chunk embedding to mitigate temporal hallucination.
For trading data, daily OHLCV histories are grouped into month-level, ticker-specific chunks, enabling temporal alignment between the question date and the relevant market evidence.

\myNum{3} \smartparagraph{Dual-Track Retrieval.}
% We use a dual-track retrieval architecture to handle the asymmetric structure of financial evidence; see Figure~\ref{fig:rag_architecture}.~\emph{Track A} indexes SEC filings using parent--child chunking: smaller child chunks are embedded for retrieval, while larger parent sections are returned to preserve document-level context. 
% Retrieval over this track combines dense embeddings (\texttt{BAAI/bge-large-en-v1.5}), BM25 lexical matching, and cross-encoder re-ranking (\texttt{ms-marco-MiniLM-L-6-v2}).~\emph{Track B} indexes market data as time period-aligned price chunks retrieved via an auxiliary temporal query mechanism; these chunks bypass cross-encoder re-ranking, as semantic relevance models tend to underweight structured time-series evidence.
% At query time, we dynamically merge the outputs from both tracks: after ticker detection, the system retrieves evidence independently per track, applies source-specific quotas, filters by temporal relevance, and removes duplicate parent contexts. We assemble the final prompt under a global token budget, balancing long-form financial texts containing company fundamentals, with short-horizon market evidence needed to calculate trading signals, while preserving signal-specific retrieval strengths.
We use dual-track RAG to handle the asymmetric structure of financial evidence; see Figure~\ref{fig:rag_architecture}. In~\emph{Track A}, for each ticker, we build a FAISS dense vector index using \texttt{BAAI/bge-large-en-v1.5} embeddings and maintain a BM25 lexical index for sparse matching. At query time, the system first detects explicit tickers, company aliases, or market groups, and extracts temporal constraints such as the target year. It retrieves SEC evidence using dense similarity followed by cross-encoder reranking with \texttt{ms-marco-MiniLM-L-6-v2}. In ~\emph{Track B}, trading-signal chunks are retrieved through a separate temporal query path that prioritizes ticker and date alignment. This separation is crucial because semantic rerankers trained primarily on natural language may undervalue structured numerical evidence. Retrieved SEC and trading contexts are merged under source-specific quotas and global context budget, with parent-level deduplication to avoid redundancies.

\myNum{4} \smartparagraph{Generation and self-selection.}
\label{sec:teler_prompting}
% In the generation phase, we use the TELeR taxonomy \citep{karmaker2023teler} to generate six distinct prompts per question, ranging from simple directives (L1) to complex RAG-aware reasoning tasks (L6); see \S \ref{app:teler} Table~\ref{tab:teler_taxonomy}. This reduces reasoning errors associated with any single prompt structure. A self-selection module evaluates these candidates against the retrieved context to identify the best RAG and best No-RAG response per model, consistent with the self-filtering approach used during benchmark construction; see \S\ref{subsec:phase1}.
Given the retrieved evidence, each model generates multiple candidate answers using TELeR-inspired prompt variants ranging from simple direct answering (L1) to evidence-grounded, step-by-step reasoning (L6); see \S \ref{app:teler} Table~\ref{tab:teler_taxonomy}. The model then performs intra-model self-selection among its own candidates against the retrieved context to identify the best RAG and best No-RAG response per model, consistent with the self-filtering approach used during benchmark construction; see \S\ref{subsec:phase1}. This design reduces sensitivity to a single prompt template while avoiding cross-model preference leakage.
% FinTradeBench uses a domain-aware dual-track retrieval architecture because
% financial evidence has two structurally different forms.

% \myNum{i} \emph{SEC track.}
% Regulatory filings are parsed from HTML; non-content elements are
% removed, financial tables are converted into text-preserving
% representations, and filing sections are indexed hierarchically.
% Retrieval combines semantic and lexical matching followed by reranking,
% while larger parent sections are returned for generation to preserve
% financial context.
% \myNum{ii} \emph{Market-data track.}
% Historical OHLCV observations are retrieved through a separate temporal
% path prioritizing ticker and date alignment rather than semantic
% similarity. This prevents natural-language retrieval models from
% underweighting structured numerical evidence.
At inference time, evidence from the two tracks is merged under
source-specific quotas and a shared context budget. 
% Ticker resolution,
% temporal filtering, parent--child chunking, embedding models, rerankers,
% chunk sizes, and retrieval hyperparameters are provided in
% Appendix~\ref{app:rag-details}.
% We might need to keep exact chunk sizes, embedding model, BM25, reranker, metadata fields,  retrieval quotas, and context budgets here rather than in the main paper.

% \subsection{Generation Budgets and Self-Selection}
% \label{app:generation-details}

% The General-QA study uses the following generation protocol:
% No-RAG generates four TELeR candidates and RAG generates two, with the same
% model selecting its final answer within each condition. This asymmetry is why
% RAG--No-RAG is reported as a benchmark comparison. In contrast, the
% Indicator-RAG, Tool-Use, and Oracle controlled conditions use a single
% deterministic final generation; Tool-Use additionally invokes a deterministic
% question-to-tool router.

\subsection{Evidence Conditions}
\label{app:condition-definitions}

Table~\ref{tab:evidence-conditions} details the evidence provided for each evaluation pipeline.
\begin{table*}[t]
\centering
\small
\begin{tabular}{p{2.2cm} p{3.9cm} p{4.0cm} p{4.0cm}}
\toprule
Condition & Fundamental \emph{(F)} & Trading \emph{(T)} & Hybrid \emph{(FT)} \\
\midrule
No-RAG &
Question only &
Question only &
Question only \\
Raw-RAG &
Point-in-time SEC evidence &
Point-in-time raw OHLCV &
Point-in-time SEC + raw OHLCV \\
Indicator-RAG &
SEC + precomputed fundamental indicators &
Precomputed trading indicators &
SEC + fundamental + trading indicators \\
Tool-Use &
SEC evidence &
Raw OHLCV + deterministic tool routing &
SEC + raw OHLCV + deterministic tool routing \\
Oracle &
Gold-selected indicator evidence &
Gold-selected indicator evidence &
Gold-selected indicator evidence \\
\bottomrule
\end{tabular}
\caption{\small{\textbf{Evidence conditions.} Indicator-RAG does not use
golden-indicator annotations for selection. Tool routing depends only on the
question and declared tool inventory. Oracle supplies gold-selected evidence,
not the benchmark reference response.}}
\label{tab:evidence-conditions}
\end{table*}

\subsection{Tool-Use Implementation}
\label{app:tool-details}

Tool-Use exposes a fixed inventory of deterministic financial computations.
The router is deterministic and conditions only on the benchmark question and
the available tool descriptions. It does not receive the reference answer,
golden-indicator labels, or judge scores.

\subsection{Time-Series Specialist Pipelines}
\label{app:timeseries-details}

The time-series evaluation contains 257 eligible Core400 questions:
122 Trading and 135 Hybrid. ChatTS-8B reasons directly over the historical
adjusted-close series. The TimesFM~2.5 pipeline uses the latest 128
adjusted-close observations available at the point-in-time cutoff to generate
a five-step forecast; forecast-derived features are then provided to
GPT-5.4-mini for natural-language synthesis. Hybrid questions additionally
receive point-in-time SEC evidence.

The comparison is between complete pipelines. It is not a pure architectural
ablation because the downstream language components differ.

\subsection{Evidence-Aware Judge and Common-Evidence Re-Judging}
\label{app:evaluation-prompts}

The final evaluator is Claude Sonnet~4.5 with deterministic structured output.
Each answer is scored from 1--5 on Accuracy, Completeness, Relevance, and
Clarity. The benchmark reference response and
\texttt{answer\_conclusion\_summary} are never supplied to the judge as factual
ground truth. Rubric scoring is anchored to point-in-time evidence.

For the General-QA comparison, both No-RAG and RAG answers are re-judged
against the same condition-independent canonical evidence package. This removes
evaluator-side evidence asymmetry while leaving the historical generation
protocol unchanged.

Golden-indicator precision, recall, and F1 are ancillary diagnostics. They are
not used as the primary response-quality score and should not be interpreted as
proof that an answer using a different but defensible evidence set is wrong.
\begin{figure*}
% \centering
\includegraphics[width=\linewidth]{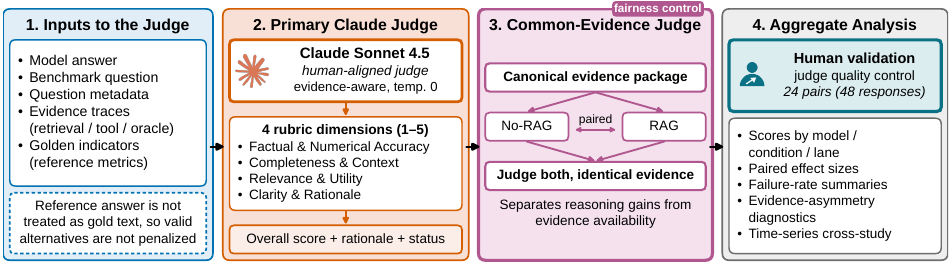} 
% \vspace{-5.5mm}
\caption{\textbf{Controlled-condition evaluation.} The question
set, cutoff rules, model, answer budget, and judge are held fixed; only the
evidence pathway changes. Indicator-RAG retrieves precomputed indicator and
fundamental tables, Tool-Use lets the model query indicator values on demand,
and Oracle supplies gold evidence as a probable upper bound, separating failures of
evidence access from failures of reasoning.}
\label{fig:judge}
% \vspace{-7.5mm}
\end{figure*}

% or treats reference metrics as factual truth.
% Include the actual frozen final prompt/schema from the released judge code.

\subsection{Evaluation Metrics and Statistical Testing}
\label{app:metrics}

For response $i$, let the four rubric scores be
$A_i,C_i,R_i,L_i \in \{1,\ldots,5\}$ for Accuracy, Completeness,
Relevance, and Clarity. The primary response score is

\begin{equation}
S_i = \frac{A_i+C_i+R_i+L_i}{4}.
\label{eq:judge-mean}
\end{equation}

We report raw 1--5 scores as the primary metric. When useful for presentation,
we additionally report the normalized judge score
$100S_i/5$; this value is not called ``absolute accuracy.''

For paired conditions $X$ and $Y$, the per-question effect is

\begin{equation}
d_i = S_i^{(X)} - S_i^{(Y)},
\qquad
\bar d = \frac{1}{N}\sum_{i=1}^{N}d_i.
\label{eq:paired-delta}
\end{equation}

We compute 95\% confidence intervals for $\bar d$ using 5,000 paired bootstrap
resamples with a fixed random seed. Statistical significance is assessed with
the paired Wilcoxon signed-rank test, and $p$-values are corrected using the
Holm procedure within each study family.

Rows without valid source evidence are marked \texttt{SOURCE\_INVALID};
judge failures are recorded separately. We do not impute artificial scores for
invalid rows.

% Golden-indicator precision, recall, and F1 are computed as

% \begin{align}
% P_{\mathrm{GI}} &= \frac{|M_{\mathrm{gen}}\cap M_{\mathrm{GI}}|}
% {|M_{\mathrm{gen}}|},\notag
% \\
% R_{\mathrm{GI}} &= \frac{|M_{\mathrm{gen}}\cap M_{\mathrm{GI}}|}
% {|M_{\mathrm{GI}}|},\notag
% \\
% F1_{\mathrm{GI}} &=
% \frac{2P_{\mathrm{GI}}R_{\mathrm{GI}}}{P_{\mathrm{GI}}+R_{\mathrm{GI}}}.
% \label{eq:gi-f1}
% \end{align}
Metric Extraction Precision ($\text{P}(a_m^\star)$), Recall ($\text{R}(a_m^\star)$), and F1 ($\text{F1}(a_m^\star)$) quantifies the overlap between the set of financial metrics cited in a generated response and the expert-defined reference set. For a self-selected response $a_m^\star$, let $M_{\text{gen}}(a_m^\star)$ denote the set of financial metrics mentioned in the response and $M_{\text{ref}}$ the ground-truth reference set. Precision, recall, and F1 are then given by:
% \begin{align}
% \text{P}(a_m^\star) &= \frac{|M_{\text{gen}} \cap M_{\text{ref}}|}{|M_{\text{gen}}|},\notag \\
% \text{R}(a_m^\star) &= \frac{|M_{\text{gen}} \cap M_{\text{ref}}|}{|M_{\text{ref}}|},  \notag \\
% {\rm and}\;\;\text{F1}(a_m^\star) &= \frac{2 \cdot \text{P}(a_m^\star) \cdot \text{R}(a_m^\star)}{\text{P}(a_m^\star) + \text{R}(a_m^\star)},
% \label{eq:f1}
% \end{align}

\begin{equation}
\left\{
\begin{aligned}
    \text{P}(a_m^\star) &= \frac{|M_{\text{gen}}\cap M_{\text{ref}}|}{|M_{\text{gen}}|}, \\
    \text{R}(a_m^\star) &= \frac{|M_{\text{gen}}\cap M_{\text{ref}}|}{|M_{\text{ref}}|}, \\
    \text{F1}(a_m^\star) &= \frac{2 \cdot \text{P}(a_m^\star) \cdot \text{R}(a_m^\star)}{\text{P}(a_m^\star) + \text{R}(a_m^\star)} 
\end{aligned}
\right\}
\label{eq:f1}
\end{equation}

respectively. The macro-averaged F1 across all $K$ models is:
\begin{align}
\overline{\text{F1}} = \frac{1}{K}\sum_{m=1}^{K}\text{F1}(a_m^{\star}),
\label{eq:f1_avg}
\end{align}
where $K$ is the total number of evaluated models and $a_m^\star$ is the self-selected best response from model $M_m$.

These overlap metrics are secondary diagnostics rather than the primary
measure of answer quality.

\smartparagraph{Statistical Analysis.}
\label{subsec:statistics}

All condition comparisons are performed at the question level using
paired observations wherever both outputs are valid. We report the
paired mean difference in raw judge Likert score together with a
non-parametric 95\% bootstrap confidence interval based on 5,000 paired
resamples.

Statistical significance is assessed using the paired Wilcoxon
signed-rank test, with $p$-values corrected using the Holm procedure
within each study family.

We do not assign artificial scores to generation, retrieval, or evidence
failures. Such cases are marked as \texttt{SOURCE\_INVALID} or judge
errors and are omitted from a paired comparison when a valid response is
unavailable.

\subsection{Compute Resources and Reproducibility}
\label{app:compute}

% KEEP the useful old compute paragraph, but UPDATE providers/models to the

Open-weight inference was run on GPU-enabled university cluster resources,
while hosted systems and the automated judge were accessed through their
corresponding cloud APIs. 
% % The released artifact should report exact model
% identifiers, provider/deployment identifiers where reproducibly shareable,
% prompt templates, retrieval settings, random seeds, condition routing logic,
% and analysis scripts. Because API-hosted models can change over time,
% versioned model identifiers and frozen output files should be preserved with
% the benchmark release.
\begin{table*}[t]
\centering
\scriptsize
\renewcommand{\arraystretch}{1.15}
\setlength{\tabcolsep}{2.8pt}

\resizebox{\textwidth}{!}{
\begin{tabular}{p{2.8cm} p{2.2cm} p{4.2cm} p{4.0cm} p{7.2cm}}
\toprule
\textbf{System} & \textbf{Study} & \textbf{Exact model identifier} & \textbf{Provider / endpoint} & \textbf{Inference configuration} \\
\midrule

DeepSeek-V4-Flash~\citep{deepseekai2026deepseekv4}
& General QA
& \texttt{DeepSeek-V4-Flash}
& Azure AI Foundry, OpenAI-compatible v1; deployment \texttt{ftb-deepseek-v4-flash}
& Max output 1,200 tokens. The endpoint is invoked without custom temperature or top-$p$. Visible \texttt{<think>} blocks are removed before scoring. API timeout 900\,s with up to six retries. \\

GPT-5.4-mini~\citep{openai2026gpt54mini}
& General QA
& \texttt{gpt-5.4-mini}
& Azure AI Foundry, OpenAI-compatible v1; deployment \texttt{ftb-gpt54mini}
& Max completion 1,200 tokens; \texttt{reasoning\_effort=medium}. No custom temperature or top-$p$ is supplied to the endpoint. API timeout 900\,s with up to six retries. \\

Gemini-3.8-Flash~\citep{google2026gemini38flash}
& General QA
& \texttt{gemini-3.8-flash}
& Google Vertex AI v1 (\texttt{google-genai}); location \texttt{global}
& Max output 1,200 tokens; \texttt{thinking\_level=MEDIUM}. Deprecated temperature/top-$p$ controls are not supplied. API timeout 900\,s with up to six retries. \\

Llama-4-Scout~\citep{meta2025llama4}
& General QA + Controlled
& \texttt{Llama-4-Scout-17B-16E-Instruct}
& Azure AI Foundry, OpenAI-compatible v1; deployment \texttt{ftb-llama4-scout}
& General QA: max output 10,000 tokens; candidate generation uses temperature $0.1$, top-$p=0.9$; self-selection is deterministic (temperature $0$, top-$p=1$). Controlled conditions use a 10,000-token answer ceiling; Tool-Use router ceiling 1,024 tokens. \\

Qwen3.8-27B~\citep{qwenteam2026qwen38}
& General QA + Controlled
& \texttt{Qwen3.8-27B}
& Local vLLM~\citep{kwon2023vllm}, OpenAI-compatible loopback endpoint
& General QA: max output 1,200 tokens; candidate generation temperature $0.1$, top-$p=0.9$; deterministic self-selection; seed 42; thinking explicitly disabled. Controlled: max output 1,200 tokens; deterministic generation; router ceiling 120 tokens. Serving: BF16, max model length 32,768, GPU-memory utilization 0.90. \\

Ministral-3-8B~\citep{liu2026ministral3}
& General QA
& \texttt{Ministral-3-8B-Instruct-2512}
& Local vLLM~\citep{kwon2023vllm}, OpenAI-compatible loopback endpoint
& Max output 1,200 tokens; candidate generation temperature $0.1$, top-$p=0.9$; deterministic self-selection; seed 42. Serving: BF16, max model length 32,768, GPU-memory utilization 0.90, Mistral tokenizer/config/load format. \\

Qwen2.5-32B~\citep{qwen2025qwen25}
& Controlled
& \texttt{Qwen2.5-32B-Instruct}
& Local Hugging Face / Transformers~\citep{wolf2020transformers}
& 4-bit model loading; max output 1,200 tokens; deterministic greedy generation (\texttt{do\_sample=False}); seed 42; Tool-Use router ceiling 120 tokens. \\

R1-Distill-Qwen-32B~\citep{deepseekai2025deepseekr1}
& Controlled
& \texttt{DeepSeek-R1-Distill-Qwen-32B}
& Local Hugging Face / Transformers~\citep{wolf2020transformers}
& 4-bit model loading; max output 1,200 tokens; deterministic greedy generation (\texttt{do\_sample=False}); seed 42; Tool-Use router ceiling 120 tokens. \\

Llama-3.1-8B~\citep{grattafiori2024llama3}
& Controlled
& \texttt{Llama-3.1-8B-Instruct}
& Local Hugging Face / Transformers~\citep{wolf2020transformers}
& 4-bit model loading; max output 1,200 tokens; deterministic greedy generation (\texttt{do\_sample=False}); seed 42; Tool-Use router ceiling 120 tokens. \\

Phi-4~\citep{abdin2024phi4}
& Controlled
& \texttt{phi-4}
& Local Hugging Face / Transformers~\citep{wolf2020transformers}
& 4-bit model loading; max output 1,200 tokens; deterministic greedy generation (\texttt{do\_sample=False}); seed 42; Tool-Use router ceiling 120 tokens. \\

GPT-5.4-mini~\citep{openai2026gpt54mini}
& Controlled
& \texttt{gpt-5.4-mini}
& Azure AI Foundry, OpenAI-compatible v1; deployment \texttt{ftb-gpt54mini}
& Max completion 3,000 tokens; \texttt{reasoning\_effort=none}; no custom temperature/top-$p$; Tool-Use router ceiling 120 tokens. \\

ChatTS-8B~\citep{xie2024chatts}
& Time series
& \texttt{bytedance-research/ChatTS-8B}
& Local Hugging Face / Transformers~\citep{wolf2020transformers}
& FP16 CUDA inference; deterministic decoding (\texttt{do\_sample=False}); 128 adjusted-close observations per series; minimum usable history 64 points; final-answer cap 2,000 tokens; chunk-analysis cap 600 tokens. \\

TimesFM~2.5~\citep{das2024timesfm,google2025timesfm25} + GPT-5.4-mini~\citep{openai2026gpt54mini}
& Time series
& \texttt{TimesFM 2.5} + \texttt{gpt-5.4-mini}
& TimesFM via BigQuery \texttt{AI.FORECAST}; GPT via Azure AI Foundry, deployment \texttt{ftb-gpt54mini}
& TimesFM receives the latest 128 point-in-time daily adjusted-close observations and forecasts the next 5 trading steps. Forecast-derived features are passed to GPT-5.4-mini; GPT max completion 3,000 tokens; \texttt{reasoning\_effort=none}. \\

Claude Sonnet~4.5~\citep{anthropic2025sonnet45}
& Automated judge
& \texttt{claude-sonnet-4-5@20250929}
& Google Vertex AI; region \texttt{us-east5}
& Temperature $0$; no extended thinking; structured output; max judge output 2,000 tokens; maximum evidence payload 70,000 characters. The benchmark reference answer is not supplied as factual ground truth. \\

\bottomrule
\end{tabular}
}
\caption{\small{\textbf{Model identifiers, serving backends, and inference
configuration.}
General-QA models generate four No-RAG candidates and two RAG candidates before
intra-model self-selection. Controlled Indicator-RAG, Tool-Use, and Oracle
conditions use a single final generation; Tool-Use additionally invokes the
deterministic router. Llama-4-Scout is an open-weight Meta model evaluated
through Azure AI Foundry. TimesFM~2.5 is used only to generate forecast
features and does not directly answer FinTradeBench questions. At the time of research, \emph{Gemini 3.8 Flash leads vals Finance Agent v2 with 61.44 $\%$ accuracy} \cite{finance_agent_v2}}}
\label{tab:model-implementation}
\end{table*}

\begin{table*}[t]
\centering
\scriptsize
\renewcommand{\arraystretch}{1.15}
\setlength{\tabcolsep}{3.0pt}

\resizebox{\textwidth}{!}{
\begin{tabular}{p{3.5cm} p{3.2cm} p{4.5cm} p{3.4cm} p{3.0cm}}
\toprule
\textbf{System / run} &
\textbf{Execution mode} &
\textbf{Inference hardware} &
\textbf{Client allocation} &
\textbf{Recorded completing-job wall time} \\
\midrule

Qwen3.8-27B~\citep{qwenteam2026qwen38}
& Local vLLM~\citep{kwon2023vllm}
& 1$\times$ NVIDIA H100 80GB HBM3
& 16 CPU cores, 256\,GB RAM
& 21:58:41$^{\dagger}$ \\

Ministral-3-8B~\citep{liu2026ministral3}
& Local vLLM~\citep{kwon2023vllm}
& 1$\times$ NVIDIA H100 80GB HBM3
& 16 CPU cores, 256\,GB RAM
& 02:27:38 \\

Llama-4-Scout~\citep{meta2025llama4}
& Azure AI Foundry API
& Provider-managed; accelerator not exposed
& 16 CPU cores, 64\,GB RAM
& 07:11:05$^{\dagger}$ \\

GPT-5.4-mini~\citep{openai2026gpt54mini}
& Azure AI Foundry API, General QA
& Provider-managed; accelerator not exposed
& 16 CPU cores, 64\,GB RAM
& 06:03:29$^{\dagger}$ \\

DeepSeek-V4-Flash~\citep{deepseekai2026deepseekv4}
& Azure AI Foundry API
& Provider-managed; accelerator not exposed
& 16 CPU cores, 64\,GB RAM
& 02:19:38$^{\dagger}$ \\

Gemini-3.8-Flash~\citep{google2026gemini38flash}
& Vertex AI API
& Provider-managed; accelerator not exposed
& 16 CPU cores, 64\,GB RAM
& 08:45:51$^{\dagger}$ \\

Qwen2.5-32B~\citep{qwen2025qwen25}, R1-Distill-Qwen-32B~\citep{deepseekai2025deepseekr1}, Llama-3.1-8B~\citep{grattafiori2024llama3}, Phi-4~\citep{abdin2024phi4}
& Local controlled jobs~\citep{wolf2020transformers}
& 1$\times$ NVIDIA H100 80GB HBM3 per job
& 8 CPU cores, 256\,GB RAM per job
& See note$^{\ddagger}$ \\

GPT-5.4-mini~\citep{openai2026gpt54mini}
& Azure AI Foundry API, controlled conditions
& Provider-managed; accelerator not exposed
& 16 CPU cores, 64\,GB RAM
& 01:55:03 \\

ChatTS-8B~\citep{xie2024chatts}
& Local inference~\citep{wolf2020transformers}
& 1$\times$ NVIDIA H100 80GB HBM3
& 16 CPU cores, 256\,GB RAM
& 00:59:37 \\

TimesFM~2.5~\citep{das2024timesfm,google2025timesfm25} feature generation
& BigQuery \texttt{AI.FORECAST}
& Provider-managed; accelerator not exposed
& 8 CPU cores, 64\,GB RAM
& 00:06:01 \\

GPT-5.4-mini~\citep{openai2026gpt54mini} + TimesFM~2.5~\citep{das2024timesfm,google2025timesfm25}
& Azure API synthesis
& Provider-managed; accelerator not exposed
& 8 CPU cores, 64\,GB RAM
& 00:39:41 \\

Claude Sonnet~4.5~\citep{anthropic2025sonnet45}: controlled-model judging
& Vertex AI API
& Provider-managed; accelerator not exposed
& 4 CPU cores, 24\,GB RAM
& 11:02:54 / 23:47:21$^{\S}$ \\

Claude Sonnet~4.5~\citep{anthropic2025sonnet45}: time-series judging
& Vertex AI API
& Provider-managed; accelerator not exposed
& 4 CPU cores, 24\,GB RAM
& 02:33:58 \\

Claude Sonnet~4.5~\citep{anthropic2025sonnet45}: common-evidence RAG judging
& Vertex AI API; parallel per-condition jobs
& Provider-managed; accelerator not exposed
& 2 CPU cores, 16\,GB RAM per job
& $\leq$02:23:57 per job$^{\P}$ \\

\bottomrule
\end{tabular}
}

\caption{\small{\textbf{Compute configuration and recorded runtime.}
Times are Slurm wall-clock durations for the completing jobs associated with
the frozen outputs; smoke tests and failed attempts are excluded.
$^{\dagger}$These runners support resumption and may reuse rows generated by
earlier jobs, so the reported time is the wall time of the completing job rather
than cumulative compute expenditure.
$^{\ddagger}$The four locally controlled baseline models were dispatched as
12 independent model$\times$condition jobs with generic Slurm job names
(\texttt{ftb403\_resume}). Their exact per-model wall times cannot be
reconstructed reliably from the accounting export, so we do not assign
potentially incorrect runtimes.
$^{\S}$The two values correspond to the controlled-model judge batches for the
two model groups.
$^{\P}$Common-evidence RAG/No-RAG responses were judged in parallel
model--condition jobs; the longest observed completing job was 02:23:57.
For API-hosted models, the listed CPU/RAM values describe the Slurm client,
not the provider-side inference hardware.}}
\label{tab:compute-runtime}
\end{table*}

% ==========================================================
\sectionNoBarrier{Addendum to Results and discussions}
\label{app:results}

\subsection{The frozen judge closely tracks held-out human scores.}
\label{subsec:results-human-validation}
At the response level, the mean four-dimension score has an MAE of
$0.214$ Likert points relative to the human reviewer and Spearman
correlation $\rho=0.938$. Across all 192 individual rubric judgments,
exact agreement is $76.6\%$, $96.9\%$ of scores fall within one Likert
point, and quadratic-weighted $\kappa=0.884$. Accuracy shows the strongest dimension-level agreement
($\rho=0.952$, $83.3\%$ exact agreement), while Relevance has the
largest residual discrepancy. The automated judge scores Relevance
approximately $0.31$ points higher on average than the human reviewer,
suggesting mild positive calibration bias on this dimension. Pairwise evaluation leads to the same qualitative conclusion.
Strict A/B/Tie agreement is $70.8\%$. The human reviewer expresses a
directional preference on 15 of the 24 pairs; the automated judge agrees
with all 15, with no directional reversals. Most disagreement therefore
comes from tie calibration: the human reviewer sometimes considers two
answers equivalent when the judge assigns a small numerical difference.
 
\begin{figure}[t]
    \centering
\includegraphics[width=.7\linewidth]{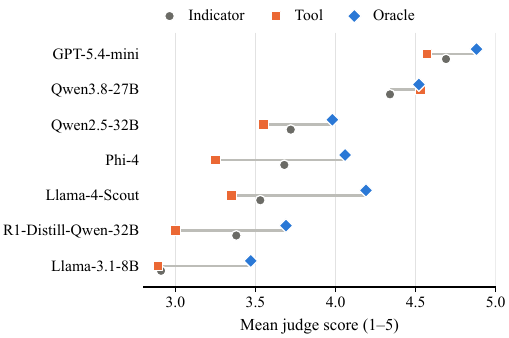}
    \vspace{-1mm}
    \caption{\small{\textbf{Oracle--Indicator effects across seven
    controlled models.} Positive values indicate improved performance
    when question-specific gold-indicator evidence replaces the broader
    Indicator-RAG evidence-selection process.}}
\label{fig:contolled_experiment_result}
    \vspace{-3mm}
\end{figure}
% ====================================================================
\subsection{Robustness to Final Core400 Curation}
\label{subsec:results-core400-sensitivity}

As a sensitivity check, we recompute the analyses before and after the
item-level Core audit. Excluding the three non-retained Core403 items leaves
all controlled-condition and common-evidence RAG/No-RAG means and paired
effects unchanged because those items were already invalid in the
corresponding automated evidence pipelines.

The time-series comparison is the only analysis affected numerically: the
eligible set changes from 259 to 257 questions and the paired
TimesFM+GPT--ChatTS effect changes from approximately $+1.08$ to $+1.09$
Likert points. The substantive conclusions are unchanged. Full Core403--Core400 sensitivity results are
reported in~\S\ref{app:core400-sensitivity}.

\subsection{Discussion}
\label{subsec:results-discussion}

Our experiments show that financial reasoning errors
cannot be reduced to a single retrieval problem.

First, retrieval remains useful: under a shared factual evaluation
anchor, RAG consistently improves Trading and Hybrid reasoning across
the six-model General-QA set. However, the weak and inconsistent Fundamental
effects show that providing more evidence alone is insufficient.

Second, the controlled conditions identify two distinct downstream
bottlenecks. Deterministic tools are most useful when explicit numerical
computation is required, whereas gold-selected evidence improves every
model and is especially beneficial on Trading tasks. This indicates that
both \emph{computation} and \emph{evidence selection} contribute to
financial reasoning failures, with their importance varying by question
type.

Third, specialized representations matter. The strong performance of
the forecast-feature pipeline suggests that summarizing numerical
time-series structure before language reasoning can be more effective
than requiring a language-oriented model to infer all relevant market
structure directly from the raw sequence.

Overall, FinTradeBench therefore functions not only as a model
comparison benchmark but as a diagnostic framework for locating failure
along the path from evidence access to computation, evidence selection,
and final financial synthesis.

%
% ============================================================
\section{Human Review and Evaluator Validation}
\label{app:validation}

Human involvement occurs at four distinct stages, each serving a different methodological purpose:
\myNum{1} expert-guided seed construction and development-time evaluator
calibration;
\myNum{2} construction-time fatal-flaw screening of scaled responses;
\myNum{3} full item-level quality auditing of the support-screened Core403;
and
\myNum{4} held-out validation of the frozen automated judge.
These stages differ in both their \emph{unit of review} and their
\emph{role in the pipeline}. In particular, the first three stages may
affect benchmark construction or evaluator development, whereas the
final held-out study is performed only after the judge configuration is
frozen and does not modify the evaluator. Table~\ref{tab:human-review-stages} summarizes these distinctions, and the
reviewer instruments are reproduced in \S\ref{app:annotation-instruments}.

\begin{table*}[t]
\centering
\small
\renewcommand{\arraystretch}{1.12}
\begin{tabular}{p{3.0cm} p{3.0cm} p{5.2cm} p{3.0cm}}
\toprule
\textbf{Stage} &
\textbf{Unit reviewed} &
\textbf{Human task} &
\textbf{Role in pipeline} \\
\midrule

Seed construction and
development calibration
&
Seed questions and generated responses
&
Experts author seed questions and golden indicators; generated
responses are subsequently scored to calibrate the development-time
automated evaluator.
&
Benchmark construction and evaluator development \\

\addlinespace

Scaled-set fatal-flaw screening
&
Scaled question--answer pairs
&
Screen for catastrophic generation failures such as wrong entity,
incomplete answers, or abrupt truncation; ambiguous cases are escalated.
&
Construction quality control \\

\addlinespace

Core403 item-level audit
&
Benchmark items
&
Inspect the question, reference answer, golden indicators, canonical
point-in-time evidence, and entity/time alignment.
&
Final benchmark curation:
Core403 $\rightarrow$ Core400 \\

\addlinespace

Held-out judge validation
&
Anonymous model responses
&
Score responses using the frozen four-dimension rubric and record
blinded A/B/Tie preferences.
&
Post-freeze evaluator validation only \\

\bottomrule
\end{tabular}
\caption{\small{\textbf{Distinct human-review stages in FinTradeBench.}
The stages operate on different objects and serve different purposes.
The held-out judge-validation labels are not used to modify the frozen
evaluation procedure.}}
\label{tab:human-review-stages}
\end{table*}

% ------------------------------------------------------------
\subsection{Expert Seed Construction and Development-Time Evaluator Calibration}
\label{app:dev-calibration}

The first stage combines two closely related development activities.
First, a domain expert constructs the seed portion of FinTradeBench,
including the benchmark questions and their expert-defined golden
indicators. These annotations establish the intended financial evidence
needed to reason about each seed question and provide the foundation for
subsequent benchmark scaling.

Second, during evaluator development, human experts score filtered
model responses to seed questions on the same response-quality
dimensions used by the automated evaluator. These annotations are used
to inspect and calibrate the evaluation prompt during development.
The corresponding human rubric includes factual and numerical accuracy,
completeness and context, relevance and utility, and clarity and
rationale, together with a binary check of agreement with the
construction-time numerical audit; see
Table~\ref{tab:human-rubric-calibration}.

On this development sample, the human and automated scores have an
overall MAE of approximately $0.40$ on the 1--5 scale. Because the
filtered development responses are heavily concentrated toward high
scores, rank-based agreement measures are less informative for this
sample. We therefore treat these measurements as evidence of
\emph{development-time calibration}, rather than as the final validation
of evaluator reliability.

Crucially, the development annotations are not the held-out validation
set used to support the final experiments. Reliability of the frozen
judge is evaluated separately in
\S\ref{app:judge-alignment}.

% ------------------------------------------------------------
\subsection{Construction-Time Fatal-Flaw Screening of Scaled Responses}
\label{app:scaled-screening}

After scaling the benchmark, the generated question--answer pairs
undergo a separate human quality-control screen. The purpose of this
stage is deliberately narrower than response grading: the reviewer is
not asked to re-evaluate the full financial reasoning quality of every
answer. Instead, the screen targets clear construction failures that
would make an item unsuitable for release.

Examples include answering for the wrong company or ticker, failing to
answer a required part of the question, using an inconsistent historical
period, or terminating before the requested analysis is complete.
Responses that fail this screen are regenerated before inclusion in the
broader benchmark, while ambiguous cases are escalated for
financial-expert review. In the original scaling process, 42 of the
1,400 generated responses (3\%) required regeneration.

The operational screening categories are summarized in
Table~\ref{tab:fatal-flaw-screen} and sample bad responses are provided in Table \ref{tab:screening}. We emphasize that this step is a
\emph{construction quality-control gate}, not an independent Likert-scale
evaluation and not the same procedure as the later Core403 item-level
audit.

% ------------------------------------------------------------
\subsection{Full Item-Level Human Quality Audit of Core403}
\label{app:core-human-audit-validation}

A substantially stricter human review is performed after the
canonical-data support screen. At this stage, the unit of review is the
\emph{benchmark item itself}, rather than a model-generated response.
All 403 retained Core candidates are examined individually.

For each item, the reviewer inspects four primary artifacts:
the question, the benchmark reference answer, the expert-defined golden
indicators, and the canonical point-in-time evidence. The reviewer then
assesses question clarity, reference-answer correctness, evidence
support, golden-indicator adequacy, and entity/time alignment. Fatal
flaws and the final disposition of each item are recorded separately.
The full categorical rubric is reported in
Table~\ref{tab:core403-audit-rubric}.

This audit is intentionally different from model-response evaluation.
For example, a question can fail because its reference answer is not
sufficiently supported, because its historical cutoff is ambiguous, or
because its golden indicators omit critical evidence even if a model
could nevertheless produce a plausible answer.

The reviewer accepts 400 of the 403 items into the final evaluation set.
Three items---FTB-0061, FTB-0135, and FTB-0149---are excluded, yielding
the final Core400. The resulting set contains 143 Fundamental,
122 Trading, and 135 Hybrid questions.

% ------------------------------------------------------------
\subsection{Held-Out Human Validation of the Frozen Judge}
\label{app:judge-alignment}

The final human-review stage evaluates the reliability of the
\emph{frozen} automated judge. This study is conducted only after the
judge prompt, scoring schema, and evaluation procedure have been fixed.
Its human labels are therefore used solely for validation and do not
modify the judge.

We construct a blinded sample of 24 paired questions, corresponding to
48 anonymous responses. The sample spans all three FinTradeBench
reasoning lanes and four experimental comparison families:
RAG--No-RAG, Oracle--Indicator, Tool--Indicator, and
TimesFM+GPT--ChatTS. The reviewer is blinded to model identity,
experimental condition, and automated judge score.

Each anonymous response is independently scored from 1--5 on
Factual and Numerical Accuracy, Completeness and Context,
Relevance and Utility, and Clarity and Rationale, using the rubric in
Table~\ref{tab:heldout-human-rubric}. After both responses in a pair have
been scored independently, the reviewer additionally records an
A/B/Tie preference and a 1--5 confidence score
(Table~\ref{tab:heldout-human-rubric}).

\begin{table}[t]
\centering
\small
\setlength{\tabcolsep}{3.5pt}
\begin{tabular}{lrrrrr}
\toprule
Dimension & MAE & $\rho$ & Exact & $\pm1$ & QWK \\
\midrule
Accuracy      & 0.17 & 0.95 & 83.3\% & 100.0\% & 0.945 \\
Completeness  & 0.29 & 0.88 & 75.0\% & 95.8\%  & 0.882 \\
Relevance     & 0.40 & 0.86 & 68.8\% & 91.7\%  & 0.809 \\
Clarity       & 0.21 & 0.86 & 79.2\% & 100.0\% & 0.857 \\
\midrule
All rubric ratings
              & 0.27 & 0.89 & 76.6\% & 96.9\% & 0.884 \\
\bottomrule
\end{tabular}
\caption{\small{\textbf{Held-out human--judge agreement.}
Statistics are computed over 48 responses and 192 individual rubric
ratings. Exact and $\pm1$ denote the proportion of automated ratings
identical to, or within one Likert point of, the human rating.
QWK denotes quadratic-weighted Cohen's $\kappa$.}}
\label{tab:human-judge-agreement-app}
\end{table}

At the response level, the arithmetic mean across the four scoring
dimensions yields MAE $=0.214$ and Spearman $\rho=0.938$ between the
human reviewer and the frozen judge. The mean Claude-minus-human
difference is $+0.099$ Likert points. The largest dimension-level mean
difference occurs for Relevance ($+0.313$), whereas Accuracy and
Completeness exhibit little mean bias.

Pairwise agreement provides a complementary test because it asks whether
the judge preserves the human reviewer's relative preference between two
responses rather than merely reproducing absolute scores.

\begin{table}[t]
\centering
\footnotesize
\setlength{\tabcolsep}{2pt}
\begin{tabular}{lrrrr}
\toprule
Comparison & Pairs & Decisive & Strict & Direct. \\
\midrule
Overall                    & 24 & 15 & 70.8\% & 100.0\% \\
Oracle vs.\ Indicator      & 8  & 4  & 62.5\% & 100.0\% \\
RAG vs.\ No-RAG            & 8  & 6  & 87.5\% & 100.0\% \\
TimesFM+GPT vs.\ ChatTS    & 4  & 3  & 75.0\% & 100.0\% \\
Tool vs.\ Indicator        & 4  & 2  & 50.0\% & 100.0\% \\
\bottomrule
\end{tabular}
\caption{\small{\textbf{Pairwise preference agreement.}
\emph{Decisive} denotes pairs for which the human reviewer selected A
or B rather than Tie. Strict agreement requires identical A/B/Tie
labels. Directional (Direct.) agreement is computed only over the 15
human-decisive pairs.}}
\label{tab:human-pairwise-app}
\end{table}

Strict A/B/Tie agreement is 70.8\% overall. The lower strict agreement
is primarily associated with pairs for which the human reviewer selects
Tie while the automated scores imply a small numerical preference.
Among the 15 pairs for which the human reviewer expresses a directional
A/B preference, the judge agrees on all 15, with no preference reversal. Moreover, the sample is intentionally stratified across
reasoning lanes, comparison families, and score-gap regimes to study evaluator behavior rather than to estimate a
population-level model preference rate.

% ============================================================
\section{Evaluation Integrity and Sensitivity Checks}
\label{app:evaluation-integrity}

Beyond human review, we perform additional checks on evaluation
coverage, evidence consistency, and sensitivity to the final Core403
to Core400 curation step. These analyses verify that the reported
effects are not artifacts of evaluator-side evidence asymmetry,
context truncation, or the final three benchmark exclusions.

\subsection{Evaluation Coverage and Evidence QC}
\label{app:evaluation-qc}

For the General-QA RAG--No-RAG comparison, both responses in each pair
are scored against the same condition-independent canonical
point-in-time evidence package. This removes evaluator-side evidence
asymmetry between the two generation conditions. The final
common-evidence and time-series judge inputs contain no evidence
truncation.

The time-series study is restricted by design to questions requiring
market time-series reasoning. Its eligible Core400 subset therefore
contains 257 questions: 122 Trading and 135 Hybrid items. Full
file-level coverage, evidence-size diagnostics, source-validity counts,
and ancillary metric-overlap analyses are included with the released
analysis artifacts.

\subsection{Sensitivity to the Core403--Core400 Transition}
\label{app:core400-sensitivity}

The final human audit excludes FTB-0061, FTB-0135, and FTB-0149.
These items already lack valid paired observations in the controlled
condition and common-evidence RAG--No-RAG analyses. Consequently,
transitioning from Core403 to Core400 does not alter their reported
means or paired-effect estimates.

The time-series study is the only analysis for which the exclusions
change the numerical summaries, because excluded items overlap its
eligible Trading/Hybrid subset. As shown in
Table~\ref{tab:core400-sensitivity}, the resulting changes are small and
do not alter the substantive conclusion.

\begin{table}[t]
\centering
\small
\setlength{\tabcolsep}{4pt}
\begin{tabular}{lrrr}
\toprule
Quantity & Core403 & Core400 & Change \\
\midrule
ChatTS overall          & 3.0386 & 3.0321 & $-$0.0065 \\
GPT+TimesFM overall     & 4.1158 & 4.1235 & +0.0077 \\
Paired time-series effect
                        & 1.0772 & 1.0914 & +0.0142 \\
\bottomrule
\end{tabular}
\caption{\small{\textbf{Sensitivity to the final human-audit exclusions.}
The final three exclusions produce only small numerical changes in the
time-series study and do not change its substantive conclusion.}}
\label{tab:core400-sensitivity}
\end{table}

% ============================================================
\section{Complete Quantitative Results}
\label{app:full-results}

The main paper reports compact summaries. This section contains the full
paired effects, bootstrap intervals, and Holm-adjusted Wilcoxon tests.

\subsection{Common-Evidence RAG vs.\ No-RAG}
\label{app:rag-full}

Table~\ref{tab:rag-common-core400} reports the per-model, per-lane effect of retrieval
under the common-evidence judge.

\begin{table}[htbp!]
\centering
\small
\setlength{\tabcolsep}{2.6pt}
\begin{tabular}{lrrcrr}
\toprule
Lane & $N$ & $\Delta$ & 95\% CI & W/T/L & $p_{\mathrm{Holm}}$ \\
\midrule
\multicolumn{6}{l}{\textit{DeepSeek-V4-Flash}} \\
\hspace{0.5em}All & 399 & \textbf{+0.23} & [.15, .30] & 53/17/31 & $<$.001 \\
\hspace{0.5em}Fund. & 143 & $-$0.06 & [$-$.19, .07] & 39/15/46 & .683 \\
\hspace{0.5em}Trad. & 121 & \textbf{+0.46} & [.35, .57] & 69/16/16 & $<$.001 \\
\hspace{0.5em}Hybrid & 135 & \textbf{+0.32} & [.21, .44] & 53/19/27 & $<$.001 \\
\midrule
\multicolumn{6}{l}{\textit{Llama-4-Scout}} \\
\hspace{0.5em}All & 399 & \textbf{+0.23} & [.18, .28] & 55/22/23 & $<$.001 \\
\hspace{0.5em}Fund. & 143 & +0.06 & [$-$.02, .14] & 39/26/35 & .551 \\
\hspace{0.5em}Trad. & 121 & \textbf{+0.41} & [.33, .48] & 73/18/9 & $<$.001 \\
\hspace{0.5em}Hybrid & 135 & \textbf{+0.25} & [.16, .34] & 55/22/23 & $<$.001 \\
\midrule
\multicolumn{6}{l}{\textit{Qwen3.8-27B}} \\
\hspace{0.5em}All & 399 & \textbf{+0.22} & [.15, .30] & 54/14/32 & $<$.001 \\
\hspace{0.5em}Fund. & 143 & $-$0.03 & [$-$.16, .10] & 37/20/43 & .683 \\
\hspace{0.5em}Trad. & 121 & \textbf{+0.23} & [.13, .32] & 60/13/26 & $<$.001 \\
\hspace{0.5em}Hybrid & 135 & \textbf{+0.48} & [.34, .63] & 67/8/24 & $<$.001 \\
\midrule
\multicolumn{6}{l}{\textit{Gemini-3.8-Flash}} \\
\hspace{0.5em}All & 398 & \textbf{+0.20} & [.13, .28] & 55/15/30 & $<$.001 \\
\hspace{0.5em}Fund. & 143 & $-$0.10 & [$-$.24, .03] & 40/15/45 & .551 \\
\hspace{0.5em}Trad. & 121 & \textbf{+0.33} & [.20, .45] & 62/11/27 & $<$.001 \\
\hspace{0.5em}Hybrid & 134 & \textbf{+0.42} & [.31, .54] & 65/19/16 & $<$.001 \\
\midrule
\multicolumn{6}{l}{\textit{GPT-5.4-mini}} \\
\hspace{0.5em}All & 399 & \textbf{+0.17} & [.09, .24] & 53/17/30 & $<$.001 \\
\hspace{0.5em}Fund. & 143 & $-$0.14 & [$-$.27, $-$.02] & 37/18/45 & .480 \\
\hspace{0.5em}Trad. & 121 & \textbf{+0.35} & [.25, .44] & 63/20/17 & $<$.001 \\
\hspace{0.5em}Hybrid & 135 & \textbf{+0.33} & [.21, .45] & 60/14/26 & $<$.001 \\
\midrule
\multicolumn{6}{l}{\textit{Ministral-3-8B}} \\
\hspace{0.5em}All & 399 & \textbf{+0.25} & [.20, .30] & 59/20/21 & $<$.001 \\
\hspace{0.5em}Fund. & 143 & \textbf{+0.14} & [.06, .22] & 49/22/29 & .013 \\
\hspace{0.5em}Trad. & 121 & \textbf{+0.32} & [.24, .40] & 68/17/15 & $<$.001 \\
\hspace{0.5em}Hybrid & 135 & \textbf{+0.31} & [.23, .40] & 62/21/16 & $<$.001 \\
\bottomrule
\end{tabular}
\caption{Effect of retrieval (RAG $-$ No-RAG) on the common-evidence Core400, per model and lane (Fund.\ = Fundamental, Trad.\ = Trading). $\Delta$: mean paired difference in 1--5 judge Likert score (multiply by 20 for the 0--100 scale); 95\% CI: paired bootstrap; W/T/L: \% of items where RAG wins, ties, or loses; $p_{\mathrm{Holm}}$: Wilcoxon signed-rank $p$, Holm-corrected across all 24 tests. Bold: $p_{\mathrm{Holm}} < 0.05$.}
\label{tab:rag-common-core400}
\end{table}
\subsection{Controlled Evidence Conditions}
\label{app:controlled-full}

Table~\ref{tab:controlled-effects-full} reports the full paired Tool, Oracle, and
Indicator-RAG comparisons for every model and lane.

\begin{table*}[htbp!]
\centering
\small
\setlength{\tabcolsep}{4pt}
\begin{tabular}{@{}lr rcr rcr rcr@{}}
\toprule
& & \multicolumn{3}{c}{Tool $-$ Indicator} & \multicolumn{3}{c}{Oracle $-$ Indicator} & \multicolumn{3}{c}{Oracle $-$ Tool} \\
\cmidrule(lr){3-5}\cmidrule(lr){6-8}\cmidrule(l){9-11}
Lane & $N$ & $\Delta$ & 95\% CI & $p_{\mathrm{Holm}}$ & $\Delta$ & 95\% CI & $p_{\mathrm{Holm}}$ & $\Delta$ & 95\% CI & $p_{\mathrm{Holm}}$ \\
\midrule
\multicolumn{11}{@{}l}{\textit{Qwen2.5-32B}} \\
\hspace{0.5em}All & 399 & \textbf{$-$0.17} & [$-$0.27, $-$0.06] & .004 & \textbf{+0.26} & [0.17, 0.36] & $<$.001 & \textbf{+0.43} & [0.33, 0.53] & $<$.001 \\
\hspace{0.5em}Fund. & 143 & \textbf{$-$0.74} & [$-$0.90, $-$0.58] & $<$.001 & +0.01 & [$-$0.11, 0.13] & 1.00 & \textbf{+0.75} & [0.61, 0.89] & $<$.001 \\
\hspace{0.5em}Trad. & 121 & \textbf{+0.64} & [0.44, 0.85] & $<$.001 & \textbf{+0.82} & [0.62, 1.04] & $<$.001 & +0.18 & [$-$0.05, 0.40] & 1.00 \\
\hspace{0.5em}Hybrid & 135 & \textbf{$-$0.28} & [$-$0.41, $-$0.15] & $<$.001 & +0.03 & [$-$0.10, 0.15] & 1.00 & \textbf{+0.31} & [0.15, 0.46] & $<$.001 \\
\midrule
\multicolumn{11}{@{}l}{\textit{R1-Distill-Qwen-32B}} \\
\hspace{0.5em}All & 399 & \textbf{$-$0.37} & [$-$0.46, $-$0.29] & $<$.001 & \textbf{+0.31} & [0.23, 0.41] & $<$.001 & \textbf{+0.69} & [0.59, 0.78] & $<$.001 \\
\hspace{0.5em}Fund. & 143 & \textbf{$-$0.72} & [$-$0.87, $-$0.58] & $<$.001 & \textbf{+0.16} & [0.05, 0.28] & .046 & \textbf{+0.89} & [0.72, 1.04] & $<$.001 \\
\hspace{0.5em}Trad. & 121 & +0.02 & [$-$0.12, 0.18] & 1.00 & \textbf{+0.73} & [0.53, 0.93] & $<$.001 & \textbf{+0.70} & [0.51, 0.91] & $<$.001 \\
\hspace{0.5em}Hybrid & 135 & \textbf{$-$0.35} & [$-$0.47, $-$0.24] & $<$.001 & +0.10 & [$-$0.03, 0.24] & .462 & \textbf{+0.46} & [0.30, 0.60] & $<$.001 \\
\midrule
\multicolumn{11}{@{}l}{\textit{Llama-3.1-8B}} \\
\hspace{0.5em}All & 399 & $-$0.02 & [$-$0.11, 0.07] & 1.00 & \textbf{+0.57} & [0.47, 0.66] & $<$.001 & \textbf{+0.58} & [0.48, 0.69] & $<$.001 \\
\hspace{0.5em}Fund. & 143 & \textbf{$-$0.21} & [$-$0.36, $-$0.06] & .032 & \textbf{+0.47} & [0.34, 0.59] & $<$.001 & \textbf{+0.68} & [0.52, 0.84] & $<$.001 \\
\hspace{0.5em}Trad. & 121 & \textbf{+0.34} & [0.17, 0.51] & .002 & \textbf{+0.87} & [0.67, 1.08] & $<$.001 & \textbf{+0.53} & [0.30, 0.75] & $<$.001 \\
\hspace{0.5em}Hybrid & 135 & $-$0.14 & [$-$0.28, 0.01] & .107 & \textbf{+0.40} & [0.27, 0.53] & $<$.001 & \textbf{+0.54} & [0.37, 0.69] & $<$.001 \\
\midrule
\multicolumn{11}{@{}l}{\textit{Phi-4}} \\
\hspace{0.5em}All & 399 & \textbf{$-$0.44} & [$-$0.54, $-$0.33] & $<$.001 & \textbf{+0.38} & [0.29, 0.47] & $<$.001 & \textbf{+0.82} & [0.72, 0.91] & $<$.001 \\
\hspace{0.5em}Fund. & 143 & \textbf{$-$1.03} & [$-$1.18, $-$0.90] & $<$.001 & +0.13 & [0.01, 0.24] & .462 & \textbf{+1.16} & [1.03, 1.29] & $<$.001 \\
\hspace{0.5em}Trad. & 121 & \textbf{+0.35} & [0.16, 0.55] & .034 & \textbf{+0.91} & [0.71, 1.10] & $<$.001 & \textbf{+0.56} & [0.36, 0.75] & $<$.001 \\
\hspace{0.5em}Hybrid & 135 & \textbf{$-$0.51} & [$-$0.63, $-$0.39] & $<$.001 & \textbf{+0.17} & [0.05, 0.28] & .008 & \textbf{+0.68} & [0.54, 0.81] & $<$.001 \\
\midrule
\multicolumn{11}{@{}l}{\textit{Llama-4-Scout}} \\
\hspace{0.5em}All & 399 & \textbf{$-$0.18} & [$-$0.28, $-$0.10] & .001 & \textbf{+0.65} & [0.58, 0.73] & $<$.001 & \textbf{+0.84} & [0.75, 0.92] & $<$.001 \\
\hspace{0.5em}Fund. & 143 & \textbf{$-$0.63} & [$-$0.77, $-$0.49] & $<$.001 & \textbf{+0.47} & [0.36, 0.57] & $<$.001 & \textbf{+1.09} & [0.94, 1.24] & $<$.001 \\
\hspace{0.5em}Trad. & 121 & \textbf{+0.29} & [0.13, 0.45] & .006 & \textbf{+0.67} & [0.51, 0.83] & $<$.001 & \textbf{+0.38} & [0.21, 0.55] & $<$.001 \\
\hspace{0.5em}Hybrid & 135 & $-$0.14 & [$-$0.27, $-$0.01] & .241 & \textbf{+0.83} & [0.71, 0.95] & $<$.001 & \textbf{+0.97} & [0.85, 1.09] & $<$.001 \\
\midrule
\multicolumn{11}{@{}l}{\textit{Qwen3.8-27B}} \\
\hspace{0.5em}All & 399 & +0.04$^{\dagger}$ & [$-$0.06, 0.13] & 1.00 & \textbf{+0.17} & [0.10, 0.25] & $<$.001 & +0.05$^{\dagger}$ & [$-$0.04, 0.14] & 1.00 \\
\hspace{0.5em}Fund. & 143 & +0.06 & [$-$0.08, 0.20] & 1.00 & +0.06 & [$-$0.06, 0.19] & 1.00 & +0.00 & [$-$0.13, 0.13] & 1.00 \\
\hspace{0.5em}Trad. & 121 & +0.56$^{\dagger}$ & [$-$0.15, 1.31] & 1.00 & \textbf{+0.40} & [0.26, 0.55] & $<$.001 & $-$0.25$^{\dagger}$ & [$-$1.00, 0.50] & 1.00 \\
\hspace{0.5em}Hybrid & 135 & $-$0.03 & [$-$0.13, 0.08] & 1.00 & +0.09 & [$-$0.02, 0.20] & .538 & +0.13 & [0.01, 0.24] & .070 \\
\midrule
\multicolumn{11}{@{}l}{\textit{GPT-5.4-mini}} \\
\hspace{0.5em}All & 399 & \textbf{$-$0.12} & [$-$0.19, $-$0.05] & $<$.001 & \textbf{+0.19} & [0.13, 0.24] & $<$.001 & \textbf{+0.31} & [0.25, 0.37] & $<$.001 \\
\hspace{0.5em}Fund. & 143 & \textbf{$-$0.35} & [$-$0.47, $-$0.24] & $<$.001 & \textbf{+0.13} & [0.06, 0.20] & .003 & \textbf{+0.48} & [0.38, 0.58] & $<$.001 \\
\hspace{0.5em}Trad. & 121 & +0.17 & [0.04, 0.31] & .538 & \textbf{+0.33} & [0.20, 0.48] & $<$.001 & \textbf{+0.16} & [0.06, 0.27] & .010 \\
\hspace{0.5em}Hybrid & 135 & \textbf{$-$0.14} & [$-$0.22, $-$0.06] & .004 & \textbf{+0.12} & [0.07, 0.18] & $<$.001 & \textbf{+0.26} & [0.19, 0.35] & $<$.001 \\
\bottomrule
\end{tabular}
\caption{Full paired effects for the controlled conditions on the final Core400, per model and lane (Fund.\ = Fundamental, Trad.\ = Trading). $\Delta$: mean paired difference in 1--5 judge Likert score (multiply by 20 for the 0--100 scale); 95\% CI: paired bootstrap; $p_{\mathrm{Holm}}$: two-sided Wilcoxon signed-rank $p$ on per-question paired differences, Holm-corrected within each experimental batch over all model $\times$ lane $\times$ comparison tests (48 tests for Qwen2.5-32B, R1-Distill-Qwen-32B, Llama-3.1-8B and Phi-4; 36 for Llama-4-Scout, Qwen3.8-27B and GPT-5.4-mini). Bold: $p_{\mathrm{Holm}}<0.05$. $N$ is the number of paired questions. $^{\dagger}$Qwen3.8-27B's Tool condition has incomplete coverage, so comparisons involving Tool use fewer pairs: 290 overall and only 12 on Trading.}
\label{tab:controlled-effects-full}
\end{table*}

\subsection{Time-Series Specialist Effects}
\label{app:timeseries-full}

Table~\ref{tab:timeseries-effects-full} reports the paired TimesFM+GPT versus
ChatTS effects on the time-series-eligible Core400 subset.

\begin{table}[hbtp!]
\centering
\small
\begin{tabular}{lrrr}
\toprule
Lane & $N$ & $\Delta_{(T+G)-C}$  & 95\% CI \\
\midrule
Overall & 257 & +1.091 & [0.944, 1.245] \\
Trading & 122 & +0.525 & [0.277, 0.775] \\
Hybrid & 135 & +1.604 & [1.472, 1.739] \\
\bottomrule
\end{tabular}
\caption{\small{Paired time-series specialist effects on the Core400
eligible subset. Here, $\Delta_{(T+G)-C}$ represents $(TimesFM+GPT)-ChatTS$ values. All three comparisons remain significant after Holm
correction.}}
\label{tab:timeseries-effects-full}
\end{table}

% ============================================================
\sectionNoBarrier{Qualitative Case Studies}
\label{app:case-studies}

Tables~\ref{tab:qual-gpt54mini-fund}--\ref{tab:qual-gpt54mini-hyb} show one
GPT-5.4-mini seed question per reasoning lane, evaluated under every
evidence condition. The qualitative examples illustrate how different evidence representations
can change a model's reasoning path. They are not used to establish
population-level effects; the controlled experiments in
\S\ref{subsec:results-controlled} provide the quantitative evidence.

% \subsection{Hybrid Evidence Ablation}

% Preamble (once): \usepackage[most]{tcolorbox}  \usepackage{dashrule}  \usepackage{xcolor}
% \cond{condition}{score}{excerpt}{judge note}: one condition's response; a starred score uses the common-evidence judge.
% Qualitative GPT-5.4-mini examples: one seed question per lane, every condition evaluated on it.
% Preamble (once): \usepackage[most]{tcolorbox}  \usepackage{xcolor}
% Excerpts are verbatim from the judged Final_Answer / golden answer (Markdown removed; [...] = omitted text).

% \cond{condition}{judge score}{excerpt}{judge note (may be empty)}
\providecommand{\cond}[4]{%
  \par\noindent\hangindent=1.2em\hangafter=1
  \textbf{#1} {\color{black!60}[#2]}: #3%
  \ifx\relax#4\relax\else\ {\color{black!55}\textit{(#4)}}\fi\par\vspace{2pt}}
% \judgehdr{protocol}: small heading separating the two judging protocols
\providecommand{\judgehdr}[1]{%
  \par\vspace{2pt}\noindent{\color{black!60}\scshape #1}\par\vspace{1pt}}
% \qualbox{lane title}{content}: shared box style
\providecommand{\qualbox}[2]{%
\begin{tcolorbox}[
    enhanced,
    colback=blue!2!white,
    colframe=black!70,
    arc=2pt,
    boxrule=0.5pt,
    left=5pt, right=5pt, top=6pt, bottom=4pt,
    title=\textbf{#1},
    coltitle=black,
    colbacktitle=gray!15,
    attach boxed title to top center={yshift=-2mm},
    boxed title style={size=small,colback=gray!15,colframe=black!70}
]
\footnotesize\raggedright
#2
\end{tcolorbox}}

%% ---------------------------------------------------------------- Fundamental
\begin{table*}[tp]
\centering
\qualbox{GPT-5.4-mini Across Conditions: Fundamental}{%
\textbf{Q:} \textit{As of Q3 2025, which generates more cash per dollar of assets: Netflix or Apple?}\par\vspace{2pt}
\textbf{Golden answer:} ``As of Q3 2025, Apple generates more cash per dollar of assets than Netflix. According to the fundamental data, Apple's Cash Flow to Assets ratio is 0.0841, meaning it generates roughly 8.4 cents of cash for every dollar of assets. In contrast, Netflix's ratio is 0.0456, generating roughly 4.5 cents per dollar of assets.''\par

\judgehdr{Per-condition-evidence judge}
\cond{Indicator}{5.00}{``From the supplied indicator data at 2025-09-30: Apple (AAPL) F\_Cash Flow/Assets: 0.084065; Netflix (NFLX) F\_Cash Flow/Assets: 0.045636. [\ldots] Apple generates more cash per dollar of assets than Netflix as of Q3 2025.''}{}
\cond{Tool}{3.00}{``The context provided does not include any Netflix Q3 2025 SEC evidence or balance sheet/cash flow figures [\ldots] The supplied evidence is insufficient to confidently determine whether Netflix or Apple generates more cash per dollar of assets as of Q3 2025.''}{retrieved filings lacked Netflix data, so the model abstained}
\cond{Oracle}{5.00}{``As of Q3 2025, Apple generates more cash per dollar of assets. Apple Cash Flow/Assets: 0.084065; Netflix Cash Flow/Assets: 0.045636. [\ldots] Apple's value is higher, so Apple is the answer based on the supplied evidence.''}{}

\judgehdr{Common-evidence judge}
\cond{No-RAG}{4.50}{``As of Q3 2025, Apple generates more cash per dollar of assets than Netflix. [\ldots] Apple has historically produced very large operating and free cash flows relative to its asset base. [\ldots] As a result, cash produced per dollar of assets is typically lower than Apple's.''}{correct conclusion, but no Q3 2025 figures cited}
\cond{RAG}{2.50}{``The most recent Apple balance-sheet total assets in the context is from the 2023 filing: Total assets = \$335,038 million [\ldots]''}{divides 2025 cash flow by 2023 assets and gives no final verdict}
}
\caption{GPT-5.4-mini responses to a Fundamental seed question under every evaluated condition (TimesFM2.5 is evaluated only on time-series-eligible Trading and Hybrid questions). Bracketed numbers are judge scores (1--5 Likert mean). Scores are comparable only within a judging protocol: the per-condition-evidence judge grades each answer against the evidence that condition supplied, the common-evidence judge grades both RAG and No-RAG against identical canonical evidence. Excerpts are verbatim apart from removed Markdown formatting; [\ldots] marks omitted text. Gray notes summarize the judge's rationale.}
\label{tab:qual-gpt54mini-fund}
\end{table*}

%% ---------------------------------------------------------------- Trading
\begin{table*}[tp]
\centering
\qualbox{GPT-5.4-mini Across Conditions: Trading}{%
\textbf{Q:} \textit{As of September 2025, which stocks have the strongest volume confirmation for their price moves?}\par\vspace{2pt}
\textbf{Golden answer:} ``The stocks with the strongest volume confirmation are TSLA, INTC, and LRCX. [\ldots] TSLA exhibits the strongest absolute volume confirmation, with nearly double the OBV of the next closest ticker, validating its 35\% upward move. INTC shows strong volume confirmation for a recovery rally, despite weak underlying fundamentals. LRCX represents the healthiest volume confirmation, where high OBV and price momentum are aligned with strong fundamental operating metrics.''\par

\judgehdr{Per-condition-evidence judge}
\cond{Indicator}{5.00}{``The clearest positive confirmations are: TSLA: PxChg +0.350255, OBVchg +1.223B; INTC: PxChg +0.385791, OBVchg +642.828M; MU: PxChg +0.412219, OBVchg +306.601M [\ldots]''}{}
\cond{Tool}{4.75}{``Using the supplied evidence, the best proxy for that is On-Balance Volume (OBV) change over the month [\ldots] TSLA: OBV\_WINDOW\_CHANGE +1,223,051,798; price rose from 329.36 on 2025-09-02 to 444.72 on 2025-09-30; NVDA: +910,704,556; price rose from 170.78 to 186.58 [\ldots]''}{}
\cond{Oracle}{5.00}{``Based on the largest OBV increases from 2025-09-02 to 2025-09-30: TSLA -- OBV\_WINDOW\_CHANGE: +1,223,051,798; NVDA -- OBV\_WINDOW\_CHANGE: +910,704,556; INTC -- OBV\_WINDOW\_CHANGE: +642,827,861 [\ldots]''}{}
\cond{+\,TimesFM2.5}{2.50}{``[T]he strongest volume-confirmed price moves for September 2025 would be the stocks showing both: Meaningful forecasted price drift [\ldots] Large forecast uncertainty bands / implied move ranges [\ldots]''}{no volume data supplied; forecast uncertainty used as a proxy}

\judgehdr{Common-evidence judge}
\cond{No-RAG}{1.50}{``[\ldots] led by: NVIDIA (NVDA); Broadcom (AVGO); Meta Platforms (META); Microsoft (MSFT); Amazon (AMZN). If you want the strongest raw volume spikes in more speculative movers, then Super Micro Computer (SMCI), Palantir (PLTR), and AMD also stand out.''}{generic picks with no data, several outside the question's stock universe}
\cond{RAG}{3.00}{``As of September 2025, the strongest volume-confirmed price moves in the provided data appear to be: INTC -- the clearest confirmation, with a strong price rise and an extreme volume surge. MU -- very strong upward move with heavy and rising turnover. LRCX -- sizable price appreciation with robust volume support. [\ldots]''}{plausible ranking, but its daily price/volume figures are absent from the judge's canonical evidence and were treated as unsupported}
}
\caption{GPT-5.4-mini responses to a Trading seed question under every evaluated condition. Notation as in Table~\ref{tab:qual-gpt54mini-fund}.}
\label{tab:qual-gpt54mini-trad}
\end{table*}

%% ---------------------------------------------------------------- Hybrid
\begin{table*}[tp]
\centering
\qualbox{GPT-5.4-mini Across Conditions: Hybrid}{%
\textbf{Q:} \textit{Has Meta bottomed out after its decline in August 2025?}\par\vspace{2pt}
\textbf{Golden answer:} ``Based on the August 2025 data, META has not clearly bottomed out after its -1.51\% decline. The stock closed below key moving averages (MA\_20 at 762.04 and EMA\_20 at 750.93), with bearish short-term momentum (negative 5D/20D momentum, MACD below signal, neutral RSI at 39.23), indicating ongoing downtrend risks rather than reversal confirmation. [\ldots] Investors should monitor for price stabilization above EMA\_20 and RSI improvement above 50 before confirming a bottom.''\par

\judgehdr{Per-condition-evidence judge}
\cond{Indicator}{5.00}{``The trading indicators for 2025-08-29 still show weakness: Adj.\ Close: 738.194; MA\_20: 762.043, so price is below the 20-day moving average; RSI: 39.23 [\ldots] the evidence does not support a confident `yes.' ''}{}
\cond{Tool}{5.00}{``Using only the supplied evidence, I cannot conclude confidently that Meta had `bottomed out.' [\ldots] Meta fell from an August 2025 high of \$790.00 on 2025-08-12 to \$738.70 on 2025-08-29. [\ldots] There was a rebound from \$739.10 on 2025-08-21 to \$754.79 on 2025-08-22, but that move did not hold [\ldots]''}{}
\cond{Oracle}{5.00}{``RSI = 39.23: This is below 50, indicating weak momentum, but it is not extremely oversold [\ldots] Momentum\_5D = -0.0213 and Momentum\_20D = -0.0151: Both short-term momentum measures are still negative [\ldots] there is not enough evidence to conclude that Meta has bottomed out after its August 2025 decline.''}{}
\cond{+\,TimesFM2.5}{4.50}{``Probably not enough evidence to say META has clearly bottomed out in August 2025. [\ldots] The TimesFM forecast is only mildly positive from the last actual close of 738.194 to about 742.7 over the next 5 trading observations, which suggests stabilization, not a strong rebound.''}{}

\judgehdr{Common-evidence judge}
\cond{No-RAG}{2.50}{``If the stock stopped making lower lows and began consolidating after the August selloff, then the decline likely marked at least a near-term trough. But unless price action has clearly turned upward [\ldots] it is more accurate to describe the move as a potential bottom in progress [\ldots]''}{conditional framework; no August data cited}
\cond{RAG}{4.25}{``The stock fell from 790.00 on Aug.\ 12 to the 739 area by Aug.\ 21--29 [\ldots] the most defensible conclusion is possible bottoming, not confirmed bottoming.''}{correct, but inside a $\sim$8k-character answer}
}
\caption{GPT-5.4-mini responses to a Hybrid seed question under every evaluated condition. Notation as in Table~\ref{tab:qual-gpt54mini-fund}.}
\label{tab:qual-gpt54mini-hyb}
\end{table*}

% \subsection{Curated-Context Illustrations}

% Some qualitative examples use manually curated or precomputed context.
% To avoid conflating these examples with the formal Oracle condition, we refer
% to them as \emph{curated-context illustrations} unless their evidence is
% constructed by the exact Oracle procedure. These examples are qualitative and
% are not treated as formal Oracle evaluations.

% KEEP at most one Hybrid example and one Trading example.
% sectors." The final common-evidence analysis no longer supports that old
% headline interpretation.

% ============================================================
\sectionNoBarrier{Data Transparency, Responsible Use, Maintenance, and AI Disclosure}
\label{app:Dt_and_all}

\subsection{Data Transparency, Privacy, and Reproducibility}
\label{app:data-transparency}

FinTradeBench is constructed from public company-level financial information,
including SEC EDGAR filings and historical market data. The benchmark is not
constructed from private investor records, user conversations, material
non-public information, or personal financial data. Questions are written at
the company/ticker and historical-period level.

The released benchmark will include the 1,400-question collection, explicit
origin metadata, the human-validated Core400, the Core403 audit/exclusion log,
golden-indicator metadata, derived signals, frozen prompts, and evaluation
scripts. Raw source-data redistribution remains subject to the terms of the
original providers; where redistribution is restricted, reconstruction scripts,
ticker/date identifiers, and derived features should be released instead.

All reported final results are tied to a frozen benchmark version and
frozen model-output/judge files so that future changes to hosted APIs or source
data do not silently alter published numbers.

\subsection{Safety, Ethics, and Broader Impact}
\label{sec:safety}

FinTradeBench is a research benchmark for historical financial reasoning. It is
not a trading system, portfolio-management system, suitability engine, or
investment-advisory product. Benchmark performance should not be interpreted as
evidence that a model is safe or profitable for live financial deployment.

Real-world financial use would require independent numerical verification,
human oversight, risk controls, source attribution, uncertainty disclosure, and
appropriate regulatory/compliance review. Systems capable of producing fluent
financial narratives may also amplify misleading or manipulative claims;
therefore, high benchmark performance should not substitute for deployment
governance.

\subsection{Data Maintenance}
\label{sec:data-maintenance}

FinTradeBench will be released upon acceptance and maintained as a versioned benchmark. Each release will
freeze the question set, reference answers, golden indicators, derived signals,
point-in-time metadata, prompts, judge version, and scoring scripts. Corrections
to ambiguous questions, ticker-period mappings, evidence, or golden indicators
will be documented in a changelog while prior releases remain available for
reproducibility.

Future versions may broaden market coverage, extend the temporal window, and
add additional evidence sources. Such expansions will be released as new
versions rather than silently modifying \emph{Core400}.

\subsection{Disclosure of Substantive AI Assistance}
\label{sec:ai_disclosure}
In accordance with the ACL and ARR Policy on Substantive AI Assistance, we disclose the ways in which artificial intelligence tools were employed during the preparation of this work:

    \smartparagraph{Writing and Editorial Assistance:} We used large language models as an editorial assistant to polish draft prose, shorten complex phrasing (e.g., streamlining the abstract to meet length requirements), and improve readability in select sections. The authors conceived, verified, and finalized all substantive technical arguments, mathematical formalisms, experimental analyses, and conclusions.
    
    \smartparagraph{Code Development and Debugging:} The authors used AI coding assistants interactively to support pipeline implementation, specifically to debug data ingestion routines, troubleshoot Slurm batch dispatch scripts, and diagnose edge cases in evaluation parsing code. All generated scripts were reviewed, tested, and audited against ground-truth outputs by the research team prior to benchmark execution.

\smartparagraph{Figure and Visual Asset Generation:} We generated or edited stylized icons for system architecture overviews (e.g., Figure~\ref{fig:calibration_and_scaling} and Figure~\ref{fig:rag_architecture}) with generative AI assistance and integrated them into human-designed vector schematics.
    
    \smartparagraph{Experimental Evaluation Artifacts:} We clarify that the use of language models as research objects (e.g., the evaluated baseline models detailed in Table~\ref{tab:model-implementation}), candidate generators in the calibration pipeline (\S\ref{sec:benchmark-design}), and Claude Sonnet~4.5 as an automated evaluation judge (\S\ref{sec:experimental-setup}, \S\ref{app:evaluation-prompts}) constitute core scientific artifacts and methodology of this study rather than external authorial assistance. The automated judge configuration was systematically aligned with human expert rubrics and independently validated against blinded human evaluations (\S\ref{subsec:results-human-validation}).

\section{Annotation Instruments}
\label{app:annotation-instruments}

This section reproduces the instruments used in the four human-review stages
of \S\ref{app:validation}: the development-time annotation rubric
(Table~\ref{tab:human-rubric-calibration}), the construction-time fatal-flaw
screening guide (Table~\ref{tab:fatal-flaw-screen}), the Core403 item-level
held-out response-scoring rubric with its pairwise fields
(Table~\ref{tab:heldout-human-rubric}), and the Core403 item-level audit
rubric with its record fields (Table~\ref{tab:core403-audit-rubric}).

\begin{table*}[p]
\centering
\footnotesize
\renewcommand{\arraystretch}{1.0}
\begin{tabular}{p{3.6cm} p{2.0cm} p{9.0cm}}
\toprule
\textbf{Criterion} & \textbf{Scale} & \textbf{Reviewer instruction / anchors} \\
\midrule

Audit Validation Agreement
& 0--1
& Does the independent human review agree with the automated numerical
audit's \texttt{is\_numerically\_accurate} flag?
\textbf{1}: Agree---the audit conclusion is correct.
\textbf{0}: Disagree---the audit missed an error or incorrectly flagged
a correct claim. \\

\addlinespace[1pt]
Factual \& Numerical Accuracy
& 1--5
& Based on the reviewer's own assessment, together with the numerical
audit, assign the final accuracy score.
\textbf{5}: 100\% correct.
\textbf{1}: Contains significant, misleading numerical errors. \\

\addlinespace[1pt]
Completeness \& Context
& 1--5
& Does the answer fully address the question and correctly use and
contextualize the golden indicators?
\textbf{5}: Excellent; uses required metrics in a deep, integrated
analysis.
\textbf{1}: Superficial; misses most required metrics or necessary
context. \\

\addlinespace[1pt]
Relevance \& Utility
& 1--5
& Is the included information relevant, and does the response avoid
fluff or potentially misleading tangents?
\textbf{5}: High precision; no fluff or harmful information.
\textbf{1}: Cluttered with irrelevant or misleading content. \\

\addlinespace[1pt]
Clarity \& Rationale
& 1--5
& Is the answer clear and well structured, and does it explain its
reasoning?
\textbf{5}: Exceptionally clear and well reasoned.
\textbf{1}: Confusing, poorly written, or reasoning is opaque. \\

\bottomrule
\end{tabular}
\caption{\small{\textbf{Development-time human annotation rubric.}
This rubric was used during calibration of the construction-time
automated evaluator. Reviewers additionally had optional fields for a
human-written golden answer and free-text notes. Intermediate values
between the endpoint anchors were available on the 1--5 scale but were
not assigned additional verbal anchors in the original annotation
instructions.}}
\label{tab:human-rubric-calibration}
\end{table*}

\begin{table*}[p]
\centering
\footnotesize
\renewcommand{\arraystretch}{1.0}
\begin{tabular}{p{4.0cm} p{8.0cm} p{2.5cm}}
\toprule
\textbf{Screening category} & \textbf{Operational interpretation} &
\textbf{Action} \\
\midrule

Wrong entity / company
& Response analyzes a different company, ticker, or proxy rather than
the entity requested by the question.
& Regenerate \\

\addlinespace[1pt]
Wrong period / ticker--period mismatch
& Response is inconsistent with the requested historical entity or
period.
& Regenerate \\

\addlinespace[1pt]
Unsupported material claims
& Response contains material claims that are not supported by the
benchmark evidence available for the item.
& Regenerate / escalate if ambiguous \\

\addlinespace[1pt]
Conclusion inconsistent with supplied evidence
& Final conclusion conflicts materially with the information or
indicators associated with the item.
& Regenerate / escalate if ambiguous \\

\addlinespace[1pt]
Incomplete answer
& Response fails to answer a required part of the question; for example,
a cross-company question is answered using only one company.
& Regenerate \\

\addlinespace[1pt]
Abrupt truncation / sudden break
& Generation terminates before completing the reasoning or answering the
main question.
& Regenerate \\

\bottomrule
\end{tabular}
\caption{\small{\textbf{Construction-time fatal-flaw screening guide.}
This stage was a release-quality screen rather than an independent
financial-reasoning score. Responses exhibiting catastrophic failures
were regenerated and retained only after passing the screen; ambiguous
cases were escalated for financial-expert review.}}
\label{tab:fatal-flaw-screen}
\end{table*}

\begin{table*}[p]
\centering
\scriptsize
\renewcommand{\arraystretch}{1.0}
\setlength{\tabcolsep}{3pt}
\begin{tabular}{@{}p{2.2cm} p{2.45cm} p{2.45cm} p{2.45cm} p{2.45cm} p{2.45cm}@{}}
\toprule
\textbf{Dimension} & \textbf{5} & \textbf{4} & \textbf{3} & \textbf{2} & \textbf{1} \\
\midrule
Factual \& Numerical Accuracy
& Fully supported by supplied evidence; calculations and entity/time references are sound.
& Mostly correct with a minor non-material issue.
& Mixed: useful core analysis but meaningful errors or unsupported details.
& Substantial factual or numerical problems.
& Materially wrong, contradicted by evidence, or wrong entity/period. \\
\addlinespace[1pt]
Completeness \& Context
& Fully answers the question and covers evidence needed for the conclusion.
& Substantively complete with a minor omission.
& Partially complete; misses a meaningful aspect.
& Major omissions.
& Largely non-responsive. \\
\addlinespace[1pt]
Relevance \& Utility
& Directly addresses the question with useful synthesis.
& Relevant with small digressions.
& Useful but noticeably generic, padded, or indirect.
& Weakly relevant.
& Mostly irrelevant or non-responsive. \\
\addlinespace[1pt]
Clarity \& Rationale
& Concise, clear, and well justified.
& Clear with minor presentation issues.
& Understandable but wordy, repetitive, or mechanical.
& Difficult to follow or poorly justified.
& Confusing or disjointed. \\
\midrule
\multicolumn{6}{@{}l}{\textit{Pairwise fields}} \\
\multicolumn{2}{@{}p{4.7cm}}{Pairwise preference (A / B / Tie)}
& \multicolumn{4}{p{10.2cm}}{Entered only after all four ratings were completed independently for both responses.} \\
\multicolumn{2}{@{}p{4.7cm}}{Pairwise confidence (1--5)}
& \multicolumn{4}{p{10.2cm}}{Integer confidence in the preference.} \\
\multicolumn{2}{@{}p{4.7cm}}{Pair notes (free text)}
& \multicolumn{4}{p{10.2cm}}{Optional.} \\
\bottomrule
\end{tabular}
\caption{\small{\textbf{Held-out human response-scoring rubric and pairwise fields.}
Reference metrics were presented only as helpful context and were
explicitly \emph{not} treated as a mandatory checklist. For subjective
investment-style questions, a conclusion differing from another
response could still receive a high score when it was supported by the
supplied evidence and appropriately qualified.}}
\label{tab:heldout-human-rubric}
\end{table*}

\begin{table*}[p]
\centering
\footnotesize
\renewcommand{\arraystretch}{1.0}
\begin{tabular}{p{3.2cm} p{3.3cm} p{8.1cm}}
\toprule
\textbf{Dimension} & \textbf{Labels} & \textbf{Rubric definition} \\
\midrule

Question Clarity
&
Pass; Minor issue; Major issue
&
\textbf{Pass}: clear and answerable at the historical point in time.
\textbf{Minor issue}: small wording or precision issue that does not
change the task.
\textbf{Major issue}: ambiguous, inconsistent, based on an unanswerable
or false premise, or not answerable from benchmark evidence. \\

\addlinespace[1pt]
Reference Answer Correctness
&
Correct; Minor correction; Major incorrect; Unverifiable
&
\textbf{Correct}: material claims and conclusion are supported.
\textbf{Minor correction}: main conclusion remains valid but a small
correction is needed.
\textbf{Major incorrect}: material claims or conclusion conflict with
evidence.
\textbf{Unverifiable}: supplied evidence is insufficient.
For subjective investment-style questions, the reviewer was instructed
not to require a unique recommendation; the conclusion instead had to be
defensible and appropriately qualified. \\

\addlinespace[1pt]
Evidence Support
&
Full; Partial; Unsupported; Evidence missing
&
\textbf{Full}: evidence is sufficient for the material claims and
conclusion.
\textbf{Partial}: supports the main conclusion but not all secondary
claims.
\textbf{Unsupported}: material claims are unsupported or contradicted.
\textbf{Evidence missing}: no usable canonical evidence. \\

\addlinespace[1pt]
Golden-Indicator Adequacy
&
Adequate; Minor issue; Missing critical; Incorrect; N/A
&
\textbf{Adequate}: captures the key evidence.
\textbf{Minor issue}: basically correct with a non-critical omission or
extraneous item.
\textbf{Missing critical}: an essential indicator is absent.
\textbf{Incorrect}: materially mismatched.
\textbf{N/A}: an indicator checklist is not meaningful for the item.
Golden indicators were explicitly treated as a minimal evidence set,
not a mandatory wording checklist. \\

\addlinespace[1pt]
Entity / Time Alignment
&
Pass; Unclear; Fail
&
\textbf{Pass}: entity and cutoff are aligned and no post-cutoff evidence
is required.
\textbf{Unclear}: alignment cannot be verified confidently.
\textbf{Fail}: wrong entity or historical period, or future information
is required or used. \\

\midrule
\multicolumn{3}{@{}l}{\textit{Record fields}} \\
\addlinespace[1pt]
Fatal flaw
& No; Yes
& If Yes, the reviewer was required to select a fatal-flaw type. \\
\addlinespace[1pt]
Fatal-flaw type
& Wrong entity/ticker; wrong period or future leakage; ambiguous or
unanswerable question; reference answer incorrect or unsupported; golden
indicators incorrect; canonical evidence missing; duplicate; Other
& Used to make exclusion and adjudication decisions auditable rather than
silently modifying problematic items. \\
\addlinespace[1pt]
Overall disposition
& Accept; Accept minor edit; Needs adjudication; Exclude
& Final item-level decision. \\
\addlinespace[1pt]
Human notes
& Free text
& Optional for \textbf{Accept}; strongly recommended for all other
dispositions. \\
\bottomrule
\end{tabular}
\caption{\small{\textbf{Core403 item-level human quality-audit rubric and record fields.}
The unit of annotation is a benchmark item, not a model response.}}
\label{tab:core403-audit-rubric}
\end{table*}